\documentclass[5p, authoryear]{elsarticle}
\usepackage[colorlinks]{hyperref}

\usepackage{amssymb}

\newcommand{\gaia}{\textit{Gaia}}

\journal{New Astronomy Reviews}

\begin{document}

\sloppy

\begin{frontmatter}



\title{Galactic Archaeology with \gaia}


\author[1]{Alis J. Deason}
\ead{alis.j.deason@durham.ac.uk}
\address[1]{Institute for Computational Cosmology \& Centre for Extragalactic Astronomy, Department of Physics, Durham University, South Road, Durham, DH1 3LE, UK}
\author[2]{Vasily Belokurov}
\ead{vasily@ast.cam.ac.uk}
\address[2]{Institute of Astronomy, University of Cambridge, Madingley Road, Cambridge CB3 0HA, UK}

\begin{abstract}
The \gaia\ mission has revolutionized our view of the Milky Way and its satellite citizens. The field of Galactic Archaeology has been piecing together the formation and evolution of the Galaxy for decades, and we have made great strides, with often limited data, towards discovering and characterizing the subcomponents of the Galaxy and its building blocks. Now, the exquisite 6D phase-space plus chemical information from \gaia\ and its complementary spectroscopic surveys has handed us a plethora of data to pore over as we move towards a quantitative rather than qualitative view of the Galaxy and its progenitors. We review the state of the field in the post-\gaia\ era, and examine the key lessons that will dictate the future direction of Galactic halo research.
\end{abstract}




\end{frontmatter}


\section{Introduction}
\label{sec:intro}

Galaxies like our own Milky Way digest and destroy hundreds of lower-mass systems over their lifetimes. Some of these `subhaloes' have no stars at all, and just contribute to the dominating dark matter halo. Most of the more massive systems do contain stars, and their contributions create an extended halo of stars surrounding the Galaxy \citep[e.g.][]{BJ2005, Cooper2010}. The long orbital timescales in the halo lead to the destroyed stellar systems reflecting their initial progenitor orbits, especially in Energy and Angular momentum \citep[][]{HdZ2000}. Thus, the `stellar halo' is a fossil record of the past accretion history of the Milky Way, and as such has been studied extensively for several decades \citep[see earlier reviews by e.g.][]{Belokurov2013, Bland-Hawthorn2016, Helmi2020}.

The stellar halo not only encodes an archaeological record of the Milky Way's assembly history, but it also provides a tool to study (now destroyed) high redshift dwarf galaxies. Moreover, the great extent (out to $> 100$ kpc) of the stellar halo makes it one of our best tracers of the dark matter halo. In past years, our study of the Galactic halo has mainly been hampered by limitations in the data. Our sample sizes have been too small, only limited areas of the sky have been covered, we have incomplete velocity information, hardly any chemical information, etc, etc. However, these limitations (excuses?) are no longer, as we are now deep into the era of \gaia\ --- the most ambitious Galactic science mission of all time.

\gaia\ is an astrometric mission with the lofty goal of producing an all-sky phase-space map of the Galaxy \citep{Gaia_mission}. Positions on the sky, parallax, and proper motions are measured for all stars down to a limiting magnitude of $G \approx 21$. In addition, a (bright) subset of stars has measured radial velocities and stellar parameters (such as effective temperature and surface gravity, see \citealt{gaia_rvs} for details of the parametrization of the \gaia\ RVS data). As of June 2024, we are now on the third \gaia\ data release with the fourth release likely arriving in 2025/26. Each subsequent data release provides data over a larger time baseline, and hence more precise astrometry, as well as additional data products, such as variability information and source classifications. To compensate for the bright limit for accurate \gaia\ spectroscopic measurements, several wide-field spectroscopic surveys have (or will) complemented the \gaia\ data to provide radial velocities and chemical measurements, e.g. SDSS/SEGUE \citep[][]{Yanny2009}, LAMOST \citep[][]{Cui2012}, APOGEE \citep[][]{Majewski2017}, GALAH \citep[][]{Buder2021}, H3 \citep[][]{Conroy2019}, DESI \citep[][]{Cooper2023}, WEAVE \citep[][]{Jin2023}, 4MOST \citep[][]{deJong2019}. These auxiliary datasets are particularly important for the halo stars, which traverse to large distances and are thus often relatively faint. The era of \gaia\ is truly data-heavy! The fruits of this labour will last for many years to come, and we are only at the beginning stages of our digestion and understanding of what this detailed mapping of the Milky Way contains.

The purpose of this review is to provide the current state of understanding of the Galactic halo and the revelations and surprises from the \gaia\ data boon. This review does not intend to provide a complete history and consensus, as the number of papers in this field has exploded in recent years. Instead, we focus on the main findings, in our humble opinion, and look ahead to the future of this field over the coming decade.
There are several articles in this special issue focused on ``\gaia\ discoveries''. As such, the in-depth exploration of stellar streams, globular clusters and Galactic dynamics with {\it Gaia} is covered elsewhere and we refer the interested reader to related reviews on those topics.

\subsection{A dominant stellar halo progenitor?}

One of the main aims of the \gaia\ mission is to piece together the building blocks of the Galactic halo. This ambitious goal was perhaps mainly expected to reveal a plethora of streams, clouds, and debris from a multitude of progenitors \citep[see e.g.][]{HdZ2000}. However, one of the clearest and most decisive discoveries from the early \gaia\ data releases was that in fact the (inner, $r \lesssim 20$ kpc) stellar halo is dominated by \textit{one} dwarf galaxy progenitor (see Section \ref{sec:gse}). As we will discuss below, in hindsight this was not a surprising discovery!

\subsubsection{Predictions from $\Lambda$CDM}

The fundamental mass function of dark matter haloes ($dn/dM$ --- i.e. how many haloes there are of different mass) follows a power-law profile $dn/dM \propto M^{-1.9}$ \citep[e.g.][]{Moore1999,Jenkins2001, Gao2004, Tinker2008}. Thus, the general expectation is that there are many more smaller lumps than bigger lumps --- a prediction that lies at the heart of the hierarchical $\Lambda$CDM paradigm. A consequence of the mass function is that haloes like our own Milky Way engulf many more smaller mass objects than more massive ones over time. For example, cosmological $N$-body simulations \citep{Fakhouri2010} predict that a Milky Way-mass halo ($10^{12}M_\odot$) typically undergoes $N \sim 30$ minor mergers (mass-ratio between 1:3 and 1:100) and $N \sim 3$ major mergers (mass-ratio $>1:3$) since redshift $z=10$.

\begin{figure*}
\includegraphics[width=\linewidth]{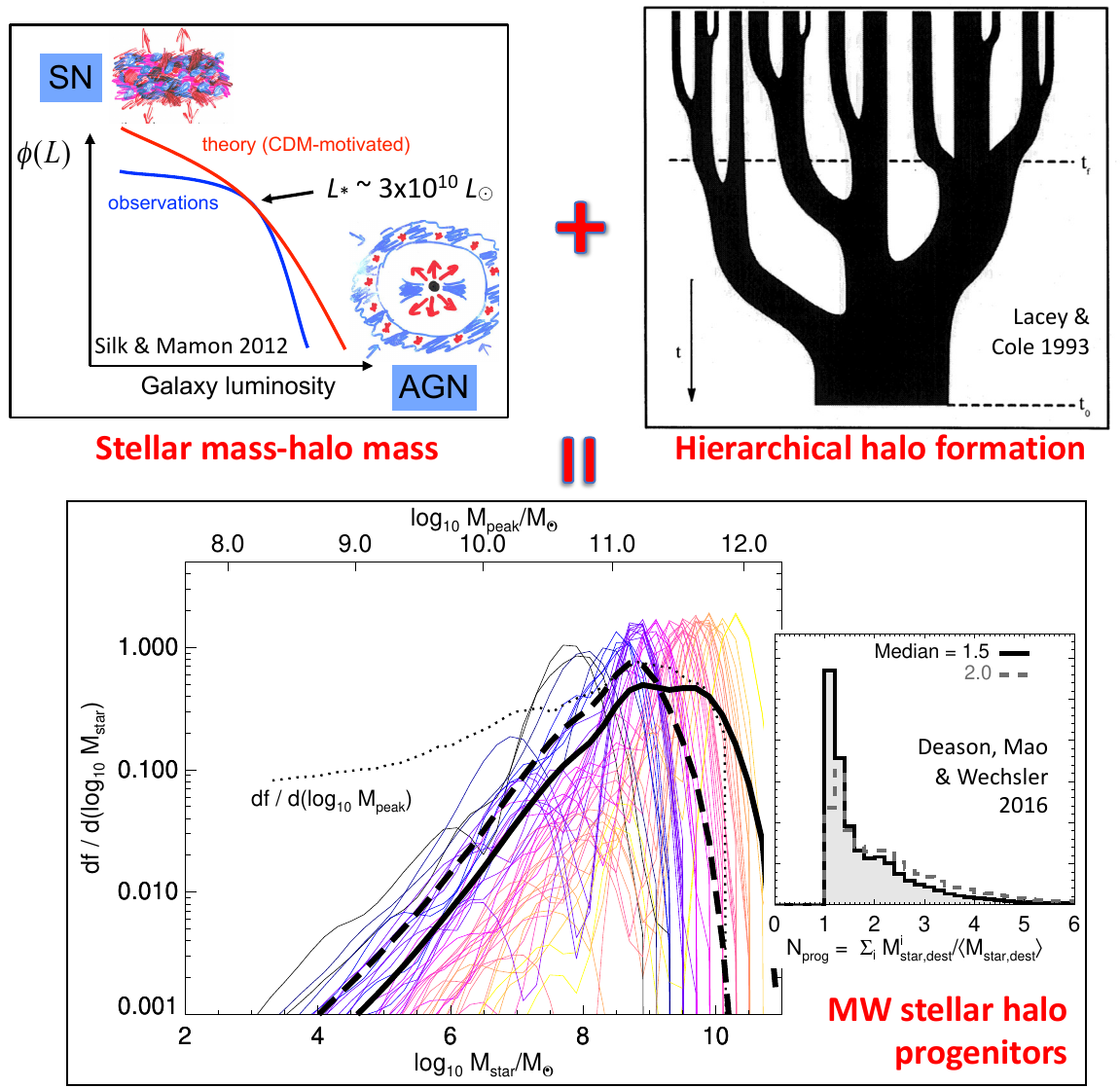}
\caption[]{\textit{Top Left:} Galaxy luminosity function --- the role of feedback \citep{Silk2012}. \textit{Top Right:} Schematic representation of a merger tree \citep{Lacey1993}. Time increases from top to bottom (present time $=t_0$, formation time $=t_f$), and the width of the branches represents the individual halo masses. \textit{Bottom panel:} The mass-weighted distribution of destroyed dwarfs that contribute to the stellar haloes of Milky Way-mass systems \citep{Deason2016}.  The thin coloured lines are for individual haloes, and the thick black line is the average.  The thick dashed line shows the average profile for `quiescent' haloes, which have not undergone a major merger since $z=2$. The dotted line shows the mass-weighted contribution of \textit{dark matter} haloes, which have a much flatter distribution. This profile differs from the stellar mass owing to the steep decline in the stellar mass--halo mass relation at low halo masses ($\lesssim 10^{11}M_\odot$). \textit{Bottom insert:} The (mass-weighted) number of dwarf progenitors that contribute to the total accreted stellar mass. Typically, $1-2$ destroyed dwarfs deposit the majority of accreted stellar mass onto the host halos [reproduced from \citealt{Silk2012, Lacey1993, Deason2016}].}
\label{fig:SHprog}
\end{figure*}

 When galaxies enter the equation, things become more complicated as the efficiency of star formation is not equal amongst haloes of different mass. The inefficiency of massive galaxies in the centres of clusters (mainly due to AGN feedback, e.g. \citealt{Croton2006,Hopkins2006,Fabian2012}) and dwarf galaxies (supernova feedback plus reionization, e.g. \citealt{Dekel1986,Efstathiou1992,Bullock2000,Pontzen2012}) at forming stars means that galaxies like the Milky Way lead the pack --- we live in the most efficient star-making galaxy environment \citep[see e.g. Figure 2 of][]{Wechsler2018}. The relation between stellar mass and dark matter mass (aptly known as the `stellar mass-halo mass' or SMHM relation), has important consequences for the growth of stellar haloes through the accretion of lower-mass objects. The SMHM has a very steep decline at low halo masses ($\lesssim 10^{11}M_\odot$, see e.g. Figure 6 of \citealt{Bullock2017}). Thus, although there may be hundreds of dark matter subhaloes accreted by the galaxy, if these have very few stars (or no stars at all!) then they have an insignificant contribution to the accreted stellar halo. This idea was recognized in \cite{Purcell2007}, who used analytic prescriptions for dark matter halo growth and galaxy formation models, and later by \cite{Deason2016} using $N$-body simulations and analytic SMHM relations. In short, the contribution of \textit{stellar mass} to the galactic stellar halo is dominated by a small number (1-2) of relatively massive dwarfs ($M_{\rm DM} \sim 10^{11}M_\odot$, $M_{\rm star} \sim 10^{8-9}M_\odot$). The contribution of lower mass dwarfs, while potentially important for dark matter growth, is relatively insignificant for the total stellar halo mass (see Fig. \ref{fig:SHprog})

These ideas are fleshed out further in more sophisticated stellar halo models \citep[e.g.][]{BJ2005, deLucia2008, Cooper2010, Monachesi2019, Fattahi2020, Horta2023}, where the spatial distribution of accretion events can be explicitly tracked. Here, the dominance of a small number of massive progenitors is retained, but they are generally confined to the \textit{inner} ($\lesssim 20-50$ kpc) halo, where they have been rapidly dragged inwards towards the centre of the host halo via dynamical friction. The small number of stellar halo building blocks and a wide range of merger configurations lead to a pronounced diversity of the resulting stellar halo density distributions \citep[see e.g.][]{Cooper2010}, in contrast to the universality of the dark matter halo densities \cite[][]{Navarro1996}. 

The low binding energy of lower mass contributors means that they can be splayed out into the outer halo in the form of streams and overdensities. Additionally, shells from massive satellites disrupted on more radial orbits can reach large galacto-centric distances as well \citep[][]{Hendel2015,Pop2018}. Thus, the outer halo can often be populated by a wider range of progenitors, and is less likely to be dominated by a small number of contributors (see. e.g. Figure 7 in \citealt{Fattahi2020}).

In summary,  the combination of the accretion of dark matter subhaloes plus the steepening of the SMHM relation leads to an intriguing scenario for the stellar haloes of Milky Way-mass galaxies. By stellar mass contribution \textit{alone} only a few progenitors really matter. So, in the real Galaxy, where are they?

\subsubsection{Pre-\gaia\ clues from observations}
\label{sec:pregaia}

\begin{figure*}
\includegraphics[width=\linewidth]{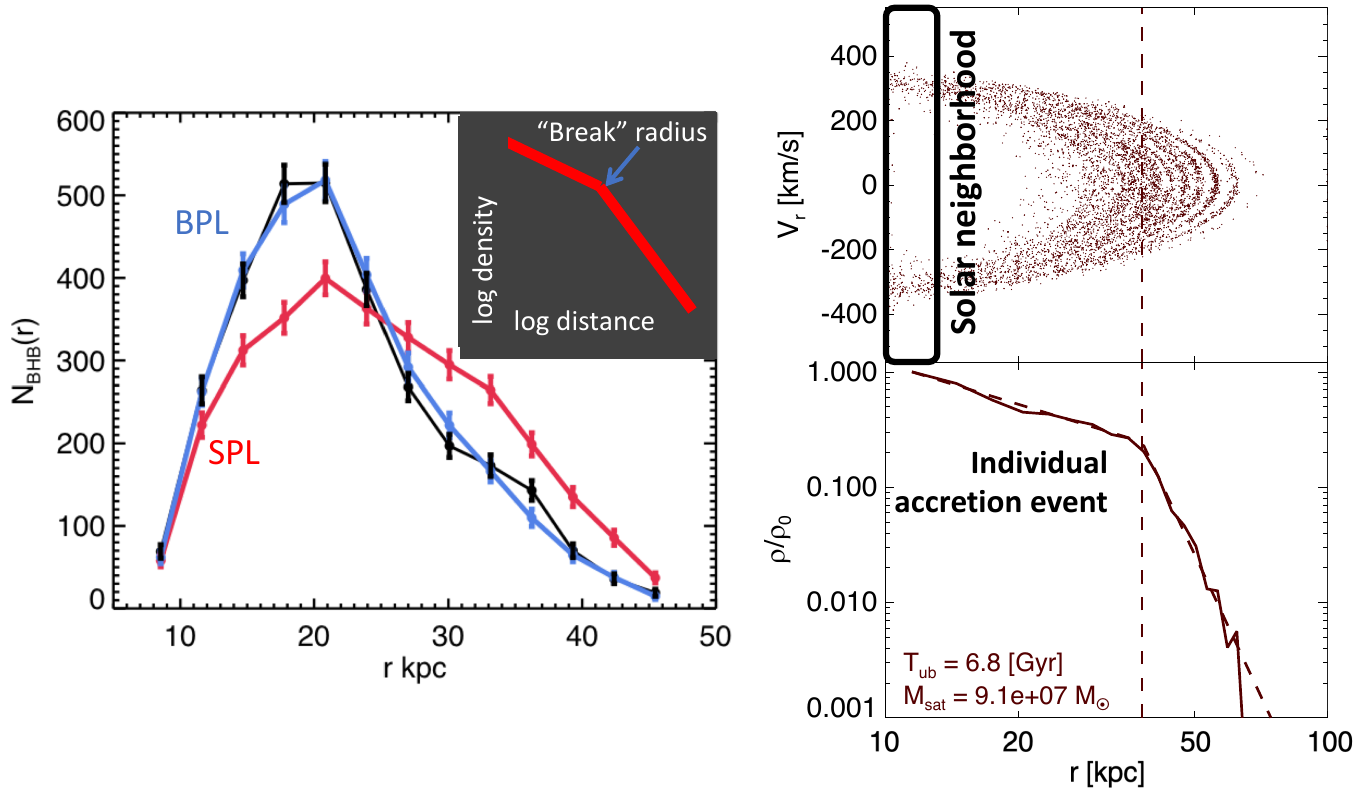}
\vspace{-25pt}
\caption[]{\textit{Left:} Number counts of BHB stars from SDSS data. The blue line is the best-fit broken power-law, and the red line is the best-fit single power-law \citep{Deason2011}. \textit{Right:} Example of an individual accretion event from the \cite{BJ2005} simulation suite. The panels show the radial velocity (top) and density profile (bottom) as a function of radius. The dashed line indicates the `break-radius' of the best-fit broken power law that coincides with the apocentre of the debris. The black box indicates the `local' region of the debris, which has a high density and significant radial velocity $v_r$ [reproduced from \citealt{Deason2013a}].}
\vspace{-12pt}
\label{fig:broken_dens}
\end{figure*}

\textit{Number counts:}
A fundamental `zeroth order' measurement of the Galactic stellar halo is to simply count the number of stars. As such, the characterization of the stellar halo density profile has been studied for several decades \citep[e.g.][]{Hartwick1987, Preston1991, Robin2000, Yanny2000, Juric2008, Deason2011, Xue2015}. In recent years, however, it has been recognized that this is not simply a counting exercise, and the form of the stellar density profile in fact encodes key information pertaining to the assembly history of the halo \citep[e.g.][]{BJ2005, Cooper2010}. Early work limited to the inner halo ($\lesssim 20$ kpc) described the stellar density as a simple power-law profile with radius ($\rho \propto r^{-\alpha}$, \citealt[e.g.][]{Robin2000, Yanny2000, Chen2001, Siegel2002, Juric2008}). When the outer parts of the halo began to be mapped (largely thanks to wide-area photometric surveys such as SDSS), it was quickly recognized that this profile does not continue to larger radii. Indeed, \cite{Deason2011} (also see \citealt{Sesar2011}) found that the stellar halo is better characterised by a `broken' power-law with a significant steepening in the power-law (from $\alpha \sim 2.5$ to $\alpha \sim 4.5$) beyond $r \sim 25$ kpc \citep[note that glimpses of the stellar halo break were seen earlier, see e.g.][]{Saha1985, Watkins2009}. In a follow-up paper, \cite{Deason2013a} provided a possible explanation for such a `broken' profile (see Fig. \ref{fig:broken_dens}). They used $N$-body simulations to argue that the debris from \textit{individual} stellar halo progenitors often follow a broken power-law, where the break radius signifies the last apocentre of the orbit. In this scenario, a broken profile in the \textit{overall} halo could imply two possible outcomes: (1) the stellar halo is dominated by one progenitor, or (2) several stellar halo progenitors all conspired to be accreted at the same time, with similar apocentres. The authors argued that the former scenario was more likely, owing to the contrived nature of the latter. However, the question remained, if the stellar halo was indeed dominated by one massive progenitor, how did we not know this already, and where is the additional observational evidence beyond the number counts?

\textit{Chemistry:}
Early studies of RR Lyrae and globular clusters (GCs) indicated that the stellar halo stars tend to have lower metallicity compared to the rest of the Galaxy \citep[see e.g.][]{Preston1959,Hartwick1976,Searle1978}. This is expected if the stellar halo is built from less massive objects that are less chemically evolved than more massive systems like the host Milky Way (i.e. following the mass-metallicity relation, \citealt[e.g.][]{Kirby2013}). However, measurements of the halo metallicity distribution function (MDF) or even its average metallicity have typically been fraught with selection biases and unknown levels of disc contamination \citep[][]{ELS1962,Pagel1975,Norris1986,Beers1992}. Thus, while the bulk of the stellar halo is relatively metal-poor ($\langle \mathrm{[Fe/H]}\rangle \sim -2.0$ to $-1.0$, \citealt[][]{Ryan1991,Ivezic2008, Carollo2010, Xue2015, Conroy2019}), whether or not this metallicity reflects a dominant massive progenitor remains inconclusive. 

A more revealing chemical analysis comes from the study of $\alpha$ elements in addition to Fe. The $\alpha$ elements are mostly produced in massive stars and dispersed into the interstellar medium by type-II (core collapse) supernovae (SNe) over short timescales (of the order of Myr) while Fe can be contributed in similar amounts by both type-II and type-Ia SNe \citep[see e.g.][]{Kobayashi2006,Kobayashi2020}. The latter occur over $\sim$Gyr timescales \citep[note however that the SNe Ia delay time distribution extends to much shorter timescales below 0.1 Gyr, see e.g.][]{Maoz2012}, and this distinction in dominant delay times leads to a characteristic `knee' in the [$\alpha$/Fe] vs [Fe/H] chemical plane \citep[][]{Tinsley1979}. Specifically, the onset of type-Ia SNe signifies a downturn from a nearly constant `plateau' in [$\alpha$/Fe] when more Fe is produced compared to $\alpha$ \citep[see examples of such behaviour in chemical evolution models of][]{Andrews2017,Spitoni2017,Weinberg2017}. Thus, by studying the [$\alpha$/Fe] vs. [Fe/H] plane we can probe the star formation history (SFH) of a stellar system. In particular, the location of the `knee' may be directly related to mass as more massive systems are able to retain and build up metals before the onset of type Ia SNe --- i.e. their `knee' is more metal-rich \citep[e.g.][]{Hendricks2014,deBoer2014, Mason2024}. While observational evidence for `knees' and their connection to the galaxy's mass remains mixed \citep[see e.g.][]{Kirby2011,Nidever2020}, different dwarf galaxies do display distinct behaviour in the $\alpha$-[Fe/H] plane \citep[][]{Tolstoy2009,Hasselquist2021}. Although the above description of the chemical evolution of a dwarf galaxy is clearly oversimplified, this chemical insight is of particular interest for the stellar halo, where we are aiming to disentangle the individual ingredients of this aggregated soup. Moreover, chemistry can also be used to assess whether or not the halo substructures are truly remnants of disrupted dwarfs \citep[e.g.][]{Horta2023b}. 

In the influential work by \cite{Venn2004} the authors compared chemical abundances of halo stars to surviving dwarf spheroidal satellite galaxies. They found that their chemical signatures are remarkably different, perhaps ruling out the common lore that the stellar halo is built up from the debris of many low-mass galaxies! However, importantly, the authors note that mergers with \textit{higher} mass dwarf galaxies cannot be ruled out, as the [$\alpha$/Fe] vs [Fe/H] plane for the halo stars exhibits a relatively metal-rich knee (although less metal-rich than the disc stars). An additional conundrum is that the chemistry of the halo stars also does not resemble the Large Magellanic Cloud (LMC) or Sagittarius (Sgr), which are two of the three most massive dwarf galaxies ($M_{\rm star} \sim 10^{8-9} M_\odot$) in the Milky Way. The $\Lambda$CDM simulation community was swift to clear this up: the stellar halo models predict that the destroyed dwarf progenitors were accreted early, and thus have star formation histories different to the late-time Sgr and LMC type objects that can continue to form stars \citep{Font2006}. \cite{Font2006} showed explicitly how the $\Lambda$CDM accretion plus chemical evolution models can match the Milky Way data. In summary, from the chemical perspective, the dominance of early, massive dwarf galaxy progenitors seems to fit the bill.

A final intriguing discovery in the chemical analysis of halo stars came from the work by \cite{Nissen2010}. These authors used precise abundance ratios to show that there are two distinct halo populations in the solar neighbourhood. One `high-$\alpha$' sequence can perhaps be attributed to an `in-situ' halo component (see Section \ref{sec:insitu}), while the other `low-$\alpha$' sequence resembles the classical `accreted' component. However, what was remarkable about this analysis was just how narrow and ordered the chemical sequence of the accreted stars was in the [$\alpha$/Fe] vs. [Fe/H] plane. Presumably, a hodge-podge of several different progenitors with different masses and SFH would produce a messier outcome. Alternatively, at least in the solar neighourhood, there really is one progenitor that dominates the accreted halo.

 \textit{Kinematics:}
 The (local) stellar halo is kinematically hot with a high overall velocity dispersion  ($\sigma \sim 100$ km/s) and little net rotation reflecting the messy build-up of a stellar halo relative to the `ordered' state of the Galactic disc \citep[e.g.][]{Norris1986,Chiba2000}. As the observational samples grew, the first hints appeared that the orbital motion in the nearby halo was not isotropic,  and in e.g. spherical polars,  the velocity dispersion in the radial direction was larger than in the other two tangential ones \citep[][]{Bahcall1986,Freeman1987,Chiba1998}. It is customary to characterise the \textit{shape} of the velocity ellipsoid using the anisotropy parameter: $\beta = 1-\left[(\sigma^2_\theta+\sigma^2_\phi)/2\sigma^2_r\right]$ \citep[][]{BT_book2008}. \cite{Smith2009} (see also \citealt{Bond2010}) used large samples of nearby metal-poor main sequence dwarf stars to show that the halo velocity ellipsoid is indeed remarkably anisotropic, with the radial component $\sigma_{r}$ dominant over the tangential components. \cite{Smith2009} and \cite{Bond2010} found $\beta \sim 0.7$ for the local samples --- did this reflect the orbital make-up of the stellar halo as a whole?

In the pre-\gaia\ era, accurate space velocity measurements for distant objects were hard to come by. As a result, outside of the Solar neighourhood, the view of the halo's orbital properties remained rather blurred. For Galactic GCs it is possible to beat down the proper motion errors by averaging over large numbers of member stars. Thus, through early heroic efforts first estimates of the Milky Way's GC orbital properties were obtained \citep[see e.g.][]{Gnedin1997,Dinescu1999}. These studies found the GC velocity ellipsoid to be broadly consistent with that measured using stellar kinematics in the Solar neighourhood. In addition, \citet{Dinescu1999} found systematic --- albeit sometimes rather subtle --- differences in the orbital properties of the GCs as a function of their age, chemistry, and location in the Galaxy. For example, in their sample, the clusters' orbital eccentricity exhibited dependence on metallicity, with the metal-poorer GCs reaching higher eccentricities and the metal-richer ones tending towards more circular orbits.

For individual distant halo stars, no useful proper motion measurements were available before \gaia\ and therefore the properties of the halo velocity ellipsoid outside of the solar neighbourhood were teased out from the distribution of the line-of-sight velocity measurements under the assumptions of relaxation and symmetry. Using models based on distribution functions \citep[similar to those described in][]{Wilkinson1999} and data from the Sloan Digital Sky Survey (SDSS), \citet{Deason2012} showed that in the range of galactocentric distances of $15<r$(kpc)$<40$, the kinematics of Blue Horizontal Branch (BHB) stars favoured a stellar halo anisotropy of  $\beta\approx0.5$. Thus, compared to the vicinity of the Sun, on scales of tens of kpc, the stellar halo's orbital anisotropy appeared less radially biased. This observation was soon confirmed by \citet{Kafle2012} who additionally reported hints of drops in the radial anisotropy profile. As a demonstration of the power of space-based astrometry and a preview of \gaia's capabilities, the Hubble Space Telescope was used to measure the 3D kinematics of a small number of stars in the distant halo by \citet{Deason2013b} and \citet{Cunningham2016}. Interestingly, these studies measured lower values of $0<\beta<0.3$ beyond $r\approx20$ kpc.

Were these early measurements of the stellar halo's orbital anisotropy consistent with models of galaxy formation, and in particular, was such a high value of $\beta$ in the solar neighbourhood expected? For example, if multiple low-mass progenitors are accreted over a prolonged time with a wide range of orbital properties, then presumably a more isotropic velocity ellipsoid is expected. This is indeed the case in simulations, where $\beta$ starts relatively low in the centres of halos $\beta<0.3$ and rises up to $0.5<\beta<1$ at the virial radius \citep[][]{Abadi2006,Sales2007,Debbatista2008,Rashkov2013}. From these trends, values of up to  $\beta\approx0.5$ are generally expected in the halo near the Sun.

A very different orbital structure forms in the stellar halo in the series of experiments discussed in the visionary work by \cite{Amorisco2017}. In their toy $N$-body models, \textit{massive} accreted satellites lose most of their angular momentum in a rapid tidal disruption, driven in part by strong dynamical friction. \cite{Amorisco2017} demonstrate (see their Figure 11) that the resulting debris has a strong radial anisotropy, reaching close to $\beta\approx1$ at the approximate location of the Sun (see also right-panel of Fig. \ref{fig:broken_dens}). This orbital radialization reported by \citet{Amorisco2017} for massive satellites went contrary to the previously established intuition where dynamical friction helped to circularize the satellite's orbit instead \citep[see e.g.][]{Jiang2000,Nipotti2017}. While hints of orbital radialization had been seen already by \citet{Barnes1988}, the mechanics of the angular momentum loss during the accretion of massive satellites were finally explained in \citet{Vasiliev2022}. Thus, this relatively new explanation of massive satellite radialization can naturally account for the high $\beta$ values found in the solar neighbourhood if the inner halo is dominated by the debris from massive systems.

\begin{figure*}
\centering
\includegraphics[width=\linewidth]{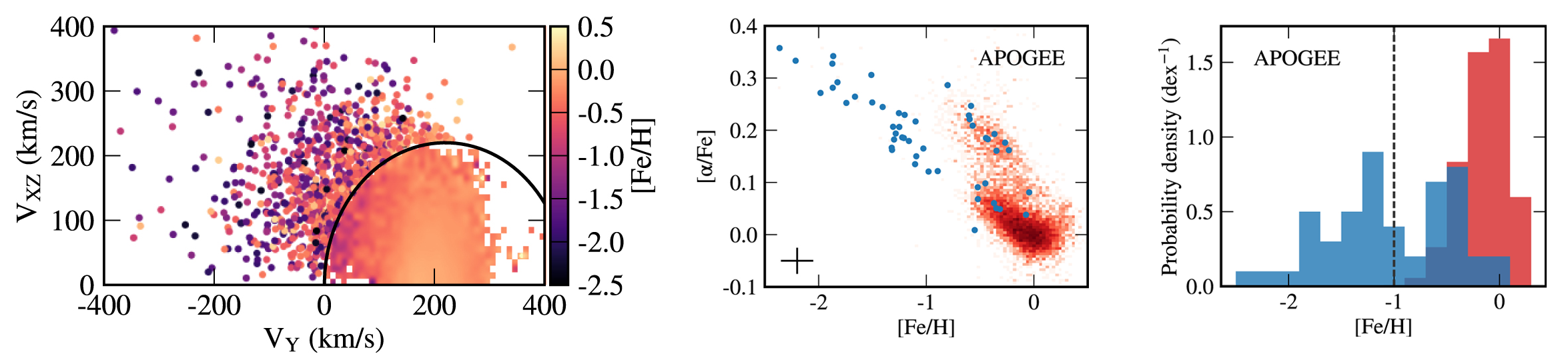}
\caption[]{\textit{Left:} Toomre diagram (velocity space spanned by $V_{\rm Y}$ aligned with Galactic rotation and $V_{\rm XZ}$, its complementary component) for stars with \gaia\ TGAS astrometry. Color represents average metallicity. \textit{Middle:} [$\alpha/Fe$]-[Fe/H] distribution of halo stars (blue points) compared to the rest of the APOGEE data (red-scale density).  \textit{Right:} Metallicity distribution function for halo (disc) in blue (red). Note a clear overlap between the disc and halo MDFs. [Adapted from \citet{Bonaca2017}].}
\label{fig:bonaca_insitu}
\end{figure*}

Finally, several trailblazing studies managed to demonstrate the principles of Galactic Archaeology in action, even with the scarce data available before \gaia. These early pioneers combine chemistry and orbital information to detect individual accretion events in the Solar neighourhood. \citet{Helmi1999} discover a stellar stream from a small dwarf galaxy passing near the Sun, in a clear example of what is to become a routine in the \gaia\ era. Equally striking are the first likely hints of the tidal debris from the last significant merger in the works of \citet{Dinescu2002} and \citet{Brook2003}. Both use the chemo-kinematic dataset by \citet{Chiba2000}. The study of \citet{Dinescu2002} finds evidence for a retrograde disruption event, while \citet{Brook2003} interprets the observations as a signal from a recent accretion of a dwarf on a polar orbit.
\\
\\
\noindent
It is worth remarking at this point that there is a lot of clarity in hindsight, and looking back at these studies in a post-\gaia\ era is particularly revealing. However, it is undeniable that putting the pieces of this puzzle together took a special mission and a game-changing observational dataset.

\section{The \gaia\ discovery of \gaia-Sausage-Enceladus}
\label{sec:gse}

\subsection{\gaia\ Data Release 1}
\label{sec:gdr1}

On September 14 2016 the very first \gaia\ Data Release was delivered to the eager, but somewhat unready, community to forever change the way Galactic Astronomy is done. However, the unprecedented all-sky $\sim$1 billion-strong stars only came with mean positional coordinates, RA and Dec --- in \gaia\ DR1, proper motions, and parallaxes were limited to a tiny $\sim$ 2 million subset of the brightest stars that had been seen previously by Hipparcos \citep[][]{Gaia_DR1}. The size of this so-called Tycho-\gaia\ Astrometric Solution \citep[TGAS, ][]{Lindegren2016} was unfortunately too small to explore in detail the behaviour of the stellar halo in the low-metallicity regime. Surprisingly, however, a significant number of metal-rich stars ([Fe/H]$>-1$) on halo-like orbits were detected in the Solar neighourhood by \citet{Bonaca2017}. These metal-rich stars comprised approximately half of the halo sample near the Sun and exhibited a clear prograde bias compared to their much more isotropic low-metallicity counterparts. Using measurements from the APOGEE survey, this metal-rich population is shown to follow the sequence of the old Milky Way disc in the $\alpha$-[Fe/H] plane (see Fig. \ref{fig:bonaca_insitu}). Weighing up the observational evidence and comparing to the numerical simulations, \citet{Bonaca2017} concluded that this metal-rich halo population must be of in-situ origin and suggested the disc heating by merger events as the key pathway to its formation. While not explicitly connecting the in-situ halo to the last massive merger,  the inventive and dexterous use of the limited DR1 data and the insightful comparison to the available simulations makes the pioneering paper by \citet{Bonaca2017} a clear highlight of the early \gaia\ exploration which helped to set up the stage for the extensive DR2 analysis. 

Aside from the TGAS, the rest of the billion stars in \gaia\ DR1 would have their detailed astrometry published later, as part of the DR2. Not prepared to wait, several groups decided to combine \gaia\ DR1 with earlier astrometric data from elsewhere thus taking advantage of the enormous, uniform, and high-quality dataset \gaia\ was providing, as well as the large temporal baseline between \gaia\ and other surveys. At least three such catalogues were made: the first two obtained proper motion estimates simply by cross-matching \gaia\ DR1 with the data from PPMXL\footnote{PPMXL itself is based on 2MASS and USNO.} \citep[HSOY, see][]{Altmann2017}, and PS1  \citep[][]{Tian2017}. The third proper motion catalogue was made by Sergey Koposov following a rather different strategy to connect \gaia\ DR1 and SDSS \citep[see][for details and the catalogue download]{Koposov_SDSS_Gaia}. Koposov realized that the \gaia\ data could be used to fix the systematic biases in the astrometric calibration of the SDSS itself before using the two to measure positional offsets. Unfortunately, the details of the SDSS re-calibration and the catalogue construction were not described in a stand-alone paper, but the tests of the catalogue performance were published in \citet{Deason2017} and \citet{deBoer2018}.

All three catalogues reported similar mean random proper motion errors between 1 and 2 mas/year. The principle difference between HSOY and PS1-\gaia\ on the one hand and the SDSS-\gaia\ on the other was the amplitude and the spatial distribution of the systematic bias. As shown in \citet{Altmann2017} and \citet{Tian2017}, the first two catalogues carried systematic proper motion offsets of the order of 1-2 mas/year, i.e. similar to their typical random error. Moreover, these systematic shifts followed an unruly patchwork-like pattern on the sky. This can be compared to $\sim$0.1 mas/year systematic uncertainty of the SDSS-\gaia\ proper motions which also exhibited no discernible large-scale pattern on the sky \citep[see][]{Deason2017,deBoer2018}. The quality of the new, large, deep, and wide-angle proper motion sample was verified with two observational tests. First, it was shown that the median proper motion of the spectroscopically confirmed QSOs was $\sim0.1$ mas/year. Second, proper motions across a wide stretch of the Sgr stellar stream were measured for the first time and were shown to be in good agreement with values predicted by simulations \citep[see][]{Deason2017}. The SDSS-\gaia\ catalogue was also used by \citet{deBoer2018} to settle the argument around the nature of the so-called Monoceros stream which had been claimed to be either an outer disc perturbation \citep[][]{Kazantzidis2008} or the tidal debris from a large dwarf galaxy disrupted on a nearly co-planar orbit \citep[see][]{Helmi2003,Penarrubia2005}. The SDSS-\gaia\ proper motions probed directly the kinematics of the Monoceros ring which appeared to exhibit the clear pattern of the remnant disc rotation \citep[][]{deBoer2018}.

\begin{figure*}
\centering
\includegraphics[width=0.95\linewidth]{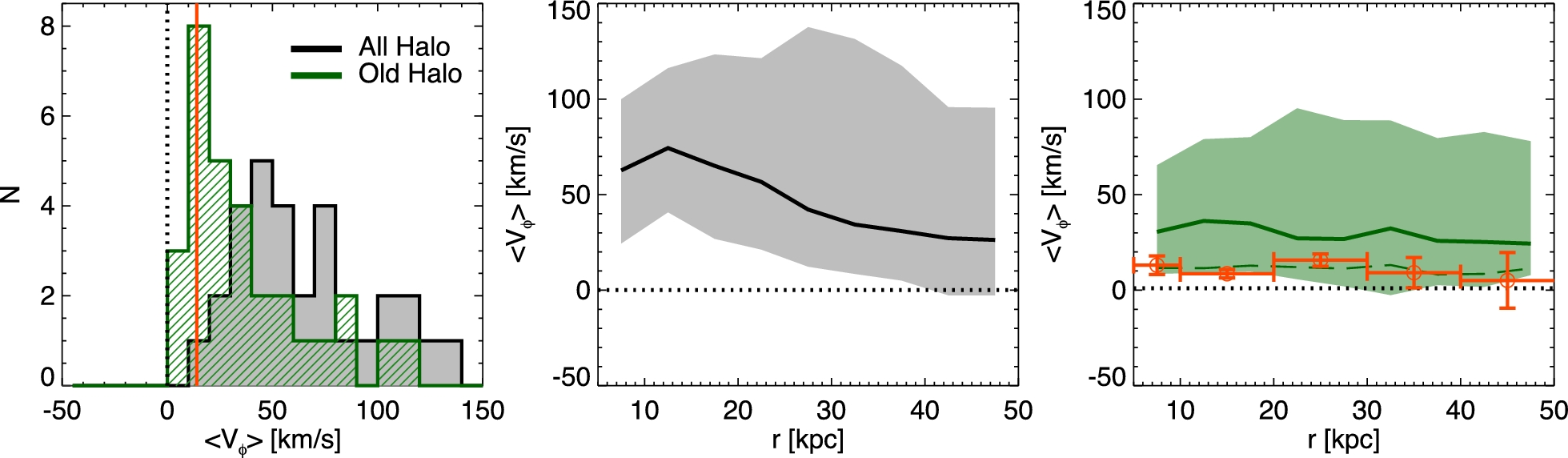}
\caption[]{\textit{Left:} Distribution of mean rotation velocities in Auriga stellar halos for all (old) accreted stars shown in black (green). Red vertical line marks the Milky Way measurement with SDSS-\gaia\ DR1. \textit{Middle:} Rotation velocity as a function of Galacto-centric radius for all accreted stars in Auriga halos. \textit{Right:} Same as middle panel but for the old halo. Red measurements with error-bars show the Milky Way data. [Reproduced from \citet{Deason2017}].}
\label{fig:sligh_spin}
\end{figure*}

\citet{Deason2017} used the SDSS-\gaia\ proper motions to study the amount of coherent rotation in the Galactic stellar halo. Previously, in the absence of ample homogeneous proper motion measurements for faint stars, the halo rotation was extracted from its projection onto the line-of-sight velocities of tracers across large swaths of the sky. These experiments had been inconclusive: only weak rotational signals had been detected \citep[see][]{Sirko2004,Smith2009,Deason2011_rotation}. However, claims had also been made of distinct coherent prograde and retrograde rotation signals in different parts of the halo by \citet{Carollo2007,Carollo2010} (although this was subsequently contested by \citealt{Fermani2013}). For the first time, \citet{Deason2017} used a sizeable proper motion dataset for a variety of stellar halo tracers  --- namely BHBs, RR Lyrae and Red Giant Branch stars --- to measure the mean azimuthal speed of the halo directly. All three tracers showed remarkable consistency: the halo's spin turned out to be minimal, $\lesssim20$ km/s. To interpret these measurements, \citet{Deason2017} studied the behaviour of the stellar halo rotation in the Auriga suite of numerical simulations of Milky Way-like galaxy formation \citep[][]{Grand2017}. Two trends became apparent: i) in Auriga, on average, stellar halos were rotating faster compared to the observed Milky Way, and ii) old and young simulated halo stars showed different rotational signatures, namely young halo stars had more spin (see Fig. \ref{fig:sligh_spin}). \citet{Deason2017} concluded that the slow rotation of the Milky Way halo based on the SDSS-\gaia\ proper motions was consistent with simulations. They noted, however, that the Galactic halo's slight spin was indicative of a particular accretion history, in which the Milky Way halo was assembled early and was not spun up by subsequent mergers. 

Inspired by the early tests of the SDSS-\gaia\ proper motion catalogue described above, \citet{Belokurov2018} used it to study the metallicity dependence of the local stellar halo velocity ellipsoid. This task required a reassessment of the conventional approach to identify a pure halo sample. In previous studies, the halo tracers had been limited to i) large heights above the disc plane and/or ii) metal-poor and old stellar populations such as BHBs and RR Lyrae. Following this approach appeared impractical at intermediate and high metallicities close to the Sun. Therefore, instead of applying hard selection cuts, \citet{Belokurov2018} chose to model the entire stellar velocity distribution. More specifically, in each metallicity bin considered, they used a Gaussian Mixture Model (GMM) implemented as part of the Extreme Deconvolution package \citep[XD, see][]{Bovy2011}. The important feature of XD is that it allows one to take the measurement's uncertainty into account and thus gauge the intrinsic properties of the distribution. While the SDSS-\gaia\ catalogues supplied the best-quality proper motions available for faint main sequence stars at the time, no geometric distances were available for the sample. Instead, \citet{Belokurov2018} relied on the power of the SDSS multi-band photometry to estimate the absolute magnitudes of their stars following the relations derived in \citet{Ivezic2008}.

\begin{figure*}
\centering
\includegraphics[width=\linewidth]{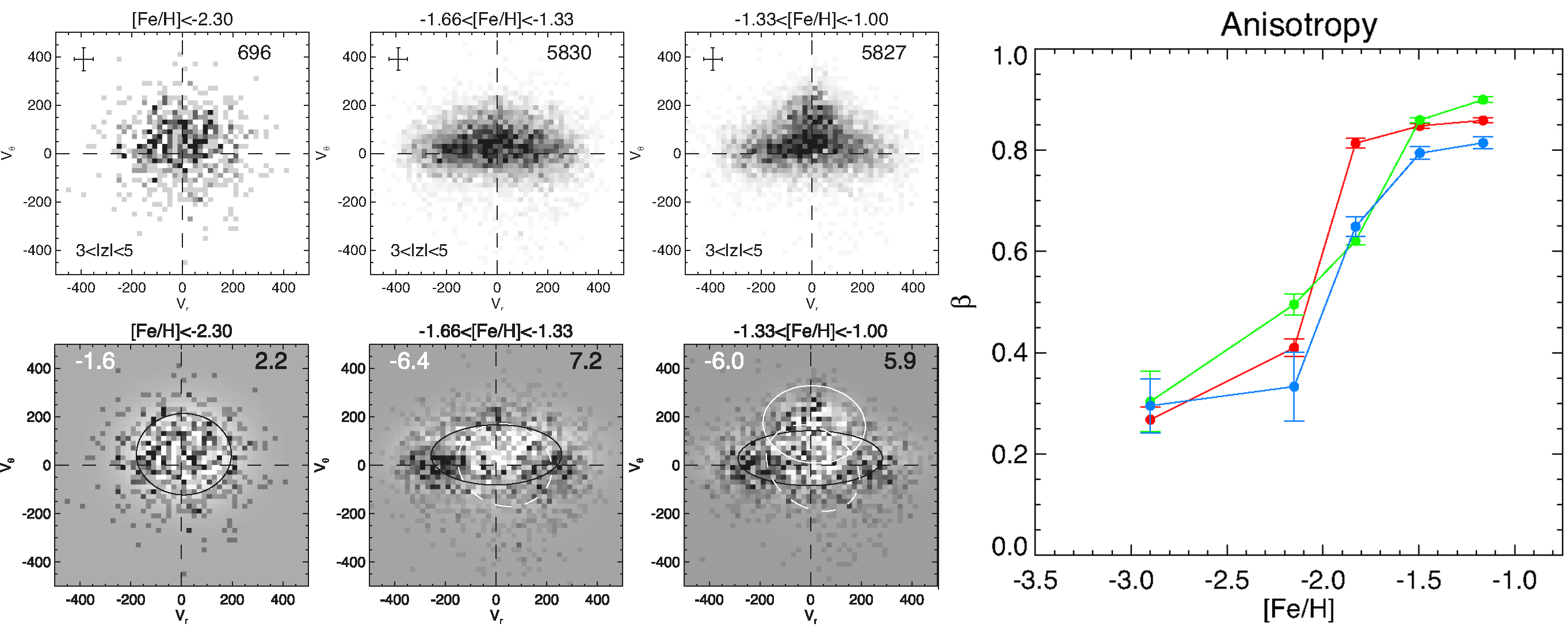}
\caption[]{\textit{Left 3x2 small panels:} Velocity distribution of local stars in the SDSS-\gaia\ catalogue. Top: greyscale gives density of stars in three metallicity bins, the numbers of stars in each bin are shown in the top right corner of each panel. Notice the pronounced change in the shape of the distribution as a function of [Fe/H]. Bottom: residuals of the Gaussian Mixture model of the data. Note the two prominent ``lobes" at high negative and positive radial velocity. \textit{Right large:} Halo orbital anisotropy as a function of metallicity for three different heights above the disc plane (red=low, green=intermediate, blue=high). [Adapted from \citet{Belokurov2018}].}
\label{fig:belokurov2018}
\end{figure*}

\citet{Belokurov2018} demonstrate that, within $\sim10$ kpc of the Sun, the stellar halo's velocity ellipsoid evolves strongly as a function of metallicity (see Fig. \ref{fig:belokurov2018}). In the low metallicity regime, i.e. [Fe/H]$<-2$, the orbital anisotropy is not too far from zero at $\beta\approx0.3$. This nearly isotropic halo component is mixed in with a large population of stars on nearly radial orbits. This metal-richer population with [Fe/H]$>-2$ reaches extremely high $\beta\approx0.9$. In velocity space, this swarm of radially biased orbits manifests itself as a sausage-like shape stretched along the radial dimension and squashed along the azimuthal one. \citet{Belokurov2018} show that the high-$\beta$ component extends to surprisingly high metallicities of [Fe/H]$\approx-1$. They also reveal that in the metallicity regime probed,  stars on highly eccentric orbits comprise some $\sim2/3$ of the local stellar halo. The na\"\i ve GMM representation of the velocity data seems to have worked well, but the residual distribution revealed two noticeable radial velocity lobes, with high positive and negative $V_r$. 

\citet{Belokurov2018} elucidate the genesis of the radially-anisotropic halo component with the analysis of high-resolution zoom-in simulations of Milky Way formation. They show that in the mock stellar halo samples at host-centric distances similar to Solar, the orbital anisotropy is a strong function of the progenitor's mass and its accretion time in agreement with earlier studies \citep[see e.g][]{Deason2013a}. Across a wide range of accretion events explored, only the most massive progenitors with dark matter masses of the order of $10^{11}M_{\odot}$ can yield high anisotropy $\beta>0.8$ provided the mergers took place at $1<z<2$. In their simulation suite, this period in the Galaxy's life also coincided with the phase of active disc growth. The growing disc seemed to help make the orbit of the in-falling massive satellite more eccentric. Most importantly, however, there is a conspicuous convergence in that many independent lines of evidence point to a particular accretion history for our Galaxy. Namely, the one in which a single massive dwarf galaxy was accreted 8-11 Gyr ago, as previously hypothesized by \citet{Deason2013a}. As predicted by \citet{Amorisco2017}, the orbit of the dwarf radialized during its interaction with the Milky Way, populating the inner halo with stellar tidal debris on nearly radial orbits. The resulting high eccentricity of the main stellar halo component also explains the low net spin measured by \citet{Deason2017}. Finally, the relatively high metallicity of the eccentric halo stars in the Solar neighbourhood agrees with the expectation for stellar contents of a massive dwarf galaxy in accordance with the established mass-metallicity relation \citep[see e.g.][]{Kirby2013}. \citet{Belokurov2018} conclude that the sausage-like feature in the local phase-space ought to be the result of an ancient accretion of a single, massive satellite.

Phase-space analysis of the kind described above is appropriate for data samples limited to the Solar neighourhood. It is also very convenient as it does not require an assumption of the gravitational potential of the Galaxy and can be re-produced easily, including in applications to mock data from numerical simulations. However, as demonstrated by e.g. \citet{HdZ2000}, the identification of merger debris is more straightforward and more efficient in the integrals-of-motion space. This is exactly what is attempted in \citet{Helmi2017} where a low-metallicity halo sample is selected using a cross-match between TGAS and RAVE \citep[see][]{Steinmetz2006}. Unfortunately, the resulting number of stars ($\sim1000$) is too low to discern clearly any global patterns, although the presence of a conspicuous retrograde population is spotted. Instead, similarly to \citet{Deason2017} and \citet{Belokurov2018}, \citet{Myeong2018} used the much larger SDSS-\gaia\ sample which contained $>10,000$ halo stars across a wide range of metallicity. They scrutinize the behaviour of the halo in the action space in different [Fe/H] bins and detect the prevalence of highly eccentric orbits at $-2<$[Fe/H]$<-1$, for which they agree the simplest explanation is the in-fall of one satellite.

\begin{figure*}
\centering
\includegraphics[width=\linewidth]{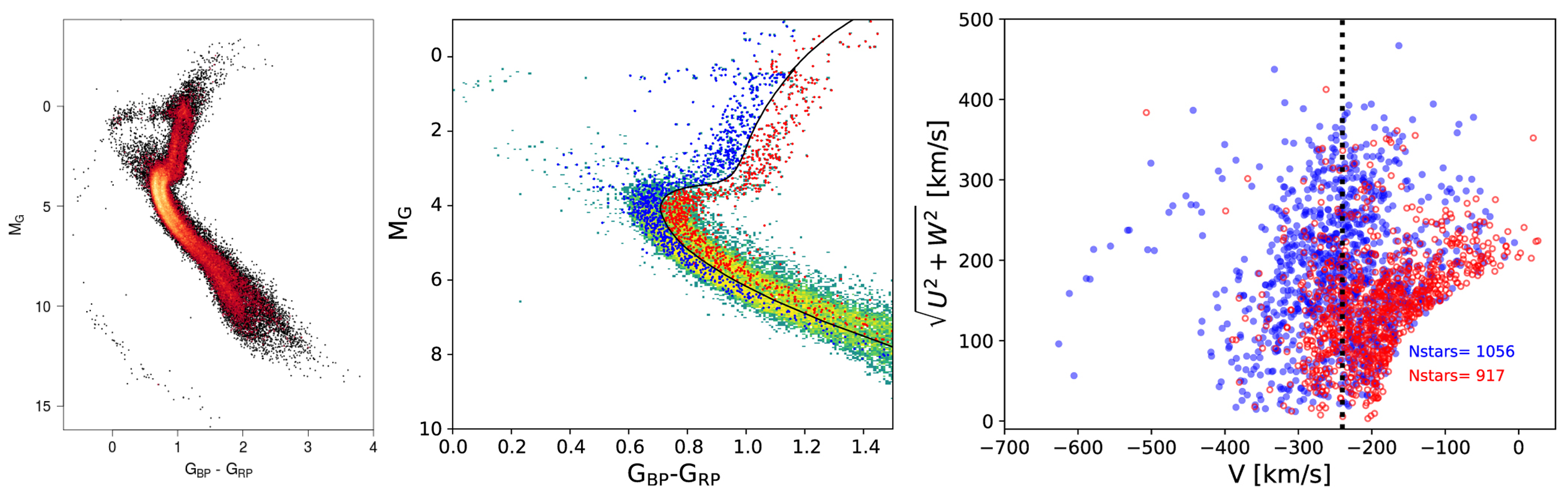}
\caption[]{ \textit{Left:} HRD for stars with high tangential velocities (typical halo kinematics) in \gaia\ DR2 \citep{Babusiaux2018}. Note the split main sequence. \textit{Middle:} High tangential velocity stars separated into blue and red populations with a model isochrone corresponding to [Fe/H]$=-0.75$ and age of 11.5 Gyr \citep{Haywood2018}. \textit{Right:} Toomre diagram (also see Fig. \ref{fig:bonaca_insitu}) for blue and red stars selected in the previous panel. [Adapted from \citet{Babusiaux2018, Haywood2018}].}
\label{fig:hrd_kinematics}
\end{figure*}

\subsection{\gaia\ Data Release 2}

The works discussed in the previous section used \gaia\ DR1 to find the first bits of irrefutable evidence of a dramatic merger event between the young Milky Way and a massive dwarf galaxy. Later, the transformational \gaia\ Data Release 2 helped to locate some of the crucial missing pieces in the puzzle of the Milky Way transformation and the stellar halo assembly. Helped by the vast number of exquisite proper motion and parallax measurements, the following five key properties of the last significant merger were quickly discovered. 

\begin{itemize}
    \item The progenitor dwarf stars follow a distinct sequence in the Hertzsprung–Russell diagram \citep{Babusiaux2018,Haywood2018}. 
    \item These stars also stand out from the rest of the halo, and, crucially, are distinct from the more metal-rich, in-situ component in the $\alpha$-[Fe/H] space \citep{Helmi2018}. \item The satellite that merged with the Milky Way came accompanied by an entourage of its own GCs \citep[][]{Myeong2018gcs}. 
    \item The motion of the stars with [Fe/H]$<-1$ within $\sim20-30$ kpc appears synchronized, which in turn explains the global spatial structure of the halo \citep[][]{Deason2018pileup,Iorio2019} 
    \item The bulk of the in-situ halo with [Fe/H]$>-1$ was produced by heating and splashing of the pre-existing Milky Way disc \citep[][]{Gallart2019,DiMatteo2019,Belokurov2020}. This is discussed further in Section \ref{sec:splash}.
\end{itemize}

\noindent Below we go chronologically (i.e. in the order they first appeared online, although some of these papers came out within a month of each other) through these individual pieces of evidence.

\vskip 0.5cm

\noindent {\it HRD dichotomy}. The striking apparent split in the Hertzsprung-Russell diagram (HRD) of the nearby stars with high tangential velocity --- i.e. on halo-like orbits --- was first presented and discussed in one of the \gaia\ DR2 science demonstration papers \citep[][see Fig. \ref{fig:hrd_kinematics}]{Babusiaux2018}. The authors notice two separate HRD sequences --- blue and red according to their $G_{BP}-G_{RP}$ colour --- most visibly disconnected around the main sequence turn-off region, which they analyse and interpret to be produced by two halo populations with distinct mean metallicities, [Fe/H]$\sim-1.3$ and [Fe/H]$\sim-0.5$.  \citet[][]{Babusiaux2018} report clear similarities with the bi-modality in the metallicity distribution of the Galactic GCs, which is known to arise due to contributions from clusters predominantly accreted from dwarf galaxies and those formed in situ in the Milky Way. They speculate that the two halo populations must have followed distinct formation channels and surmise that there may be a connection between the two halo HRD sequences and the chemical sequences highlighted earlier in \citet{Nissen2010} and \citet{Bonaca2017}. \citet{Haywood2018} take a deeper look at the \gaia\ DR2 HRD to solve the puzzle of the dichotomy of the stellar halo. They demonstrate that while both populations have orbits distinct from the disc, there is a clear difference in the kinematics of the stars in the blue and red sequences. \citet{Haywood2018} show that the red sequence is composed of stars on mostly prograde orbits with lower eccentricities and lower total energies. The blue sequence stars have little angular momentum, have higher eccentricities, and reach out further into the halo. The HRD sequences indeed correspond almost perfectly to the two chemical sequences found by \citet{Nissen2010}: the red one contains stars with higher $\alpha$-abundances and the blue one with lower $\alpha$-abundances. The connection between kinematics and detailed chemistry is also made in \citet{Helmi2018}, who together with \citet{Haywood2018} appear inspired by the earlier analysis of the halo chemistry using APOGEE  \citep[see][]{Hawkins2015,Fernandez-Alvar2018,Hayes2018}.  \citet{Haywood2018} associate the red sequence to the thick disc in agreement with \citet{Bonaca2017} and the blue sequence to the ancient massive merger as proposed in e.g. \citet{Deason2013a} and \citet{Belokurov2018}. \citet{Haywood2018} argue that the tightness of the blue sequence in the \gaia\ HRD is an argument in favour of the single dominant accretion event because contributions from many dwarf progenitors would otherwise bloat it.

\vskip 0.5cm

\begin{figure*}
\centering
\includegraphics[width=\linewidth]{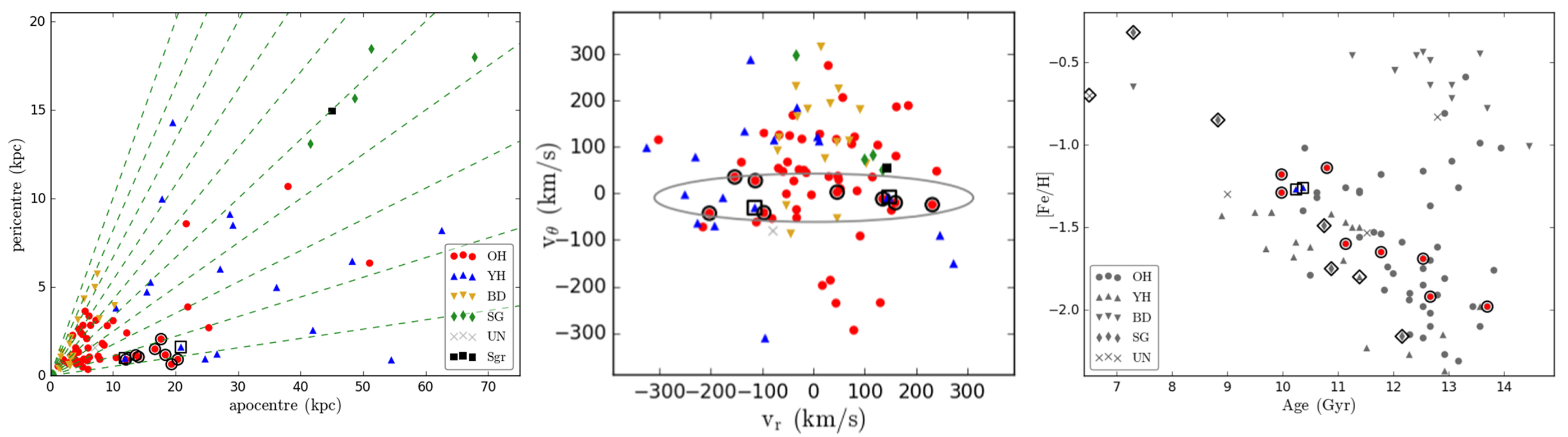}
\caption[]{ \textit{Left:} Galactic GCs in the space spanned by their peri-centric and apo-centric distances. Old halo (OH, red filled circles), young halo (YH, blue filled triangles), bulge/disc (BD, yellow triangles) and Sgr dwarf (black filled square), and its GCs (SG, filled green diamonds) are marked. The 8 GCs attributed to GS/E by \citet{Myeong2018gcs} are circled in black. \textit{Middle:} The same GCs are shown in the space of radial and azimuthal velocity components (also see Fig. \ref{fig:belokurov2018}). \textit{Right:} Age-metallicity plane for the GCs with available ages. [Adapted from \citet{Myeong2018gcs}].}
\label{fig:sausage_gcs}
\end{figure*}

\noindent {\it Member Globular Clusters.} There exists a remarkable connection between the combined mass of a galaxy's GCs and its total (dark matter) mass with more massive galaxies hosting larger GC populations \citep[see][]{Blakeslee1997,McLaughlin1999,Spitler2009,Harris2017,Dornan2023}. In the Milky Way, this correlation manifests itself very clearly: only the four largest dwarf satellites possess their own GC posses. Aside from the LMC, SMC, Sgr and Fornax, only one other dwarf galaxy (Eridanus 2) has a single member GC \citep[][]{Koposov2015,Crnojevic2016}. In archaeological studies of the Milky Way halo this GC-host mass link implies that while many tidal debris remnants are expected to be detected in the stellar halo, only the most massive of these would be accompanied by GCs on similar orbits. A comparison between the orbital properties of the halo stars and the Galactic GCs \citep[using proper motions measured by ][]{Helmi2018sats} is carried out in \citet{Myeong2018gcs} (see Fig. \ref{fig:sausage_gcs}). First, they introduce the idea of the ``critical energy'' which separates the in-situ and the accreted GCs. Then they show that amongst the accreted clusters, a tight group stands out. The GCs in this group have very similar orbital actions. More specifically, their pericentres are all remarkably low ($<2$ kpc) but their apocentres are some of the largest in the sample ($>10$ kpc). \citet{Myeong2018gcs} point out that there is most likely a link between this GC group and the eccentric stellar halo component discussed in \citet{Belokurov2018}. Both exhibit high orbital anisotropy $\beta>0.9$ and the sausage-like shape in the space spanned by velocity components in spherical polars. Consequently, they argue that such a notable stand-out GC agglomeration favours a single accretion event and propose to use the correlation discussed above to estimate the total mass of the progenitor dwarf which they roughly guess must be in the ballpark of $5\times10^{10}M_{\odot}$. \citet{Myeong2018gcs} are inspired by the works from the pre-\gaia\ era \citep[][]{Zinn1993,Mackey2004,Mackey2005,Forbes2010} but they are able to take these ideas to the next level thanks to the unprecedented quality of the \gaia\ astrometry.

\vskip 0.5cm

\begin{figure*}
\centering
\includegraphics[width=\linewidth]{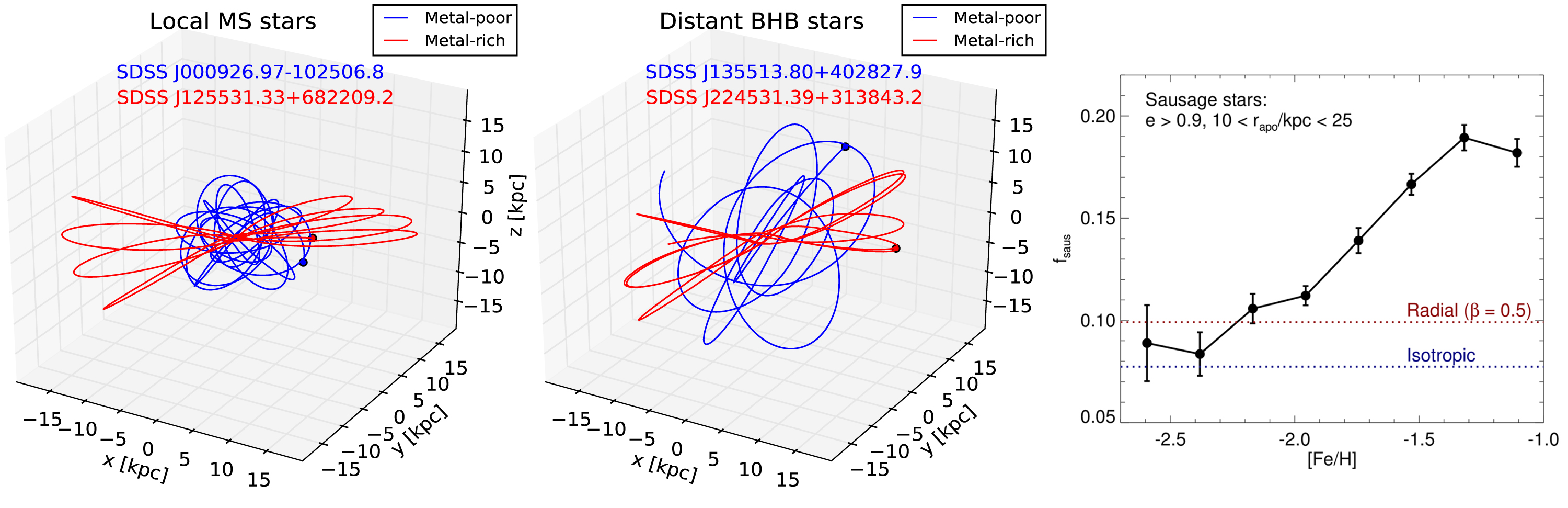}
\caption[]{ \textit{Left:} Orbit comparison for two local halo main sequence populations, one with low metallicity (blue) and one with high metallicity (red). \textit{Middle:} Orbit comparison for two distant halo BHB populations (same colour coding). Note the similarity of the orbital behaviour of the metal-rich main sequence and BHB stars: both are on highly eccentric orbits coming close to the centre of the Milky Way, and both with apocentres close to 15-20 kpc. \textit{Right:} Fractional contribution to the halo number counts of stars with high eccentricity as a function of metallicity. [Adapted from \citet{Deason2018pileup}].}
\label{fig:orbital_pileup}
\end{figure*}

\noindent {\it Orbital synchronicity}. The use of the GCs as tracers highlighted one particular difficulty of the stellar halo studies with \gaia. 
Put simply, given the limited reach of the \gaia\ DR2 parallaxes, the vast extent of the halo appeared inaccessible. These observational hindrances can be circumvented using tools of Galactic Dynamics with which one can predict the overall density distribution of the halo using only a small sample of tracer kinematics constrained to the Solar neighourhood \citep[see e.g.][]{ELS1962,Chiba2000}. This is because the spatial appearance of the Galaxy is nothing but the superposition of the orbits of the individual stars and the density in the configuration space is linked to the kinematics through the continuity of the phase-space flow \citep[][]{BT_book2008}. In a similar vein but in a much more direct experiment, \citet{Deason2018pileup} compute orbits of the metal-rich halo stars in the Solar vicinity and demonstrate that their apo-centres pile-up around Galactocentric distance $r\approx20$ kpc (see Fig. \ref{fig:orbital_pileup}). The location of this pile-up matches the break radius in the stellar halo density (see Section~\ref{sec:pregaia}). As a sanity check, they also compare apo-centres of nearby main sequence stars to the apo-centres of the distant BHB stars that were originally used to detect the break and show that stars in the two populations, distant and local, turn around on their orbits at approximately the same distance. Thus three different tracer populations have been shown to have the same orbital behaviour: the nearby main sequence stars, the distant BHBs, and the GCs. \citet{Deason2018pileup} demonstrate in a very candid way that the local metal-rich and radially anisotropic population is only a glimpse of a vast and dominant cloud of stellar debris that dictates the global properties of the stellar halo (e.g. the density break). They also provide the first constraint on the metallicity distribution function of the high-eccentricity population and show that its incidence drops by a factor of $\sim2$ going from [Fe/H]$=-1$ to [Fe/H]$=-2$.

\vskip 0.5cm

\begin{figure*}
\centering
\includegraphics[width=\linewidth]{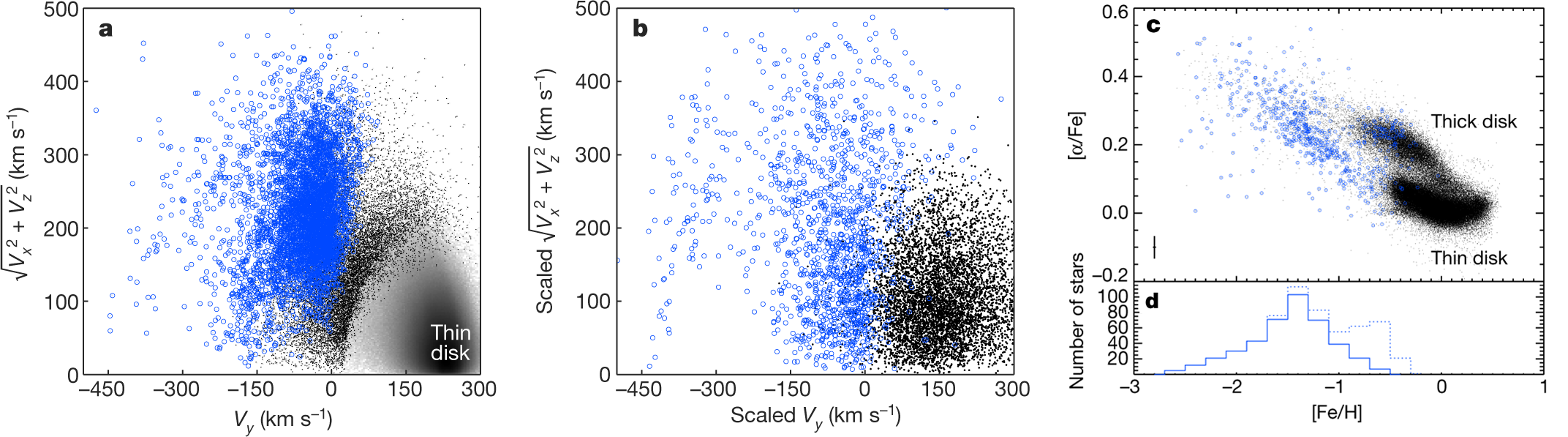}
\caption[]{ \textit{Panel (a):} Toomre diagram for stars in \gaia\ DR2 (see also Fig. \ref{fig:bonaca_insitu} and Fig. \ref{fig:hrd_kinematics}). Blue points are selected to pick out a large, mostly retrograde accreted sub-structure. \textit{Panel (b):} Similar to Panel (a) but for a simulation of a minor merger from \citet{Villalobos2008}. \textit{Panel (c):} [$\alpha$/Fe]-[Fe/H] space for stars in \gaia\ DR2 with APOGEE chemical abundance measurements (same colour-coding as in Panels (a) and (b)). {\it Panel (d):} Metallicity distribution for blue stars shown in Panel (c). Solid (dotted) lines show the distribution of [Fe/H] when high-$\alpha$ stars are excluded (included). [Adapted from \citet{Helmi2018}].}
\label{fig:helmi2018}
\end{figure*}

\noindent {\it Detailed chemistry}. \citet{Helmi2018} draw attention to the fact that the vast numbers of eccentric stars covering a huge portion of the accessible Galactic halo all appear to follow incredibly narrow sequences in both colour-magnitude and $\alpha$-[Fe/H] spaces (see Fig. \ref{fig:helmi2018}). More specifically, they demonstrate the existence of a large elongated cloud \citep[or a ``blob'', see][]{Koppelman2018} in the velocity and/or energy-angular momentum ($E, L_z$) space and isolate its stars using simple cuts in $E$ and $L_z$. They also notice a prominent retrograde extension of the debris cloud \citep[previously discussed in e.g.][]{Myeong2018,Koppelman2018}. Note that the follow-up work by \citet{Myeong2019} attributes these retrograde sub-structure to a different, but perhaps contemporaneous accretion event, the so-called Sequoia.

\citet{Helmi2018} show that on the HRD, these stars are largely confined by isochrones with $-1.3<$[Fe/H]$<-0.9$ and ages of 10-13 Gyr. \citet{Helmi2018} cross-match their selected sample to APOGEE and see two narrow sequences in the $\alpha$-[Fe/H] plane, one mostly flat with $-1<$[Fe/H]$<-0.3$ and [$\alpha$/Fe]$\approx0.25$  and one sloping down from [Fe/H]$\approx-1.3$, [$\alpha$/Fe] $\approx0.25$ to [Fe/H]$\approx-0.7$, [$\alpha$/Fe] $\approx0.1$. Following \citet{Bonaca2017}, they assign the high-$\alpha$ sequence to the in-situ population \citep[see also][]{Nissen2010}. Given that the two sequences co-exist over a range of metallicities, \citet{Helmi2018} conclude that the halo stars in the lower $\alpha$-[Fe/H] sequence must have formed in a separate, smaller galaxy. In view of the large spread of metallicity and the presence of low-$\alpha$ high-[Fe/H] stars this dwarf could not be much smaller than the LMC. Their interpretation of the two chemical sequences builds on the analysis of the halo stars in APOGEE by \citet{Hawkins2015}, \citet{Fernandez-Alvar2018} and \citet{Hayes2018}. These earlier works provide a detailed chemo-dynamical comparison of the high- and low-$\alpha$ sequences and agree on the in-situ nature of the former. Crucially, however, in the absence of the \gaia\ data, these authors refrain from making a claim as to the exact nature of the accreted population, although \citet{Fernandez-Alvar2018} report a significantly lower star-formation rate for the low-$\alpha$ population. \citet{Helmi2018}'s conclusion is not solely based on the accumulation of the observational evidence available at the time. Their insight is also the result of an earlier numerical exploration, a series of experiments simulating high mass-ratio mergers between the Milky Way and a dwarf satellite \citep[][]{Villalobos2008}. \citet{Helmi2018} find striking similarities between the \gaia\ DR2 data and one of the pre-existing simulations (once the simulation is appropriately re-scaled). \citet{Helmi2018} propose to give a name to the progenitor system (and presumably its tidal debris): {\it Gaia-Enceladus}. 

\vskip 0.5cm

Following the two papers (\citealt{Belokurov2018} and \citealt{Helmi2018}) that first identified the metal-rich highly-eccentric accreted halo population and proposed that it must be produced in a single massive accretion event, the vast tidal debris cloud is referred to as {\it Gaia} Sausage-Enceladus (GS/E).

\section{Characterization of \gaia-Sausage-Enceladus}

In the previous section, we have recounted the overwhelming evidence for the massive GS/E progenitor in the Milky Way stellar halo. Naturally, since its discovery there has been a concerted effort in the community to further characterize this colossus accretion event. 

\subsection{Selecting GS/E members}
\label{sec:gse_sel}
Perhaps the most important first step in such studies is to understand how exactly we select the GS/E stars, as any deduction of its properties (e.g. total mass, metallicity etc) will be significantly impacted by any selection effects. On the one hand, the proximity of this structure to the solar neighourhood means we have significant numbers of potential GS/E stars to study. However, this also means that the GS/E is located in a region close to the stellar disc, and is thus prone to contamination. In the \cite{Belokurov2018} discovery paper the authors use a GMM method to simultaneously model the different Milky Way components in velocity space, but even this advance over simple kinematic cuts cannot guarantee a `pure' sample of GS/E stars. Unsurprisingly, more recent works have looked to action-angle space to locate the GS/E stars \citep[e.g.][]{Feuillet2020, Lane2022}, which should result in a cleaner selection than velocity information alone. Other works have made use of the exquisite APOGEE data to perform a purely chemical selection of GS/E stars, whereby a cut in the plane of [Mg/Mn] versus [Al/Fe] is assumed to cleanly separate accreted and in-situ stars \citep[e.g.][]{Das2020, Carrillo2022,Lane2023}.

\begin{figure*}
\centering
\includegraphics[width=\linewidth]{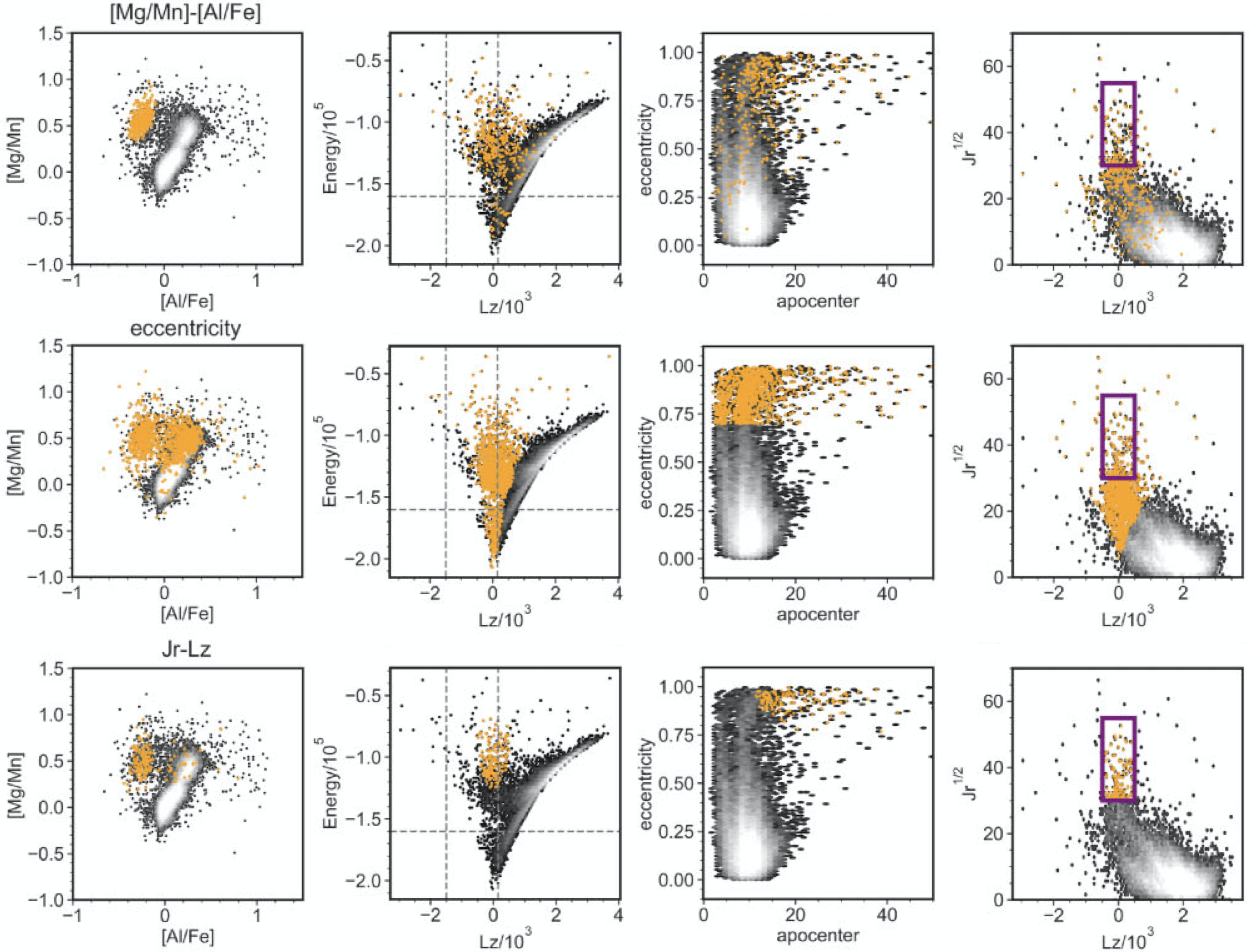}
\caption[]{Chemical and orbital properties of different GS/E selections (yellow points) compared to the rest of the APOGEE dataset (greyscale density). {\it Top row:} GS/E sample selected using [Mg/Mn] and [Al/Fe] abundance ratio space. {\it Middle row:} GS/E stars selected using an eccentricity cut (see 3rd column). {\it Bottom row:} GS/E sample created with a box in $J_r-L_z$ space (magenta box in the fourth column). All GS/E samples are compared in four spaces: [Mg/Mn]-[Al/Fe] (first column), $E$-$L_z$ (second column), eccentricity-apocentre (third column) and $J_r-L_z$ actions (fourth column). [Adapted from \citet{Carrillo2024}].}
\label{fig:carrillo}
\end{figure*}

It is clear that there is no `perfect' mode of selection, and how the GS/E stars are selected should depend on the question at hand (or the property being determined!). Comparisons of different methods are examined explicitly in several works with varying degrees of success \citep[e.g.][see Fig. \ref{fig:carrillo}]{Feuillet2021, Buder2022, Lane2022, Lane2023,Carrillo2024}. \cite{Carrillo2024} in particular uses cosmological simulations to inform the selection choices, which is likely a necessary tool moving forward in our understanding of the GS/E. Regardless of what selection mode is chosen, it is clear that quantifying any systematic influences is crucial in order to properly characterize the GS/E.  For example, \cite{Carrillo2024} shows that stellar mass estimates of the GS/E can vary by a factor of $\sim 2$ depending on the selection used. \citet{Lane2023} provide a convincing demonstration that once the APOGEE spatial, chemical, and kinematic selection effects are taken into account, the mass of the GS/E based on its red giant member stars is at the low end of the spectrum, around $M_{\rm star}\sim1.5\times10^8 M_{\odot}$, and its overall fractional contribution to the stellar halo is around $15\%-25\%$. Most mass measurements rely on the metallicity distributions of GS/E stars, and the assumption of well-known mass-metallicity relations \citep[e.g.][]{Feuillet2020, Carrillo2024}. While these measures typically favour a GS/E with $M_{\rm star} \sim 10^{8}-10^{9} M_\odot$, a more accurate determination will likely require more robust GS/E selections (in addition to a better-known redshift dependence of the mass-metallicity relation).  This is still a very active area of research, but one in which the combination of multiple observational dimensions  (i.e. action-angle plus chemistry) plus numerical calibration will likely provide the most robust quantitative measures.

\subsection{Halo kinematics and the orbital make-up of the GS/E debris}
\label{sec:gse_kin}

The GS/E debris covers a large portion of the Galactic halo, overlapping broadly with both the in-situ populations and other, smaller accreted sub-structures. However, its member stars appear to follow distinct trends in phase-space, which can be retrieved with appropriate tools.

\citet{Necib2019} and \citet{Lancaster2019} focus on the bi-modality in the Galactocentric radial velocity distribution of the local halo stars (see Fig. \ref{fig:belokurov2018}). Such a bi-modality can form when stars deposited by a progenitor on a nearly radial orbit are observed between the turn-around points in their orbits, i.e. between their peri- and apo-centres (see Fig. \ref{fig:broken_dens}). \citet{Necib2019} demonstrate that in the Solar vicinity, the halo's velocity distribution can be decomposed into two components, one approximated well by a single multi-variate Gaussian and another better described by a model with two peaks in $v_r$ space. In the latter model, the $v_r$ peaks are two identical Gaussian distributions with their separation controlled by a free model parameter. \citet{Necib2019} argue that the presence of a significant anisotropic sub-structure in the local halo can have important implications for direct dark matter detection experiments. If the progenitor dwarf galaxy managed to dump enough of its dark matter on orbits reaching the Sun's location (similar to the GS/E stars), then the resulting velocities of the dark matter particles interacting with the detectors will be different from those described by the conventional Standard Halo Model. 

Similarly,  \citet{Lancaster2019} rely on an identical two-component mixture to model a sample of previously identified BHB stars reaching far beyond the Solar neighourhood. Again, one model component is allowed to have a double-peaked $v_r$ distribution with an unconstrained separation between the peaks. They demonstrate that a single-Gaussian description of the velocity distribution of the Galactic stellar halo is a gross over-simplification. \citet{Lancaster2019} show that out to $r\approx25$ kpc from the Galactic centre, such a model does not fit the data correctly. They show that within this region, the BHB kinematics are best described by an equal contribution from a roughly isotropic halo and a radially anisotropic component, which they attribute to the GS/E merger. They also detect a decline in the separation between the radial lobes with Galacto-centric distance $r$, as expected from merger simulations (see Fig. \ref{fig:broken_dens}). \citet{Lancaster2019} determine that beyond $r\approx25$ kpc, the fractional contribution of the bi-modal $v_r$ component is consistent with zero. These findings are in agreement with the detection \citep[see e.g.][]{Sesar2011, Watkins2009,Deason2011} and the interpretation \citep[see e.g.][]{Deason2013a,Deason2018pileup} of the stellar halo density break at a similar distance. 

The appeal of BHBs is that despite being quite rare, they are a robust standard candle that can be picked out of a mix of stellar populations cleanly and easily. Compared to BHBs, K-giants are a more numerous and brighter halo tracer, at the expense of being a less accurate distance indicator. \cite{Bird2019} and \cite{Bird2021} use a large number of K-giants to probe the global anisotropy profile of the Galactic stellar halo from 5 to 100 kpc. Their single multivariate Gaussian model shows qualitatively similar behaviour to the single-component model of \citet{Lancaster2019}. However, they report a higher orbital anisotropy for K-giants compared to the BHBs. Moreover, unlike the anisotropy of the BHBs which shows a pronounced decline around $r\approx20$ kpc as discussed above, the K-giants' $\beta$ remains more or less flat across the whole range of Galactocentric distances probed.

\begin{figure}
\includegraphics[width=\linewidth]{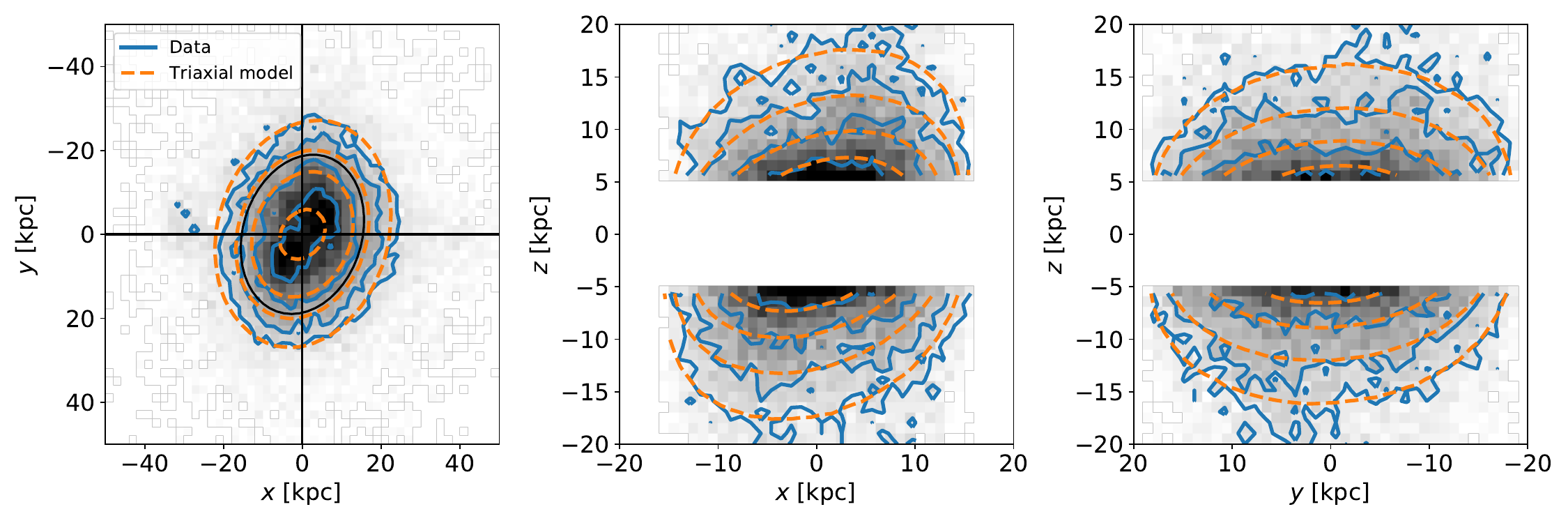}
\caption[]{Projections of RR Lyrae density in the
  inner halo. The \gaia\ DR2 data is shown as a greyscale map and blue contours. A triaxial model is indicated with the orange contours. [Reproduced from \cite{Iorio2019}].}
\label{fig:Iorio_RRL}
\end{figure}

The long-anticipated boon of the \gaia\ data releases have produced the largest catalogue of RR Lyrae, the pulsating counterpart of the BHBs, to date. However, even before the first \gaia\ variable star catalogues were provided as part of the \gaia\ DR2, \citet{Belokurov2017_lmc}, \citet{Deason2017_lmc} and \citet{Iorio2018} used the excess in \gaia\ DR1 mean flux uncertainty to identify likely variables such as RR Lyrae and Mira stars. In particular, \citet{Iorio2018} combined this rudimentary variability statistic with the 2MASS colour information to build the first all-sky RR Lyrae sample. They test a wide range of halo models and show that the best-fit is achieved by a triaxial ellipsoid whose vertical flattening varies with Galacto-centric distance. In this model, the longest axis lies in the Galactic plane but is rotated by $\approx70^{\circ}$ with respect to the Sun-Galactic centre direction. Once the \gaia\ RR Lyrae sample was released within DR2, \citet{Iorio2019} tested their earlier model and demonstrated that it was indeed a good fit, albeit to an even larger, purer, and more complete sample (see Fig. \ref{fig:Iorio_RRL}). 

\begin{figure*}
\includegraphics[width=\linewidth]{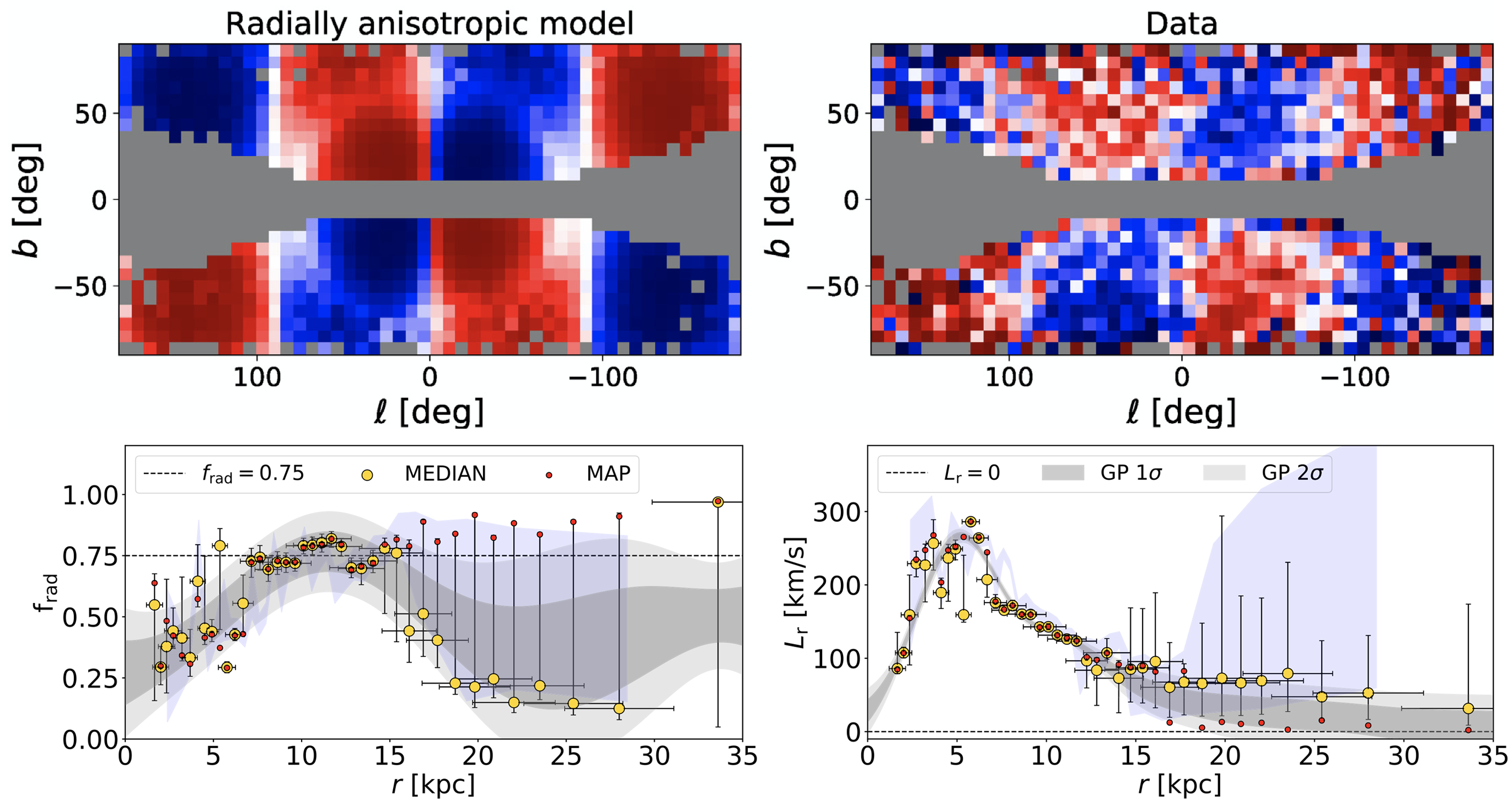}
\caption[]{{\it Top row:} Strength of correlation between the longitudinal and latitudinal components of proper motion in Galactic coordinates for a model with a strong radial anisotropy (left) and \gaia\ DR2 RR Lyrae (right) \citep{Iorio2019}. {\it Bottom row:} Fractional contribution of the radially-biased halo component in the \gaia\ RR Lyrae as a function of the Galactocentric distance (left) and the radial velocity offset (the separation between the two Gaussians is twice this value) in the kinematic model of radially-biased halo component of the \gaia\ RR Lyrae (right) \citep{Iorio2021}. [Adapted from \citet{Iorio2019, Iorio2021}].}
\label{fig:ib_kin}
\end{figure*}

A minor tweak was introduced by \citet{Iorio2019} to their earlier model \citep[][]{Iorio2018} in order to tilt the halo slightly out of the Milky Way's disc plane \citep[see also][]{Han2022c}. This non-axisymmetry is warranted by the RR Lyrae number counts and is likely related, at least in part, to the two previously identified large-scale sub-structures, the Virgo Overdensity \citep[VOD][]{Juric2008, Vivas2016} and the Hercules Aquila Cloud \citep[HAC][]{Belokurov2007_hac, Simion2014}. Curiously, VOD and HAC are shown to be closely related by \citet{Simion2019} who, based on orbital analysis, argue that both originated in the GS/E merger. A similar conclusion was reached by \cite{Balbinot2021}. Moreover, \cite{Perottoni2022} show that, in addition to their dynamical link, the VOD and HAC are chemically indistinguishable from the prototypical GS/E population. The tilt in the GS/E debris cloud with respect to the disc as exemplified by the HAC-VOD asymmetry is examined by \citet{Han2022} who point out that such an asymmetry cannot survive for a very long in an axi-symmetric potential of the Galaxy. They argue that to maintain the stellar halo's tilt over a long period of time, a similarly tilted dark matter component is required. This idea is fully fleshed out in \citet{Han2023} and \citet{Han2023b} where the misaligned dark matter halos are analysed and shown to be ubiquitous in Cosmological simulations. Such tilted dark matter halo components are damaging to the Galactic discs, causing them to warp and flare. In a similar vein, \citet{Davies2023} argue that the evolution of the GS/E debris in an axi-symmetric potential would lead to discernible striation in the integrals-of-motion space once the stellar halo is tugged by the passing LMC. The fact that no such characteristic substructure is observed \citep[although see][for the detection and interpretation of high-frequency substructure in the GS/E debris]{Belokurov2023_wrinkles} could be interpreted as evidence for a non-axisymmetric component in the Milky Way's dark matter distribution. From the analysis in \citet{Dillamore2022}, it is clear that tilts such as those proposed in \citet{Han2023} and \citet{Davies2023} are common in galaxies recovering from a significant merger.

\citet{Iorio2019} also uncover a tight kinematic bond amongst most of the inner halo RR Lyrae. They show that within $r\approx25$ kpc from the Galactic centre, the stellar motions are correlated: the kinematics of the bulk of the RR Lyrae is consistent with a nearly radial motion, one of the key characteristics of the GS/E (see also \citealt{Ablimit2022}). \citet{Iorio2021} move beyond demonstration to modelling of the kinematic properties of the \gaia\ RR Lyrae sample (see Fig. \ref{fig:ib_kin}). Their model contains three distinct components: a rotating disc-like population, a quasi-isotropic one, and finally, a strongly radially anisotropic population corresponding to the GS/E. The modelling approach is inspired by an earlier work of \citet{Wegg2019} where 5D data with missing line-of-sight velocity is modeled under the assumption of axi-symmetry to extract the properties of the RR Lyrae velocity ellipsoid. For the radially anisotropic GS/E population, \citet{Iorio2021} rely on the functional form introduced in \citet{Necib2019} and \citet{Lancaster2019} and map out the separation between the GS/E radial velocity bumps inside $r\approx30$ kpc. The separation reaches $\approx500$ km/s around $r\approx5$ kpc and drops to zero around $r\approx25$ kpc. The GS/E contribution to the halo RR Lyrae is maximal around the Sun at $\sim75\%$ but decreases to $\sim20\%$ inside $r\approx5$ kpc and outside $r\approx20$ kpc. Curiously, the properties of the isotropic component are not uniform throughout: \citet{Iorio2019} show that the central $\approx10$ kpc are dominated by a more metal-rich RR Lyrae population with a higher fraction of rare High Amplitude Short Period pulsators \citep[see][]{Fiorentino2015, Belokurov2018_rrl}. Given its properties, this centrally concentrated RR Lyrae population is likely of in-situ origin.

\subsection{Chemical trends}
\label{sec:GSE_chem}

\begin{figure*}
\includegraphics[width=\linewidth]{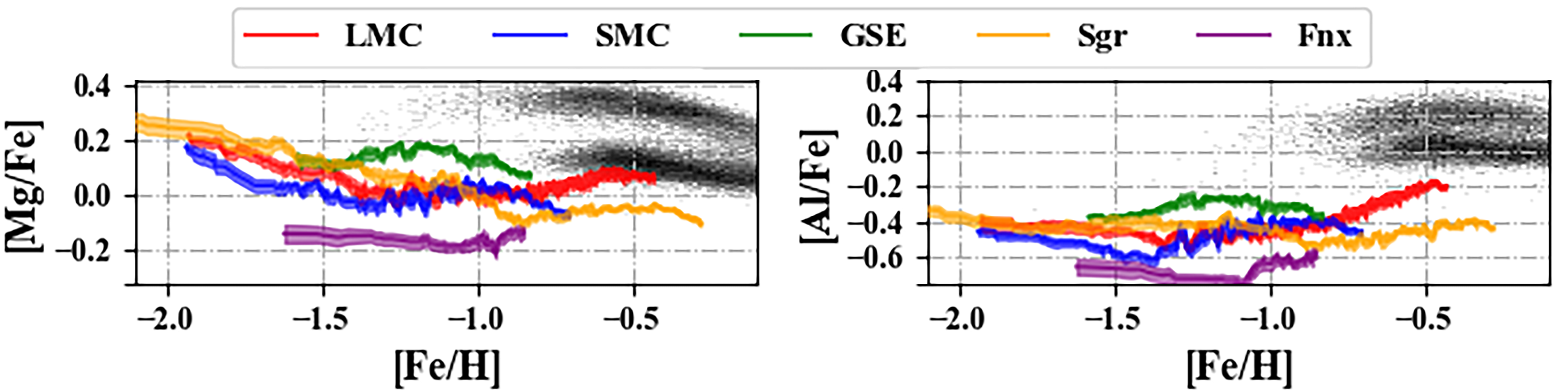}
\caption[]{Median trends in [Mg/Fe] (left) and [Al/Fe] (right) as a function of [Fe/H] for stars in GS/E (green), LMC (red), SMC (blue), Sgr (yellow) and Fornax (purple). Note that for [Fe/H]$>-1.5$,  the GS/E median trends in both [Mg/Fe] and [Al/Fe] are above those for the four massive intact satellites. [Adapted from \citet{Hasselquist2021}].}
\label{fig:Hasselquist}
\end{figure*}

Even though the GS/E progenitor is no longer intact, its pre-disruption properties can be studied \textit{now}, provided the constituent stars are identified (see Sections~\ref{sec:gse_sel} and ~\ref{sec:gse_kin}). As the GS/E stars flood the Solar neighourhood, as aptly pointed out by \citet{Hasselquist2021}, the community had been exploring the GS/E's chemical trends long before we had realised these stars were all of common origin \citep[see][]{Nissen2010,Hawkins2015,Fernandez-Alvar2018,Hayes2018}. Thus, the main trends, i.e. the downward slope of the $\alpha$-[Fe/H] track \citep[][]{Nissen2010} and the constant low [Al/Fe] ratio \citep[][]{Hawkins2015} were known before the \gaia\ data started to arrive. These trends are updated, improved upon, and contrasted with those in surviving dwarf galaxies in \citet{Hasselquist2021} (see Fig. \ref{fig:Hasselquist}). They show that compared to the four most massive current satellites of the Milky Way, namely the LMC, the SMC, Sgr (holding on for dear life), and Fornax, the GS/E progenitor attains demonstrably higher [Mg/Fe] at $-1.5<$[Fe/H]$<-1$ (but still some 0.1 dex below the disc sequence at the same metallicity). Clearly, the GS/E dwarf was more efficient at converting gas into stars compared to the present-day survivors. \citet{Hasselquist2021} hypothesize that this is due to its proximity to the Milky Way. Equally, the environment may have also played a crucial role in shaping the SFHs of the Magellanic Clouds which appear to have lived a mostly dull and lethargic life in the suburbs of the Galaxy. However, having formed next to a bigger galaxy did not end well for the GS/E progenitor: its life was cut short, although none of the models in the literature reveal signs of abrupt truncation in star formation (possibly due to the rigidity of the SFH approximation).

The chemical trends of the GS/E can also be compared to other known debris in the stellar halo. For example, \cite{Horta2023b} compare the chemical abundances of several known substructures and find that some (namely, Arjuna,
LMS-1, I’itoi, and perhaps Sequoia --- see also \citealt{Monty2020, Feuillet2021}) are chemically indistinguishable from the GS/E, perhaps suggesting that they share a common origin. Thus, similarly to the dynamical association of the VOD and HAC with the GS/E (see Section \ref{sec:gse_kin}), care must be taken in defining distinct substructures in the halo. Indeed, a massive structure like the GS/E is omnipresent, and can have several sub-components.

Curiously, the first constraints on the details of the GS/E SFH based on the behaviour of its stars in the $\alpha$-[Fe/H] plane are also pre-\gaia\ \citep[][]{Fernandez-Alvar2018}. This can be compared to the models of \citet{Vincenzo2019} and \citet{Sanders2021} that were built to reproduce a much more fine-tuned selection of the likely GS/E members. All three models agree on the relatively short SFH of the GS/E progenitor: star formation activity had to stop after 3-4 Gyr because by then the dwarf was stripped of its gas and fully disassembled by the Milky Way tides. The star formation efficiency of \citet{Fernandez-Alvar2018} and \citet{Vincenzo2019} are similar at 0.3-0.4 Gyr$^{-1}$,
whilst that in \citet{Sanders2021} is almost an order of magnitude lower at $\approx0.06$ Gyr$^{-1}$. To infer the SFH of the dwarf, the former two studies use theoretical elemental yields, while the latter simultaneously and self-consistently models a wide range of elements and pins down both the yields and the global parameters of the chemical enrichment model. The amplitude of the resulting star formation rate in the model of \cite{Vincenzo2019} is 3 $M_{\odot}$yr$^{-1}$, some 7 times that of the model in \citet{Sanders2021}. Integrating the SFR over GS/E's lifetime, \citet{Vincenzo2019} find a total stellar mass of $\approx5\times10^9M_{\odot}$. Although not quoted directly in \citet{Sanders2021}, their model would infer an order of magnitude lower total stellar mass of $\approx5\times10^8M_{\odot}$. \citet{Hasselquist2021} provide two models of the GS/E chemical evolution, one with a very broad SFH lasting more than 10 Gyr and one with a much narrower peak in agreement with the works above. Even though trained only on the [Mg/Fe]-[FeH] data, this latter model has a peak SFR much closer to the results of \citet{Sanders2021}.

Most of the chemical studies above use the exquisite APOGEE data. However, APOGEE is somewhat limited in its ability to measure heavy element abundances, and, in particular, has only scarce data on neutron-capture tracers (only Ce is listed amongst the relatively reliably measured species as part of the APOGEE DR17). The information on $s$- and $r$-process had to come from elsewhere, namely from the data releases of the GALAH survey \citep[][]{Desilva2015,Buder2018,Buder2021} and individual follow-up studies. These challenging measurements (due to a small number of weak lines) yielded a result not many had predicted: a noticeable enhancement in the $r$-process abundances in the GS/E stars compared to the rest of the halo (including the in-situ stars where available at comparable metallicities) and most of the surviving dwarf satellites \citep[][]{Aguado2021,Matsuno2021}. Measuring the rather low [Ba/Eu]$\approx0$ in their follow-up VLT spectra, \citet{Aguado2021} deduce that the neutron-capture production in the GS/E progenitor proceeded mostly via the $r$-process channel. Note that while [Eu/Fe]$\approx0.65$ and [Eu/Mg]$\approx0.3$ are at the upper end of what is observed, these values are still noticeably below the extreme $r$-process enhancement detected in one ultra-faint dwarf, namely Reticulum 2 \citep[][]{Ji2016} which reaches [Eu/Fe]$>1$, but similar to levels in other $r$-process enhanced UFDs such as Tucana 3 \citep[][]{Hansen2017_tuc} and Grus 2 \citep[][]{Hansen2020_grus}. Focusing on the tracks in [Eu/Mg]-[Mg/Fe] space for the GS/E stars with [Fe/H]$>-1.5$ in GALAH, \citet{Matsuno2021} conclude that the bulk of $r$-process in the dwarf must be due to neutron star mergers.  \citet{Naidu2022} observe GS/E stars together with targets they designate ``Kraken'' with Magellan and measure (amongst other lines) Eu, Ba, and Mg abundances. At similar [Mg/H], the GS/E has 0.3 dex higher [Eu/Mg]. \cite{Naidu2022} assume that the ``Kraken'' had a more rapid star formation activity and conclude that the enrichment due to the neutron star mergers must be delayed by some 0.5 Gyr. Note that most ``Kraken'' stars analysed in this study would also pass the cuts for the Milky Way pre-disc in-situ population (Aurora --- see \ref{sec:aurora}). Not only do these stars have low orbital energy and high Mg abundance, 16 out of 20 objects in the sample lie above the accreted sequence in the plane of [Al/Fe] vs [Fe/H] (see panels 2 and 3 in Fig. 2  of \citealt{Belokurov2022}) and along the rapidly rising in-situ track (see Fig. 2 in \citet{Horta2021}.  In our opinion, replacing ``Kraken'' with ``early Milky Way'', the arguments presented in \citet{Naidu2022} remain valid, or perhaps, become even more coherent.

\subsection{Globular clusters} 

Owing to the strong evidence for a significant GS/E contribution to the inner halo of the Galaxy, several groups have re-assessed the classification of Milky Way GCs into in-situ and accreted systems and updated the individual clusters' past GS/E membership.

\citet{Massari2019} assemble a compilation of GC ages from various sources and, following the earlier studies of \citet{Forbes2010} and \citet{Leaman2013}, point out a clear bifurcation in the Milky Way's GC age-metallicity plane. In this space, at fixed age, GCs with higher metallicities are likely to have been born in a more massive and therefore more rapidly evolving progenitor system (the Milky Way) compared to the GCs with lower metallicities that could come from disrupted dwarf galaxies. Guided by the distribution of the GCs in the age-metallicity plane, \citet{Massari2019} set aside two distinct cluster groups that have likely been born in the Galaxy proper: i) those that show clear prograde rotation and therefore are linked to the Galactic disc and ii) those that are embedded close to the nucleus of the Galaxy (inside the inner 3.5 kpc) and designated as the ``bulge'' GCs. This approach allows \citet{Massari2019} to set aside some $\approx40\%$ of all Milky Way clusters, classified as in-situ. The remaining $\approx60\%$ of clusters, hypothesized to be accreted, are divided into groups according to their approximate location in the space of integrals-of-motion (IOM). For the IOM space of choice, \citet{Massari2019} relies on total energy $E$ and the vertical component of angular momentum $L_z$. Limiting themselves to low $|L_z|$ and a broad range of $E$, they classify 26 as past members of GS/E with a further possible 6 members. While some groups of GCs appear more-or-less standalone in the IOM space (such as those belonging to GS/E and Sgr) others are less obvious (such as those assigned to the Helmi stream). Furthermore, several GCs, some at high $E$ and some at low $E$, are left without a known progenitor. The low-energy group, in particular, becomes a focus of attention for several follow-up studies \citep[see e.g.][]{Kruijssen2020,Forbes2020}. However, \citet{Horta2020gcs} see no obvious chemical difference between the GCs in the low-energy group and the rest of the in-situ clusters.

The premise of the study by \citet{Kruijssen2019} is identical to the starting point in \citet{Massari2019}, namely that in-situ and accreted populations behave differently in the age-metallicity space. Instead of linking the chrono-chemical properties of the Milky Way globulars to their orbital behaviour, \citet{Kruijssen2019} attempt to decipher the make-up of the Galactic GCs using the cluster E-MOSAICS models based on the EAGLE Cosmological numerical simulations \citep[][]{Pfeffer2018,Kruijssen2019_emosaics}. Using the E-MOSAICS framework, \citet{Kruijssen2019} isolate the 3 most significant events in the Milky Way's past, contributing the largest numbers of accreted GCs. However, without the use of the orbital information, identifying the progenitors of these systems turns out to be rather difficult. While the smallest group associated with the Sgr dwarf appears to be self-evident, the two most massive events are much more uncertain. Some 22 GCs are attributed to i) the so-called Canis Major dwarf \citep[see][]{Martin2004} and ii) the hypothetical Kraken galaxy, predicted by \citet{Kruijssen2019} to be the largest event in the Milky Way's life, bringing with it, some $\sim2\times10^9 M_{\odot}$ of stellar mass. The current consensus on the accreted nature of the Canis Major is rather sceptical given the disc origin of the various ``streams'' associated with it \citep[see e.g.][]{Xu2015,PriceWhelan2015,deBoer2018,Sheffield2018,Laporte2020}. Most importantly, however, neither of the two groups of Galactic GCs identified by \citet{Kruijssen2019} exhibits coherent orbital behaviour and thus both are instead most likely amalgamations of parts of different mergers. The analysis of \citet{Kruijssen2019} is trail-blazing but falls short due to the limited quality of the age-metallicity information for the Galactic GCs and the neglect of the orbital information. An attempt to rectify the latter is made in \citet{Kruijssen2020}, which re-uses the IOM classification made by \citet{Massari2019}. In a similar vein, \citet{Forbes2020} starts with GC groups constructed by \citet{Massari2019} and tweaks them very slightly using the age-metallicity information at hand.

\begin{figure*}
\includegraphics[width=\linewidth]{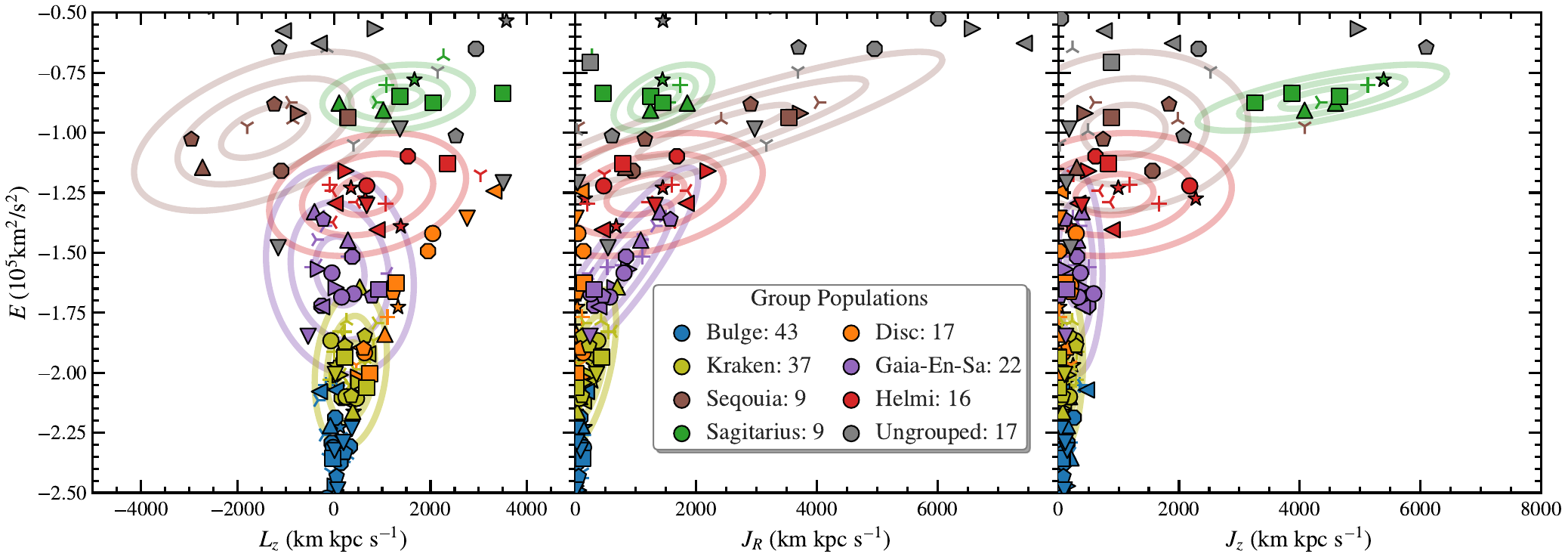}
\caption[]{Groups of GCs in action-angle space inferred from chemodynamical modelling. The coloured symbols are the observed GCs: different colours indicate different groups, with each GC coloured by its most likely group. The accreted groups of GCs are indicated with their $1-,2-,3-\sigma$ confidence intervals. [Reproduced from \cite{Callingham2022}].}
\label{fig:Callingham2022_GCgroups}
\end{figure*}

\citet{Myeong2019} focus on a group of retrograde Galactic GCs which they associate with Sequoia, an accretion event distinct from the GS/E but taking place around the same time in the Galaxy's past. Compared to the GS/E, Sequoia is of lower mass and potentially may have even been a companion to the GS/E's progenitor \citep[but see][for an alternative hypothesis in which the retrograde stars assigned to Sequoia are instead a part of the GS/E's outer disc]{Koppelman2020}. Similarly to \citet{Massari2019} and \citet{Kruijssen2019}, \citet{Myeong2019} use the age-metallicity relation to isolate groups of GCs but rely on actions to describe their orbital behaviour. The use of action space improves the detection of individual accretion events and helps \citet{Myeong2019} to update their GS/E membership from the initial 8 clusters published in \citet{Myeong2018gcs} to a total of 21. The main factor leading to this uptake in the GS/E membership is due to the inclusion of GCs with slightly lower total energies. 

\citet{Limberg2022} combine the most recent chemical and dynamical insights into the composition of the GS/E debris to update the GS/E GC membership. As several groups before them, they take advantage of the exquisite quality of the APOGEE chemical measurements to note the unique trajectory of the GS/E stars in the abundance space. In particular, \citet{Limberg2022} highlights the raised levels of [Mg/Mn] and [Al/Fe] \citep[see also][]{Hasselquist2021}. \citet{Limberg2022} combine the detailed chemistry, ages and actions \citep[similar to e.g.][]{Myeong2018gcs,Myeong2019} to identify a high-probability sample of 19 GS/E GCs. They also build a strong argument in support of the hypothesis in which Omega Cen is the surviving nuclear cluster of the GS/E progenitor dwarf \citep[see also][for a similar discussion]{Massari2019}. 

Note that caution is needed when classifying GCs using light elements like Mg and Al because these can be strongly affected by the cluster's chemical evolution \citep[see e.g.][]{Bastian2018,Gratton2019,Milone2022}. To avoid the clusters' anomalous chemical patterns when identifying their original host galaxy, the use of heavy elements has been advocated \citep[see e.g.][]{Minelli2021, Monty2023}. In particular, \citet{Monty2023} demonstrate that the two GCs, NGC 288 and NGC 362, often assigned to the GS/E based on their ages and chemo-dynamical properties \citep[][]{Helmi2018, Massari2019} are clearly distinct in the ratios of $r$- and $s$-process to $\alpha$ elements. \citet{Monty2023} point out that NGC 362 is $r$-process enhanced to levels similar to the GS/E stars while NGC 288 is not. The elevated ratio of $r$-process to $\alpha$ elements is in agreement with the trends found in GS/E stars \citep[see][as well as Section~\ref{sec:GSE_chem}]{Aguado2021,Matsuno2021}. The subtle variation in chemical fingerprints detected by \citet{Monty2023} is perhaps reflected in dramatic structural differences of these GCs.

Compared to all preceding attempts, the study of \citet{Callingham2022} stands out as the only unbiased GC classification scheme. Instead of hand-picking groups of clusters in the chosen parameter space, they devise a method similar to Gaussian Mixture Modelling. For the Milky Way GCs, ages, metallicities, energy, and three orbital actions are modeled with component mixtures to obtain objective membership probabilities for each GC (see Fig. \ref{fig:Callingham2022_GCgroups}). This innovative approach is demonstrated to work well for simulated data. Unfortunately, \citet{Callingham2022} also reports a large degree of overlap between GCs across all dimensions studied. To mitigate against the arising degeneracies, \citet{Callingham2022} decide to initialize their mixture models to the GC classes reported by \citet{Massari2019}. 

Concerned with the separation between the GCs born in-situ and those formed in dwarfs and later accreted, \citet{BK2023_nitrogen} go back to the idea of the ``critical energy'' highlighted in \citet{Myeong2018gcs}. They show that in the stellar halo too, a rather sharp boundary exists in $E, L_z$ space which divides stars into mostly accreted and mostly in-situ. This division is mapped by the clear change in the overall levels of [Al/Fe] ratio amongst the field stars. Using the $E, L_z$ boundary delineated by the [Al/Fe] levels, \citet{BK2023_gcs} classify all Milky Way GCs into two classes and find 106 out of 164 to be of in-situ origin, and 58 to be accreted. Therefore the most up-to-date GS/E membership of $\sim20$ GCs translates into $\sim1/3$ of the total accreted cluster population in the Galaxy. Note that although the proposed GC classification scheme based on their position in the $E, L_z$ plane has recently been independently verified in \citet{Chen2024}, \citet{BK2023_gcs} caution that some $\approx10\%$ of the objects classified as in-situ could be of accreted origin. This will bring down the GS/E fractional contribution to the accreted GC population in the MW.

Finally, \cite{Valenzuela2024} recently argue that the GS/E GCs themselves have a bimodal age distribution, which they suggest is the result of a ``wet'' GS/E merger that both brings in its own GCs while also forming new GCs in the process of the gas-rich merger with the Milky Way. This intriguing scenario allows for the gas mass of the GS/E progenitor to be estimated based on the properties of its GCs.

\subsection{Simulations}

As mentioned in Section \ref{sec:gse}, the use of numerical simulations has been instrumental in our understanding of the GS/E progenitor. Indeed, both discovery papers \citep{Belokurov2018, Helmi2018} relied on the use of simulations to interpret the observational results. \cite{Koppelman2020} explored the same pure N-body simulation suite used by \cite{Helmi2018} to scrutinize a particular halo that coincidentally matches the observed properties of the GS/E very closely. More specifically, they show that if the GS/E's dwarf progenitor had a metallicity gradient in its stellar disc --- as would be typical for galaxies of that mass --- the resulting tidal debris might end up showing groups with distinct chemo-kinematic properties similar to the GS/E's low-$L_z$ ``blob'' and the Sequoia's retrograde branch. \citet{Khoperskov2023} draws inspiration from the study of \cite{Koppelman2020} and detect a decrease in the overall metallicity of the GS/E debris in the APOGEE data as a function of the total energy in the Milky Way's gravitational potential. They run tailor-made pure N-body simulations of GS/E accretion to map this energy gradient into a spatial gradient within the progenitor dwarf galaxy. These estimates are then bench-marked using the HESTIA suite of hydrodynamical zoom-in simulations of Milky Way formation \citep[][]{Khoperskov2023_hestia1,Khoperskov2023_hestia2}. \citet{Khoperskov2023} find that the APOGEE measurements are consistent with 0.1 dex/kpc metallicity gradient in the progenitor before disruption.

\begin{figure*}
\includegraphics[width=\linewidth]{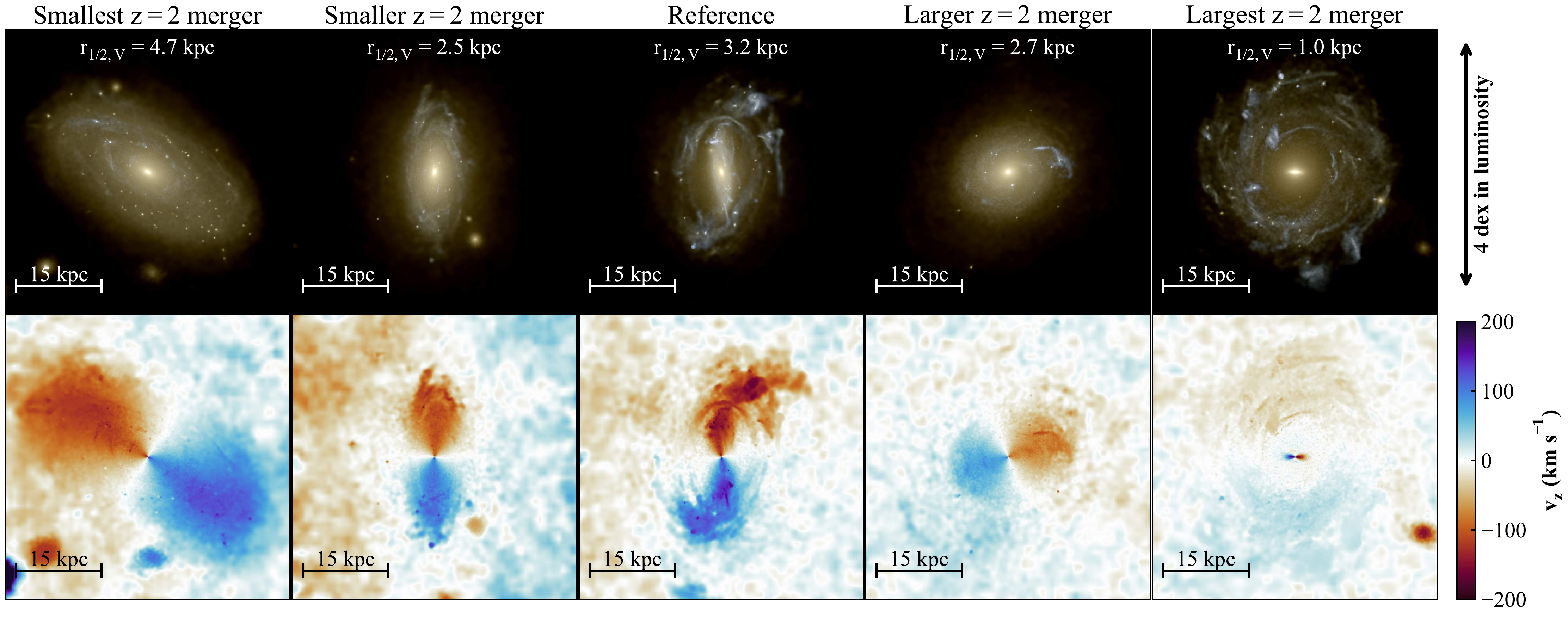}
\caption[]{Response of the galaxy to different early accretion events. From left to right shows increasing mass mergers. The top panels show the present-day stellar light, and the bottom panels show the kinematics. More massive mergers induce bulge-dominated structure, while less massive mergers result in rotationally dominated discs [Reproduced from \cite{Rey2023}].}
\label{fig:Rey2023_GM}
\end{figure*}

The study by \cite{Koppelman2020}  made the important point that the destruction of a massive dwarf galaxy leaves behind a complex chemodynamical structure. These ideas are similar but perhaps less pessimistic than the earlier insights of \citet{Jean-Baptiste2017} who show that the archaeological excavation of ancient accretion events can be made difficult or almost impossible if the incoming dwarf galaxies are massive enough \citep[see also][]{Pagnini2023}. The above works do make a good point that even though we can observe a multitude of substructures with very different properties, they can still all have the same origin (i.e. to the GS/E event). This idea was taken further by \cite{Naidu2021} who uses a suite of $N$-body simulations to confirm that a significant number of the observed properties of the (inner) stellar halo can be attributed to a single GS/E event. Note that important to both of these works is the inclusion of a disc potential in their simulations, which is essential in order to interpret any inner halo structure. 

While the above reflects a more tailored approach to the simulations of a Milky Way-GS/E merger, other works have focused on large suites of Milky Way-mass galaxies simulated in a cosmological context. In this case, an `ideal' Milky Way-GS/E system is harder to find, but such analyses are essential in order to put the Milky Way assembly history in context with the wider galaxy population. \cite{Fattahi2019} use the Auriga suite of simulations to show that approximately one-third of the (28) haloes have a dominant stellar halo component similar to the observed Milky Way. From this subset, \cite{Fattahi2019} deduce that these merger events typically occur 6-10 Gyr ago, with merging fragments of stellar mass $M_{\rm star} \sim 10^9-10^{10}M_\odot$. \citet{Orkney2023} extend the analysis of the GS/E events in  Auriga and highlight the diversity of possible progenitor systems, with a broad range of stellar, dark matter masses and rotational properties. \cite{Mackereth2019} use the EAGLE simulation suite to show that merger events with a similar pattern in chemistry and dynamics to the real GS/E are typically related to massive dwarfs with $M_{\rm star} \sim 10^{8.5}-10^9 M_\odot$. Moreover, they argue that the high eccentricity of the GS/E stars today suggests the merger would have taken place at $z \lesssim 1.5$. One particular Milky Way analogue within the EAGLE suite is additionally studied in greater detail by \citet{Bignone2019}. \citet{Elias2020} show that many of the conclusions reached through studies of GS/E-like events in suites of zoom-in simulations also hold in a much larger but lower-resolution sample from Illustris. Finally, \cite{Dillamore2022} uses the ARTEMIS simulations to again analyse GS/E-like events in cosmological simulations. In agreement with \cite{Fattahi2019} they find that typically a third of the Milky Way-mass systems contain a feature similar to the GS/E. 

Moving forward, a combined effort of tailored Milky Way-GS/E mergers \citep[e.g.][]{Koppelman2020, Naidu2021} and analyses of cosmological simulation suites \citep[e.g.][]{Fattahi2019, Mackereth2019, Dillamore2022} are needed to (a) link the observed properties of the GS/E to its initial state and co-evolution with the Milky Way, and (b) calibrate our various observational techniques to infer these properties. An important step in the right direction are the works of \citet{Amarante2022} and \citet{Rey2023} who create hydro-dynamical models of the GS/E-Milky Way encounter. \citet{Amarante2022} build tailor-made single merger models with realistic star-formation recipes and detailed chemistry. They show that many previously detected halo substructures can indeed be attributed to the single GS/E-like event. \citet{Rey2023} pioneers a new powerful and flexible way to carry out inference with Cosmological hydro-dynamical simulations. They use genetic modifications of the simulation's initial conditions \citep[see][]{Roth2016,Rey2018} to modify the properties (such as the mass ratio) of the GS/E merger. \citet{Rey2023} take a Milky Way zoom-in simulation from the VINTERGATAN suite and show how the properties of both the host and the resulting debris cloud change as the merger strength is varied. The Milky Way suffers more dramatic truncation and becomes more bulge-dominated as the GS/E mass is increased (see Fig. \ref{fig:Rey2023_GM}). In the absence of detailed chemistry, \citet{Rey2023} find a pronounced kinematic degeneracy for the stellar halo: different mergers and different mixtures of accreted and in-situ stars end up producing similar looking radially-biased ``sausage''-like structures.  

It is worth emphasizing a sobering point regarding simulated `Milky Way-like' galaxies. While they can be extremely helpful for understanding and interpreting the observational data, they are not, and never will be, the `real' Milky Way. Moreover, the selection effects present in the data are not often (and rarely fully) accounted for in the simulations. Thus, great care must be taken when using the simulations to draw conclusions about our own Galaxy.

\subsection{The puzzle of GS/E}

With the arrival of the \gaia\ data, Galactic Archaeologists find themselves struck by the epiphany of the most important merger in the Milky Way's accretion history. However, the grasping of the GS/E event is only starting and many of its crucial details remain blurry or even completely unconstrained.

\subsubsection{The timing of the GS/E}

At this stage, we are ready to move away from the assumption that the event was instantaneous and accept that it takes some time for the dwarf to fully merge. But, what was the duration of the GS/E merger? The truncation of star formation in the GS/E \citep[see e.g.][]{Bonaca2020} alludes to a $z \approx 2$ merger, but the details of the beginning and end of the Milky Way-GS/E interaction are still unclear. The pace of the interaction depends on the masses, the densities, and the initial angular momentum in the system \citep[e.g.][]{Amorisco2017, Vasiliev2022}. Can we reconstruct this information with the current data at hand, and, if not, what are we missing?

\subsubsection{The mass budget of the GS/E progenitor} 

The most recent stellar mass estimates of the GS/E are typically a few $\times 10^{8}M_\odot$ \citep[e.g.][]{Mackereth2020, Lane2023}, which is almost an order of magnitude lower than earlier estimates of $\sim 10^{9}M_\odot$ \citep[e.g.][]{Belokurov2018, Helmi2018, Feuillet2020}. This is likely due to a greater purity of GS/E stars in more recent studies, but, even now, a robust stellar mass estimate is lacking. Nonetheless, a lower mass estimate for GS/E has several important consequences. For example, it would imply less damage to the pre-existing disc --- this would agree with observations that the high-$\alpha$ disc is still intact. However, simulated analogues of the Milky Way-GS/E encounter would need to be updated in light of this seemingly preferred lower GS/E mass.

The contribution of dark matter from the GS/E to the inner parts of the Milky Way halo is expected to be small \citep[see e.g.][]{Fattahi2019}. However, the contribution could be enough to provide a non-axisymmetric component to the dark matter distribution \citep{Han2022}, which may have important implications for direct detection experiments on Earth \citep[see e.g.][]{Evans2019, Necib2019}. Moreover, a `tilted' dark matter halo
can induce a warp and flare in the Galactic disc \citep[see e.g.][]{Han2023, Han2023b}.

Finally, it is also worth considering what the gas content of the GS/E progenitor was. The early nature of the merger strongly suggests that the progenitor was gas-rich, but where and when was the gas lost? Could it contribute to (or re-start) the Galactic star formation? Does the two infall model of the Galaxy \citep{Chiappini1997} still hold weight in light of the apparent gas-rich GS/E merger? Can the GS/E GC population be used to estimate the gas mass of the progenitor \citep{Valenzuela2024}?

\subsubsection{Initial state of the progenitor}

Although much work is needed to characterize the GS/E debris \textit{today}, ultimately we also want to know what the initial state of the progenitor was (i.e. before any interactions with the Milky Way). How much information is encoded in the $z=0$ debris? For example, if the progenitor was a disc galaxy with a large satellite (Sequoia?), broadly similar to the LMC and SMC, it could produce a variety of halo substructures with distinct properties. However, even without a complex morphology and kinematics, high-mass events tend to break up into a large number of clumps of the $E, L_z$ space \citep[see e.g.][]{Jean-Baptiste2017, Belokurov2023_wrinkles}. Indeed, the `chevrons' in $r-V_r$ space uncovered by \cite{Belokurov2023_wrinkles} could result from the phase-mixing of the massive GS/E. However, \cite{Dillamore2023bar} recently showed that such stellar substructures in the local halo could also generated by bar resonances.

Finally, it is worth considering that a massive satellite such as the presumed GS/E progenitor likely had a satellite population of its own. Where are the dwarf satellites of the GS/E? Did they decouple from the GS/E dwarf before it started to radialize and sink? If so, do the GS/E satellites today (presumably ultra-faint dwarfs) resemble the original orbital plane configuration? Past work has shown that groups of satellites tend to disperse quickly after infall into a Milky Way-mass halo \citep[e.g.][]{Sales2011, Deason2015}, so it is unlikely that there is any kinematic link between the GS/E satellites and the progenitor debris today. However, are there other ways we can link known satellites to the GS/E?  And, what are the consequences for the overall Milky Way satellite population \citep[see e.g.][]{Bose2020}?

\subsubsection{Debris distribution in the Galactic halo}

Most of the analysis of the GS/E debris has been limited to the inner ($  5 \lesssim r/\mathrm{kpc} \lesssim 30$) halo. This, in part, is because here we have the richest multi-dimensional data, but also because the high eccentricity of the GS/E orbit implies there are plenty of GS/E stars at small Galactic radii. However, there is not much evidence (so far) for GS/E stars at low total $E$. How deep a satellite can sink in the halo depends not only on its mass, but also on (i) the density of the host and the satellite, and (ii) the initial amount of angular momentum \citep{Amorisco2017, Vasiliev2022}. The low $E$ regime is complicated by the overwhelming in-situ population, but there is potentially a rich amount of uncharted information about the GS/E lurking in the Milky Way's depths. 

At the other `end' of the Galaxy, there is still much to learn about the GS/E stars at large distances. Here, shells and earlier stripped GS/E stars are likely prevalent in the halo. Indeed, the distant `echoes' of the GS/E are now starting to be discovered \citep{Chandra2023}, and with increasing amounts of data at large Galactic radii, this will likely be a fruitful expedition in future studies. Mapping the GS/E debris over a wide range of radii (or energies) will be vital in order to reconstruct the full dynamical history of the Milky Way-GS/E interaction.

\subsubsection{A complete re-interpretation} 

Although the vast majority of the works described above advocate (or assume?) that the GS/E is a single, early, and massive progenitor, this is not the only interpretation in the literature. In particular, in a series of recent papers \citep{Donlon2022, Donlon2023a, Donlon2023b}, Donlon et al. argue that the GS/E in fact comprises debris from \textit{several} lower mass radial mergers, which moreover, were accreted fairly recently.  This alternative view, coupled with the large uncertainty in several of the GS/E properties (e.g. total stellar mass, time of accretion) shows that we are still far from a robust characterization of this stellar halo component. Perhaps the best way to avoid the degeneracy between total mass and the number of progenitors is the use of the well-known stellar mass-metallicity (MZR) relation (as several lower mass fragments will have lower metallicity than a single higher mass system). However, the biases associated with different selection methods \citep[see e.g.][]{Carrillo2024} coupled with uncertainties in the time evolution of the MZR has hampered this crucial line of evidence (see also Section \ref{sec:census}). Nonetheless, it is clear that a scenario accounting for \textit{all} of the chemodynamical evidence is required in order to fully solve the GS/E puzzle.

\section{The early Milky Way}
\subsection{In-situ stellar halo}
\label{sec:insitu}
\begin{figure*}
\includegraphics[width=\linewidth]{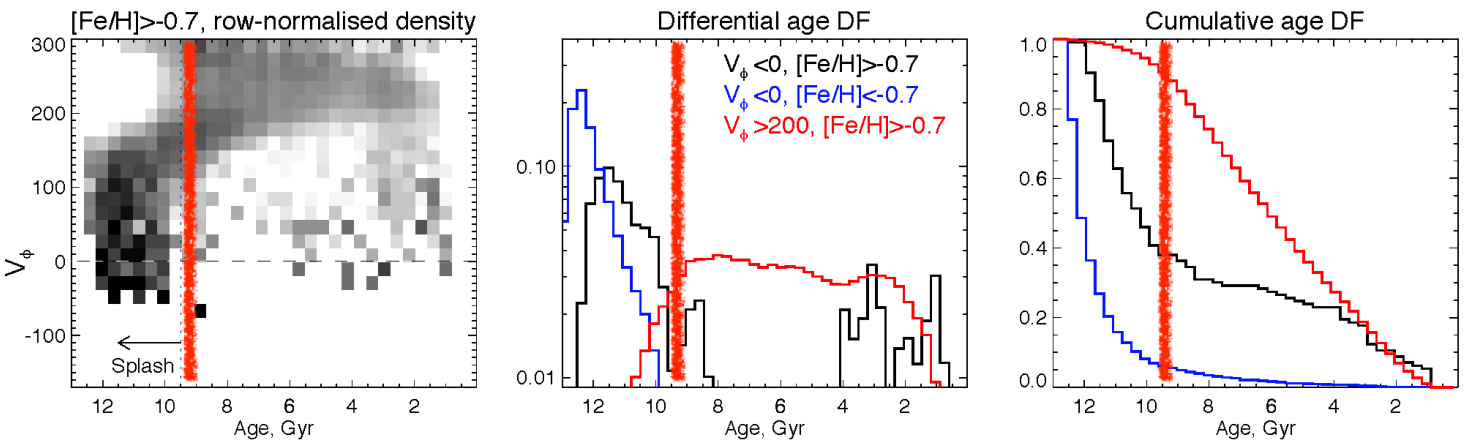}
\caption[]{\textit{Left:} Azimuthal velocity ($V_\phi$) as a function of age for local metal-rich stars selected from \gaia\ DR2. \textit{Middle:} Distribution of ages for stars in different bins of $V_{\phi}$ and [Fe/H]. Disc stars (red) are metal-rich with high $V_\phi$, Splash (black) is metal-rich with low $V_\phi$, and GS/E (blue) stars are metal-poor(er) with low $V_\phi$. \textit{Right:} Cumulative age distribution. \cite{Belokurov2020} argue that the truncated age distribution of the Splash stars (indicated by a vertical thick red line) can put a strong constraint on the time of the last major merger (if indeed the Splash is caused by this merger event)}. [Adapted from \citealt{Belokurov2020}]
\label{fig:Belokurov2020_splash}
\end{figure*}

Until now, our discussion has mainly revolved around a Galactic stellar halo solely comprised of an \textit{accreted} component. However, both observationally and theoretically, there are lines of evidence suggesting that some fraction of halo stars are born in-situ --- i.e. forming in the host halo itself \citep[e.g.][]{Zolotov2009, Font2011, Schlaufman2012, Bonaca2017}. The idea of an in-situ halo component dates back to the seminal \cite{ELS1962} work, which put forward the idea of a monolithic gas collapse model --- here, the halo and disc are drawn from the same population. While the mode of accretion from smaller (external) subcomponents gained more traction in the proceeding years, mainly thanks to the emergence of the $\Lambda$CDM paradigm, more recent work from hydrodynamical simulations revisited the idea of an in-situ halo component \citep[e.g.][]{Zolotov2009, Font2011, McCarthy2012, Tissera2013, Pillepich2015}. Indeed, many state-of-the-art cosmological simulations predict a certain fraction (sometimes quite significant) of halo stars have an in-situ origin. These stars are typically metal-rich and confined to the inner halo. However, the exact mode of how these stars form and resemble a halo-type population today (i.e. kinematically hot and pressure-supported) is unclear. Moreover, there remains uncertainty over whether or not some fraction of the in-situ halo components found in hydrodynamical simulations are artificial (i.e. owing to resolution and/or star formation prescription limitations, see e.g. \citealt{Zolotov2010, Cooper2015}).  

One interesting finding from most simulation efforts is that the in-situ halo is intimately linked to the overall assembly history \citep[e.g.][]{McCarthy2012, Cooper2015}. One key origin of an in-situ halo is the emergence of a `kicked-up' disc population, caused by a past merger event. Thus, although the accreted (or ex-situ) and in-situ halo stars can have vastly different birth sites, their origin can often be linked to the same major accretion events. As we have already discerned, we now know that the Galaxy digested the GS/E several Gyr ago --- what impact did this have on the Galactic disc, and did this event result in a residual in-situ halo?

\subsection{Splash}
\label{sec:splash}

Using \gaia\ DR2 data \cite{Belokurov2020} confirmed previous claims of a metal-rich `halo-like' component in the solar neighbourhood \citep[e.g.][]{Bonaca2017, Haywood2018}. A detailed analysis of the kinematics, abundances, and stellar ages of this metal-rich halo-like component (dubbed as the `Splash') led the authors to conclude that it is in fact linked to the thick disc. However, unlike the disc, the Splash typically has little or no net rotation, and some of the stars are even on retrograde orbits. A key part of the investigation was the availability of stellar ages --- these were computed by \cite{Sanders2018} for $\sim$ 3 million \gaia\ stars by combining astrometric, photometric, and spectroscopic datasets to calculate isochrone ages. The inclusion of `age' in the analysis allowed \cite{Belokurov2020} to deem that the Splash stars are predominantly old, but they are slightly younger than the GS/E (by $\sim 1$ Gyr) and overlap with the old age tail of the thick disc. The close connection between the disc and the Splash (they overlap both in age and in $V_{\phi}$ --- the tail of the disc runs into the Splash),  led to the hypothesis by \cite{Belokurov2020} that Splash is the population of stars originally born in the proto-disc of the Galaxy and subsequently kicked (splashed) into low-angular-momentum (high eccentricity) orbits by an accretion event that finished around $\sim 9.5$ Gyr ago. The best candidate for such an event is the GS/E merger! 

The Splash age distribution looks sufficiently different from that of the GS/E. For example, its peak is not at 12.5 Gyr but 1 Gyr later (see middle panel of Fig. \ref{fig:Belokurov2020_splash}). Even though the shapes of the age distribution of the GS/E and Splash stars are rather different, they have one particular feature in common: a truncation at 9.5 Gyr. \cite{Belokurov2020} used the synchronicity between the cessation of star formation in the GS/E and the finishing of the disc heating in the Milky Way to put a constraint on the epoch of the last major merger event (see Fig. \ref{fig:Belokurov2020_splash}). Similar `age-dating' analyses of local metal-rich and metal-poor stars were performed by other teams (e.g. \citealt{DiMatteo2019, Gallart2019, Bonaca2020}), and, although differing in the details, they confirm the accretion of the GS/E between 9-11 Gyr ago. 

The uncertainty in the precise timing of the events in the distant past of the Galaxy reflects the difficulty of obtaining reliable ages for large numbers of old stars, even in the {\it Gaia} era. Four main, distinct methods to infer stellar ages have been relied upon for archaeological studies, and each have been modernized and significantly improved thanks to the availability of the {\it Gaia} data. These four involve i) colour-magnitude diagram fitting, ii) astero-seismological mass estimates, iii) [C/N]-based mass estimates for red giants, and iv) isochronal fitting of the main sequence sub-giant and turn-off regions.

\citet{Gallart2019} demonstrate the power of the CMD fitting of large and homogeneous {\it Gaia} datasets to recover star-formation histories of the GS/E and the Splash in-situ stars. \citet{Montablan2021} determine astero-seismological ages for a handful of stars belonging to the GS/E and the Splash using the data from the {\it Kepler} space observatory. \citet{Mackereth2019_ages} infer spectroscopic ages (relying on C and N abundances) for red giant stars in the APOGEE dataset using a neural network trained on astero-seismological age measurements and employ them to study the evolution of the disc heating with age. \citet{Sanders2018} measure isochronal ages for a combined catalogue of stars with spectroscopy from APOGEE, Gaia-ESO, GALAH, LAMOST, RAVE, and SEGUE.

Irrespective of the chronological method used, there appears to be a good deal of consensus as to the synchronicity of the SFHs of the GS/E and Splash. Both the accreted and the in-situ stars are mostly old, with relatively few stars younger than 8 Gyr. However, the exact shapes of the age distributions of the nearby halo stars do vary quite noticeably between the methods applied and the samples used \citep[see][]{Gallart2019, Bonaca2020, Belokurov2020, Xiang2022}, and some works do find a tail of younger ages in the GS/E debris \citep[e.g.][]{Feuillet2021, Grunblatt2021, Horta2024b}. It is currently not clear if the ages and/or association of the intermediate stars in the GS/E are robust, but, clearly accurate ages are a vital ingredient for future studies of early-time Galactic archaeology.

\begin{figure}
\includegraphics[width=\linewidth]{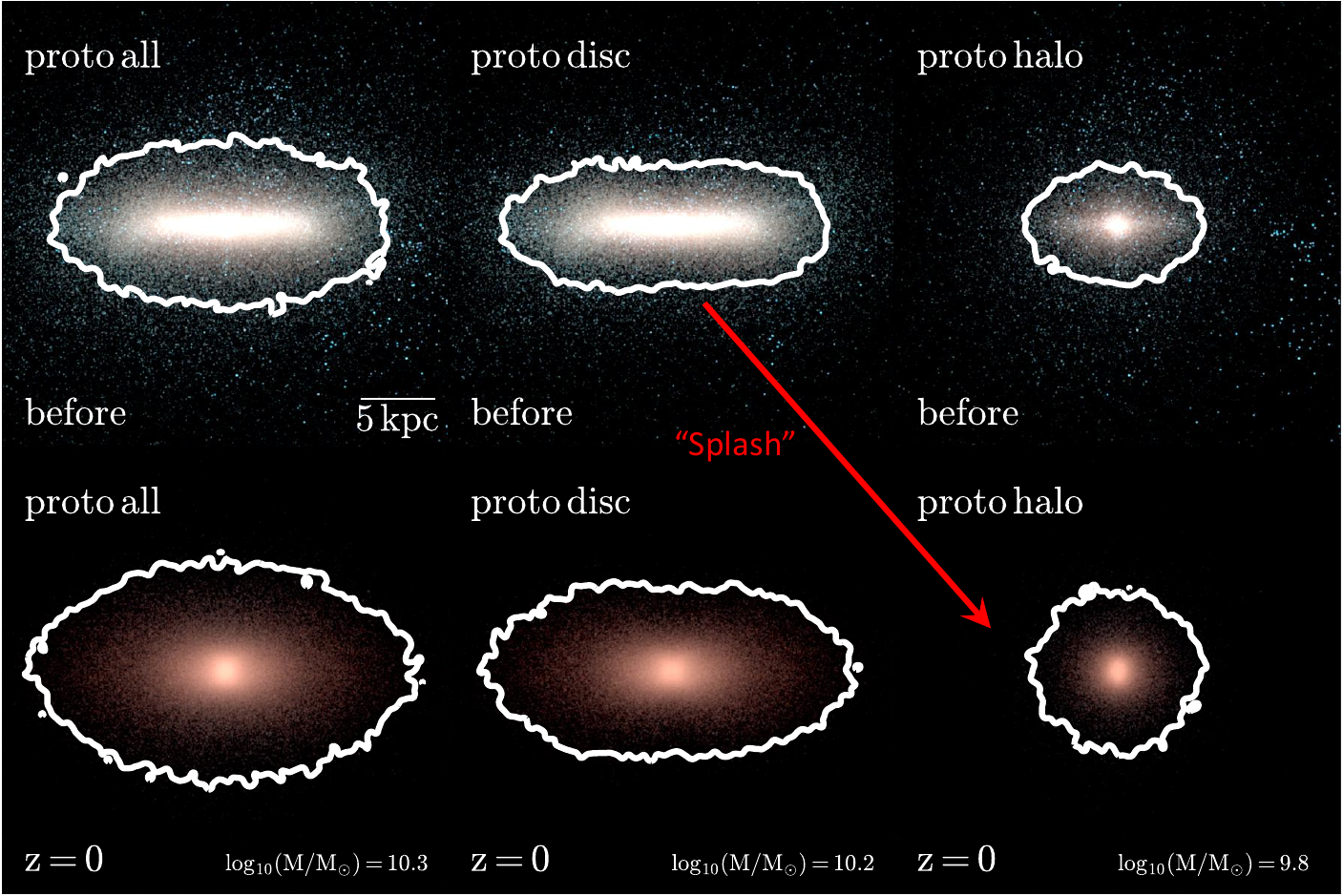}
\caption[]{Edge-on projections of the stellar light for the proto-galaxy populations in the Auriga-18 halo, which undergoes a GS/E-like merger. Each population has been kinematically decomposed into a disc (middle columns) and spheroid (right columns) component. Projections are shown before the GS/E merger (top row) and at redshift zero (bottom row). The GS/E merger kinematically heats the proto-galaxy into a more puffy configuration as some proto-disc stars are scattered into the halo (i.e. the Splash). [Adapted from \cite{Grand2020}].}
\label{fig:Grand2020_AuSplash}
\end{figure}

The link between major accretion events and `Splash-like' halo populations is also confirmed in state-of-the-art simulations of Milky Way-mass galaxies \citep[e.g.][]{Grand2020, Renaud2021}. Both \cite{Belokurov2020} and \cite{Grand2020} showed similar GS/E-Splash examples in the Auriga simulation suite. \cite{Grand2020} argue that gas-rich mergers can heat the proto-disc of the Galaxy, and scatter stars onto less circular orbits (see Fig. \ref{fig:Grand2020_AuSplash}). Such a population retains a correlation between rotation velocity and metallicity and thus contributes an `in-situ' halo component that connects the thick disc to the inner stellar halo. These findings from the latest cosmological simulations link back to earlier work  \citep[e.g.][]{Zolotov2009,McCarthy2012} that argued for a heated proto-disc origin for the in-situ halo. The intimate link between the disc, Splash, and GS/E stars presents another intriguing way to measure the properties of the GS/E progenitor. For example, \cite{Belokurov2020} discuss how the kinematic heating of the proto-disc is proportional to the impacting progenitor mass, and \cite{Grand2020} show explicitly in the Auriga simulations that the fraction of `kicked-out' stars in the local halo correlates with the progenitor mass. Thus, quantifying the fractional contribution of the in-situ halo stars provides an additional observational indicator of the GS/E progenitor mass. 

Although sustaining plenty of damage, the disc appears to have survived the interaction with the GS/E progenitor. However, both instantaneous and long-term effects of the battering on the Galactic star formation rate are not yet fully determined. Numerical simulations show that, in principle, such interactions can be both destructive and constructive, i.e. inducing a bout of star formation activity \citep[e.g.][]{Mihos1996, Springel2005, Hopkins2006, Brook2007}. The possibility that the GS/E merger could have driven a starburst in the heart of the Milky Way is explored with numerical simulations in \citet{Bignone2019}, \citet{Grand2020}, \citet{Renaud2021} and \citet{Dillamore2022}. Clues in the data for a merger-driven starburst have also been reported. A population of stars with matching properties is presented in \citet{Myeong2022} as {\it Eos}, in \citet{An2023} as {\it Galactic Starburst Sequence}, and in \citet{Ciuca2024} as {\it Great Galactic Starburst}. 

The idea that at some point in the past the Galactic reservoir was rapidly replenished with poorly-enriched gas has been entertained for a while \citep[see e.g. the review by][]{Matteucci2021}. Variants of this so-called two-infall model \citep[][]{Chiappini1997} can to some degree explain the emergence of two $\alpha$-[Fe/H] sequences observed in the Milky Way disc \citep[][]{Hayden2015}. The timing of the GS/E merger well-aligned with the Galactic high-$\alpha$ to low-$\alpha$ transition is rather fortunate for a pure coincidence. Note, however, that the $\alpha$-[Fe/H] bimodality in the Galactic disc can be reproduced with a model without mergers but with a continuous star formation and radial migration instead \citep[][]{Schonrich2009a,Schonrich2009b,Sharma2021}. Thus, while the origin of the disc sequences in the Milky Way is still under debate, there is certainly compelling evidence that the GS/E plays a major role in their early evolution.

Finally, it is worth remarking that the origin of the Milky Way's `Splash' (or metal-rich halo) is still under debate, and, with the current data, it is not possible to rule out other formation scenarios. Possible other formation channels include smoothly accreted gas and/or gas stripped from gas-rich satellites \citep{Cooper2015}, gaseous outflows \citep{Maiolino2017, Gallagher2019} and even `clump' scattering \citep{Amarante2020}. In the future, more detailed chemical abundance analysis of the Splash coupled with accurate age analysis is likely the most fruitful way to shed light on these possible origins. In any case, clearly the study of this metal-rich component, with links to both the disc and halo, is of vital importance to our understanding of the early phases of the Milky Way formation.

\subsection{Pre-disc Milky Way}

The synchronous truncation of the SFHs in the GS/E and the Splash is a compelling marker of the epoch of the last significant merger in Milky Way history. In this picture, Splash therefore contains some of the the Galactic disc's earliest forming stars \citep[see also][]{Montablan2021}. As its age distribution demonstrates, the first of these stars were in place $\approx$12 Gyr ago. Are there any hints as to the state of the Galaxy even before that? The view of the most ancient Milky Way appears blurred and fragmented due to the scarcity of age measurements for its oldest stellar populations. In the absence of reliable age estimates, stellar metallicity can be used as an age indicator. However, a metallicity-age relationship is quasi-monotonic only for stars born in the same galaxy, and under the assumption of one-zone chemical enrichment. In the Milky Way, this poses a problem in the low-metallicity halo regime where contributions from multiple accreted progenitors are expected. 

Luckily, additional chemical tags can be used to attempt to separate the accreted and in-situ born stars at [Fe/H]$<-1$. These ideas are first presented in \citet{Hawkins2015} who suggest that the accreted and in-situ halo stars attain distinct levels of [Al/Fe] abundance ratio. The stellar nucleosynthetic yield of aluminium (and sodium) is low at low metallicity but increases dramatically as more of C, N, O is pumped into the interstellar medium \citep[see][]{Kobayashi2006} which causes a delay in Al enrichment compared to conventional $\alpha$-elements. Compared to more massive systems, such as the progenitor of the Milky Way, dwarf galaxies are less efficient at retaining and re-using gas during star formation and are therefore slower to reach higher levels of metallicity when Al production becomes efficient. As a result, in a dwarf, by the time Al starts to be produced in earnest, type Ia SNe begin to dominate Fe production, thus inhibiting the growth of the [Al/Fe] ratio. In a more rapidly evolving Milky Way progenitor, Al production kicks in before the Type Ia SNe contribution rises and thus the young Galaxy can enjoy a period of over-abundance of Al relative to Fe. The systematic difference between [Al/Fe] levels in dwarf galaxies and the Milky Way proper is now well established \citep[see e.g.][]{Hasselquist2021} and is routinely taken advantage of to identify accreted stars in the nearby halo \citep[e.g.][]{Hawkins2015,Das2020,Horta2021}.

\subsubsection{Aurora}
\label{sec:aurora}
\begin{figure*}
\includegraphics[width=\linewidth]{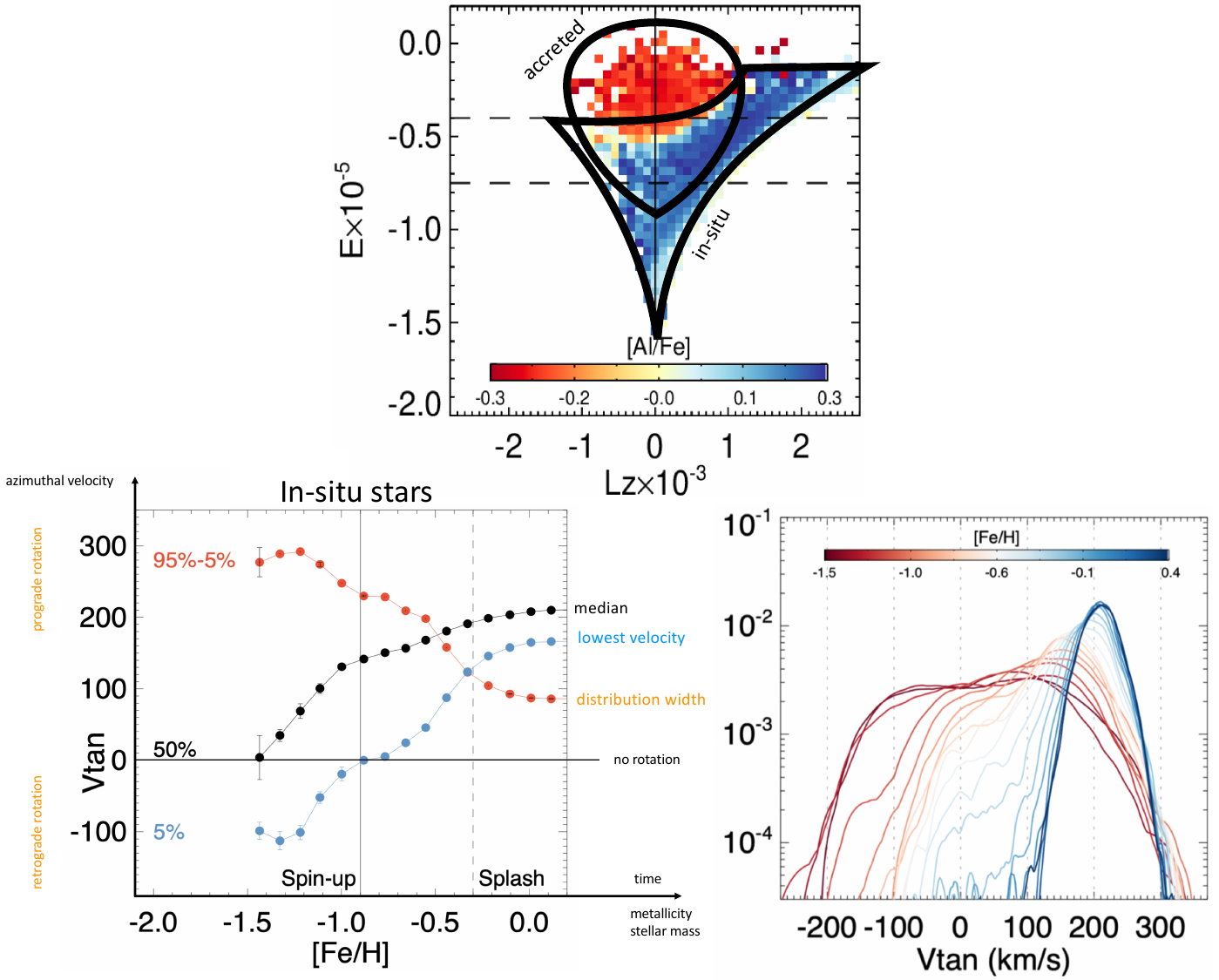}
\caption[]{\textit{Top panel:} Distribution of APOGEE DR17 + \gaia\ DR3 stars in Energy and angular momentum space coloured by median [Al/Fe]. There is a clear separation between the in-situ (high [Al/Fe]) and accreted (low [Al/Fe]) populations. \textit{Bottom-left panel:} Evolution of the in-situ azimuthal velocity distribution as a function of metallicity. The black, blue, and red points show the median, low $V_{\rm tan}$ wing, and distribution width, respectively. The solid and dashed vertical lines indicate the Spin-up (rapid increase in spin) and Splash (emergence of low $V_{\rm tan}$ wing) phases. \textit{Bottom-right panel:} Probability density function of the in-situ $V_{\rm tan}$ in bins of [Fe/H]. At low metallicities the in-situ population (i.e. Aurora) has low net spin and a wide distribution of $V_{\rm tan}$. [Adapted from \cite{Belokurov2022}].}
\label{fig:BelokurovKravtsov2022}
\end{figure*}

Instead of focusing on the accreted debris, \citet{Belokurov2022} combine [Al/Fe] measurements published as part of the APOGEE DR17 and the \gaia\ EDR3 astrometry to create a pure Milky Way in-situ sample, including stars at low metallicities (see top panel of Fig. \ref{fig:BelokurovKravtsov2022}). They show that while there may be some amount of false negatives, i.e. genuine in-situ stars identified as accreted, the in-situ selection is largely uncontaminated. Having access to a pure set of stars born only in the Milky Way means that the metallicity can now be used as the age proxy. Studying the behaviour of azimuthal velocity of the Milky Way stars as a function of [Fe/H], \citet{Belokurov2022} identify several key epochs in the life of the Galaxy, marked by noticeable changes in the stellar kinematics. For example, focusing on stars with retrograde motion, they see an edge to the Splash population at [Fe/H]$\approx-0.35$ in agreement with earlier studies \citep[e.g.][]{Bonaca2017,Haywood2018,Helmi2018,DiMatteo2019, Gallart2019, Belokurov2020}. Across $-1<$[Fe/H]$<-0.35$, Splash co-exists with the high-$\alpha$ (``thick") disc which, while heated by the interaction with the GS/E, appears largely intact, i.e. it retains a relatively high amplitude of rotation. However, going further back in time to [Fe/H]$<-1$, the median azimuthal velocity drops precipitously low and the spread of the azimuthal velocity distribution reaches its largest value.  In other words, starting instead from high redshift, the early Milky Way began without coherent rotation and then went through a rapid ``spin-up" phase at $-1.3<$[Fe/H]$<-1$ from which point onward it has been dominated by a fast-rotating disc (see Fig. \ref{fig:BelokurovKravtsov2022}). 

\citet{Belokurov2022} demonstrate that, as a function of [Fe/H], these dramatic kinematic transformations are accompanied by clear changes in chemical abundances. More precisely, they show that the abundance scatter in most elements considered evolves as a function of metallicity in sync with $V_{\phi}$. The scatter is largest in the pre-disc era, then shrinks during the ``spin-up" phase and stays approximately constant thereafter. Four elements, Al, N, Si and O show a particularly prominent evolution with metallicity, reaching much higher values of scatter in the pre-disc phase compared to the accreted halo population at fixed [Fe/H]. \citet{Belokurov2022} chose to give the pre-disc in-situ population the name {\it Aurora} appealing to its connection to the dawn of star formation in the Milky Way.

To place their findings in context and reveal a plausible close-up view of the young Milky Way as predicted by current state-of-the-art models, \citet{Belokurov2022} compare the APOGEE+\gaia\ observations to numerical simulations of galaxy formation, namely FIRE \citep[][]{Wetzel2023} and Auriga \citep[][]{Grand2017}. In both sets of simulations, Aurora (the state the Milky Way starts in) is characterized by a messy, turbulent spatial stellar distribution with a low overall angular momentum. Interestingly, some small amount of net prograde rotation is present at early epochs in most simulated Milky Way-like galaxies. Therefore, the in-situ population born before the disc appears to have angular momentum similar in amplitude and direction to the accreted halo \citep[see][]{Deason2017}. At the time of formation, Aurora stars can be seen close to the centre of the proto-Galaxy but lack order in both spatial and kinematic behaviour. As time passes and the Milky Way grows, Aurora settles and phase-mixes into a slightly flattened, quasi-spheroidal distribution with a very steep radial density profile. The Galactic spin-up phase is ubiquitous across the two simulation suites but the disc emergence typically happens at higher metallicities (and later times) compared to the observations. This view of the early Milky Way remains unchanged when other numerical simulation suites are considered, such as Illustris \citep[][]{Semenov2023} or ARTEMIS \citep[][]{Dillamore2023}. As a result of the Aurora discourse, recent works have paid more specific attention to the proto-disc galaxy populations in cosmological simulations \cite[see e.g.][]{Horta2024, McCluskey2024}.
 
\begin{figure}
\includegraphics[width=\linewidth]{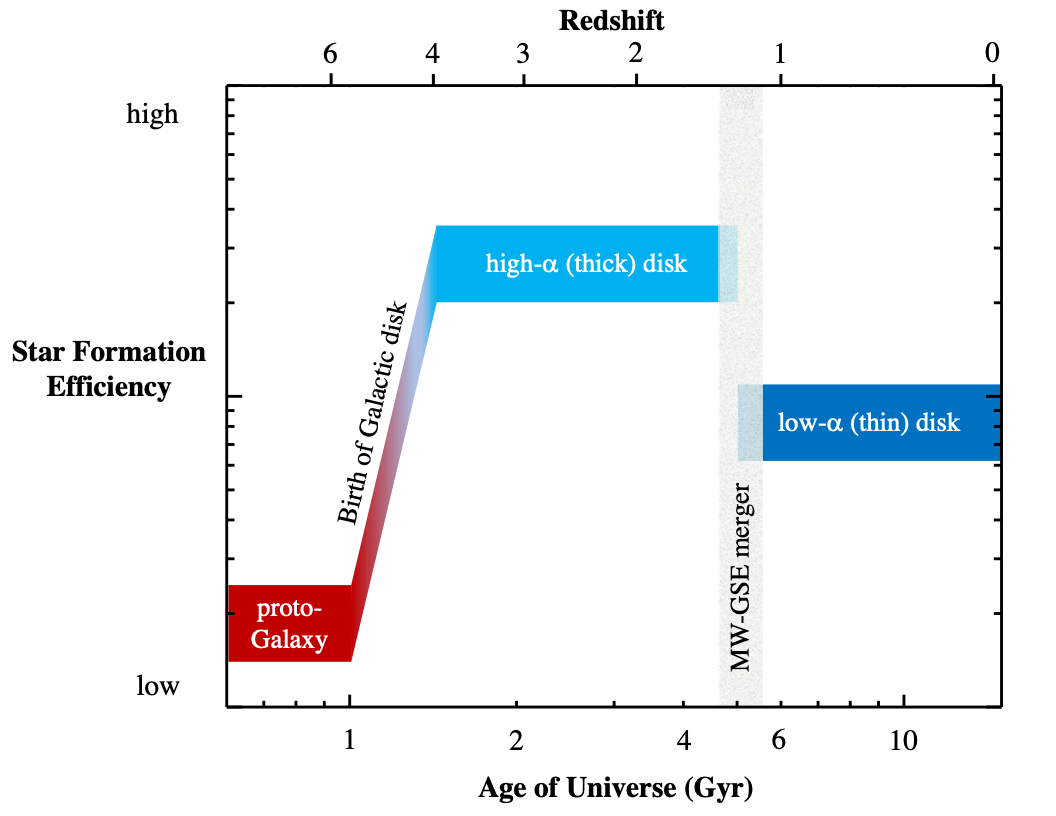}
\caption[]{A schematic overview of the Galactic star formation efficiency (SFE) over cosmic time.  [Reproduced from \cite{Conroy2022}].}
\label{fig:Conroy2022}
\end{figure}

Given the earlier studies of the GS/E interaction and the Splash formation, Aurora must certainly be older than $\approx$10 Gyr, and likely formed earlier than $\approx12$ Gyr ago. The study of \citet{Belokurov2022} however did not involve age estimates; moreover, it was largely limited to [Fe/H]$>-1.5$ as below this metallicity, the [Al/Fe]-based separation of stars into in-situ and accreted halo becomes much less effective. This deficiency is alleviated in the works of \citet{Conroy2022} and \citet{Rix2022} that both attempt to gain a view of the Galaxy at metallicities below [Fe/H]$\approx-2$. \citet{Conroy2022} use [Fe/H], [$\alpha$/Fe], and age estimates together with radial velocity measurements provided by the H3 survey to split the low-metallicity population into a non-rotating component (mostly GS/E) and a slowly rotating subset with a large velocity dispersion which the authors identify with the nascent disc of the Galaxy. Not only do these two samples have distinct kinematics, but their [$\alpha$/Fe] and age distributions are also different. At low metallicity, the slowly rotating in-situ component is clearly older compared to the GS/E: many stars have age estimates $>$12 Gyr, while the GS/E stars range from 8 to 12 Gyr. In the $\alpha$-[Fe/H] plane, the GS/E stars delineate a downward-sloping track or $\alpha$-knee consistent with earlier studies \cite[e.g.][]{Hasselquist2021}. The in-situ population shows a non-monotonic behaviour with an inflection around $-1.3<$[Fe/H]$<-1$ which the authors explain with changes in star formation efficiency. \citet{Conroy2022} interpret their low-metallicity, old, kinematically-hot, high-$\alpha$ population as the earliest phase of the Milky Way formation and find agreement with the properties of Aurora at slightly higher metallicity. An illustration of the Galactic star formation efficiency over cosmic time, including major events such as the proto-Milky Way and GS/E merger, is provided by \cite{Conroy2022} and reproduced here in Fig. \ref{fig:Conroy2022}. 

\citet{Rix2022} use overall metallicity [M/H] inferred from the \gaia\ DR3 low-resolution XP spectro-photometry to provide a panoramic view of the low-metallicity Milky Way. Importantly, this work is not limited to the solar neighourhood, which is the vicinity where Aurora was originally discovered. \cite{Rix2022} detect clear differences in the orbital and therefore spatial properties of their stars as a function of [M/H]. Above $-1.3<$[M/H]$<-1$, the Milky Way is dominated by a rapidly rotating disc, although accreted (GS/E) stars are also present. At [M/H]$<-1.3$ the rotation quickly disappears and eventually goes to zero below [Fe/H]$=-2$. \citet{Rix2022} also detect a change in the slope of the metallicity distribution around [M/H]$\approx-1$ which they surmise must be connected with the onset of the disc formation 12 Gyr ago in agreement with age determinations of \citet{Xiang2022}. \citet{Rix2022} conclude that the Milky Way's ``poor old heart" that they see in the \gaia\ XP data is most likely the pre-disc proto-Galaxy and is the extension into the central Galactic regions of the same population which had been seen by \citet{Belokurov2022} and \citet{Conroy2022} in the Solar neighourhood. 

A clearer picture of the orbital structure in the inner metal-poor halo emerges in the analysis of the Pristine Inner Galaxy Survey (PIGS) data \citep[][]{Arentsen2020,Arentsen2020b}. As 
\citet{Arentsen2024} demonstrate, below [Fe/H]$\approx-1$ the typical size of the orbital apo-centre is small, $<5$ kpc. However, at lower metallicities, i.e. [Fe/H]$<-2$ it starts to grow: only $\approx60\%$ of stars remain confined within 5 kpc. Thus, while \citet{Arentsen2024} agree that the Milky Way hosts a significant, centrally concentrated, metal-poor component with a modest prograde rotation, likely connected to the pre-disc proto Galaxy (i.e. Aurora), they also see clear evidence of a more extended halo component in the very metal-poor regime, i.e. at [Fe/H]$<-2$, which may be of accreted nature. The above PIGS study concurs with the earlier work of \citet{Lucey2021} who saw a similar pattern in the change of the apo-centre with metallicity in their smaller COMBS survey sample. Larger samples of (very) metal-poor stars from upcoming spectroscopic surveys (e.g. 4MOST) are on the horizon, and thus there is great scope to shed further light on the earliest stages of the Galaxy's evolution in the coming years.

\subsubsection{Kraken, Koala, and Heracles}
\begin{figure*}
\includegraphics[width=\linewidth]{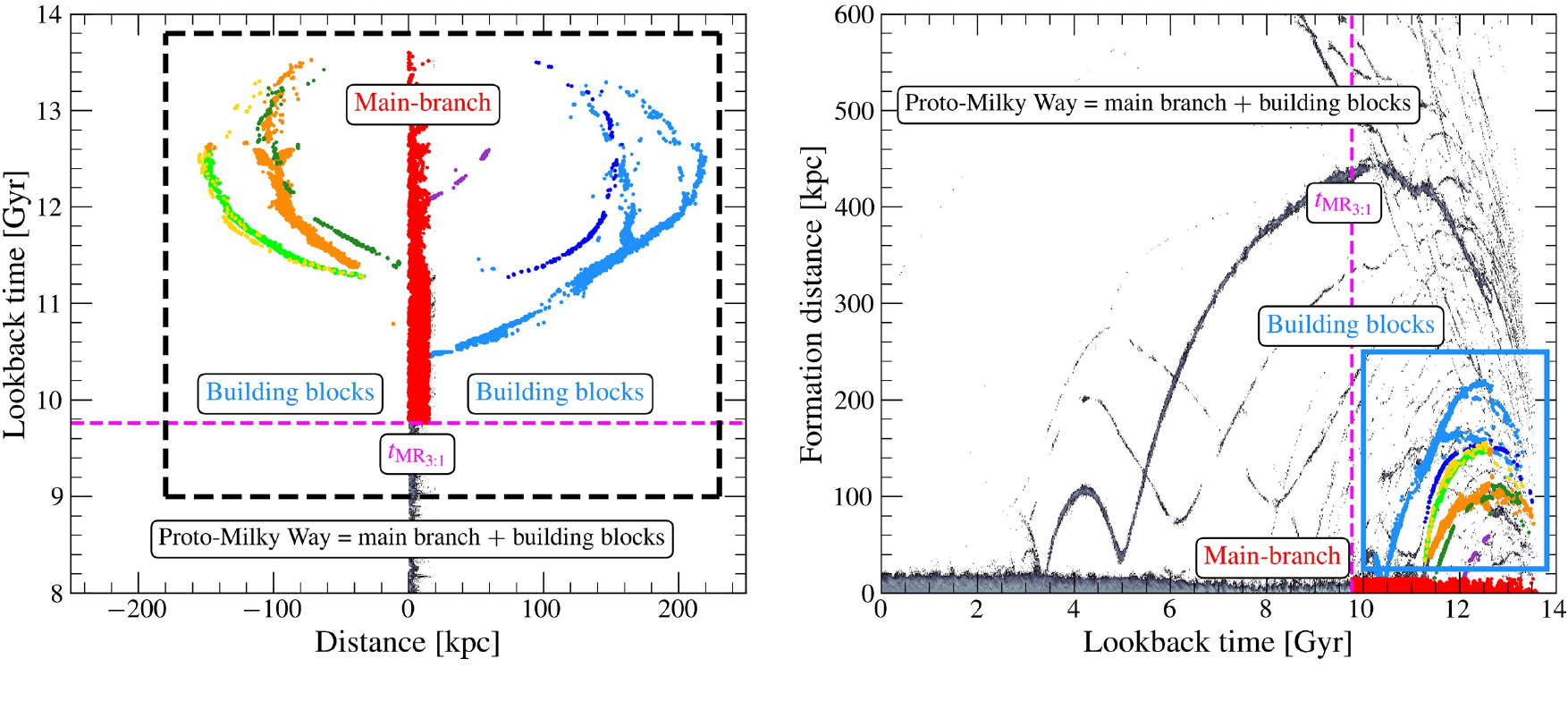}
\caption[]{\textit{Left panel:} The merger tree of the m12b simulation in Latte. The `proto-Milky Way' is defined as the main branch halo (red) plus all the building blocks that merge (other colours) before $t_{\rm MR_{3:1}}$ (the time at which the main branch reaches a stellar mass ratio of 3:1 with the second most massive luminous subhalo). \textit{Right panel:} Formation distance of star particles as a function of lookback time. At early times (i.e. before  $t_{\rm MR_{3:1}}$) the main branch and its building blocks can have very similar chemical properties. [Reproduced from \cite{Horta2024}.]}
\label{fig:Horta2024_bb}
\end{figure*}

The total mass of the Milky Way's stellar halo \citep[$M_{\rm star} \approx 1-2 \times 10^9M_\odot$, ][]{Deason2019, Mackereth2020} puts a limit on the number of significant accretion events the Galaxy could have experienced. These mergers ought to be massive enough to sink deep in the Galactic potential and thus populate the Solar neighourhood with their tidal debris. Equally, their progenitor galaxies are likely amongst the main contributors to the accreted Milky Way GCs. The latter idea is explored in \citet{Kruijssen2019}, who compare the age-metallicity properties of the simulated and the observed Galactic GCs to infer the total number of large mass ratio accretion events in the Milky Way. Working with a suite of 25 zoom-in simulations, \citet{Kruijssen2019} conclude that there have been at least two --- but more likely three --- main mergers in the Galaxy. The one with the lowest mass of these three is linked to the Sgr dwarf. The other two are more mysterious, in particular, the ancient interaction with the so-called `Kraken' galaxy. As noticed quickly by \citet{Massari2019}, the Kraken GCs identified by \citet{Kruijssen2019} are not dynamically coherent and could therefore originate from different progenitor systems. In dividing up the Galaxy's GCs, \citet{Massari2019} find a group of 24 unassigned clusters at low total energy. These unclaimed low-energy GCs are classified by \citet{Forbes2020} as a single progenitor named Koala, while \citet{Kruijssen2020} advocate that these GCs are part of Kraken. An alternative hypothesis is proposed by \citet{BK2023_gcs} and \citet{BK2023_nitrogen} who show that the bulk of the low-energy group originally highlighted by \citet{Massari2019} can instead be of in-situ origin and thus linked to the pre-disc Galactic population Aurora. 

The properties of the Kraken/Koala group of GCs are difficult to reconcile with the accretion of a single galaxy. These clusters have a wide range of eccentricities $0.2<e<0.8$ and an unusually broad spread of metallicity $-2.1<$[Fe/H]$<-0.5$. Overall, these GCs are rather similar to the rest of the in-situ clusters in \citet{Massari2019}, which come in two flavours: disc and bulge. If, for example, the bulge boundary used was slightly larger than 3.5 kpc, many of the low-energy clusters would be included. Indeed, as pointed out by \citet{Callingham2022}, there is a large degree of overlap between low-energy (Kraken/Koala) GCs and the in-situ population. The quality of the age and metallicity measurements is not sufficient to make an unambiguous classification of these GCs either. Only 12 out of 24 have ages and of these, half have the ages of the oldest GC in the Milky Way. The other 6 are contemporaneous with the GS/E clusters (for example, in both groups, the youngest GC are $\sim10$ Gyr old). Finally, it is also unclear if the Milky Way's disc could withstand the combined force from two massive accretion events around $1.5<z<2$. 

Could instead a Kraken-like merger take place earlier in the history of the Galaxy, before its disc comes into existence? \citet{Horta2021} find chemo-dynamical evidence for such a primeval event in the central portion of the Milky Way using data from APOGEE and \gaia. Importantly, this work (unlike others which are limited to the solar neighourhood) investigates the nature of halo stars within 5 kpc from the Galactic centre, and thus, presumbably, in the vicinity where the majority of the mass from proto-Galaxy populations reside. Selecting stars with low [Al/Fe], indicative of the accreted population, they identify a gap in the total energy distribution, which separates the GS/E debris with high total energies from the more tightly bound, low-energy structure they dub `IGS'. They also give a name to the progenitor of the IGS --- Heracles. \citet{Horta2021} demonstrate that the properties of the IGS are clearly distinct from those of the GS/E both in the chemical abundance space and in orbital space (although, as later demonstrated by \citealt{Lane2022}, the energy gap used for the IGS identification may instead be the imprint of the APOGEE selection function). However, differences between the IGS and the in-situ stars are somewhat less clear. \citet{Horta2021} show that at low metallicity and early epochs, the Milky Way in-situ chemical track passes through the region of low [Al/Fe], thus overlapping with the IGS selection. They conclude that the in-situ ``bulge'' population can have chemical properties identical to those of the IGS. 

It is perhaps not impossible for the main progenitor of the Milky Way and a large, early-accreted dwarf galaxy to have similar chemistry if their masses were comparable. The dwarf's star formation history would be truncated abruptly as it merged with the main progenitor of the Galaxy, but in the beginning (at lower metallicity), its chemical track and that of the Milky Way may look indistinguishable. This scenario is explored and quantified in \citet{Horta2024} where the FIRE simulations are used to compare the Milky Way's main progenitor and its building blocks (see Fig. \ref{fig:Horta2024_bb}). The authors propose the following working definition of a ``building block'': it is a dwarf galaxy that merges with the Milky Way during the period when the main progenitor is not yet clearly dominant within the simulation volume (typically around redshift $z\approx3$). In the 13 galaxies within the FIRE suite, the main progenitor is easily identifiable: it contributes the most mass to the forming Milky Way, from $\sim70\%$ to $\sim95\%$. \citet{Horta2024} find that in 5 out of 13 galaxies considered, a building block exists that contributes between 15\% and 30\% of the total mass of the proto-Milky Way. While these building blocks have 2 to 5 times lower mass than the main progenitor, their chemical fingerprints are almost identical. Maybe, Heracles --- the parent galaxy of the IGS --- was such a building block and thus its chemistry and that of Aurora are indiscernible? 

Sometimes Kraken/Koala and Heracles are implicitly assumed to be the same entity. However, there may be considerable tension between the observational definitions of the two. Kraken is posited to be an accreted dwarf based on half a dozen GCs with age measurements that are located on the accreted branch in the age-metallicity space. At fixed age of $\approx10$ Gyr, these clusters are some $\approx0.7$ dex more metal-poor than the Milky Way in-situ GCs. Thus, the Kraken stellar population must be distinct from that of the early Milky Way, in contrast to the proposed building block, Heracles.

\subsection{Implications of the GS/E and the early disc formation}
\textit{An atypical assembly history?}
As we discussed in Section \ref{sec:intro}, the accretion of a small number of massive progenitors is not atypical for Milky Way-mass galaxies. However, what is perhaps more unusual in the case of the Milky Way is the \textit{lack} of such events since the GS/E merger ($\sim 10$ Gyr ago). \cite{Evans2020} explored this explicitly and used the large-volume EAGLE simulations to quantify how `unusual' the Milky Way accretion history is (at least with regard to massive progenitors). The authors find that only 5\% of the Milky Way-mass haloes ($M_{200c} = 0.7-2 \times 10^{12}M_\odot$) in EAGLE have accreted a GS/E-like progenitor around 8-11 Gyr ago, but have had no subsequent massive merger events. This fraction lowers to $<1$ \% of haloes if the late-time accretion of the (surviving) LMC is also used as a constraint on the assembly history. An interesting consequence of the lack of other major mergers since the GS/E is that the haloes satisfying the Milky Way assembly history constraints in EAGLE are biased towards lower mass haloes within the `Milky Way-mass' range, i.e. the haloes are typically $\leq 1 \times 10^{12}M_\odot$. This bias is due to a long period of quiescence during which the haloes did not accrete as much mass as a ‘typical’ Milky Way galaxy.

Another repercussion of this `quiescent' period in the Milky Way's history is that the inner halo as we see it today provides a window to the high redshift Universe. In particular, the GS/E debris is a relic of an ancient, massive dwarf galaxy. Dwarfs with similar mass that are still intact today are very different (e.g. the LMC) as they have vastly different star formation histories. As shown by \cite{Evans2022}, galaxies resembling the GS/E should be observable with JWST beyond redshift $z \sim 2$. Thus, our Galactic archaeological dig of GS/E with \gaia\ data can be compared with the properties of similar mass dwarf galaxies at high redshift found with JWST, which presents a fascinating way to connect the local Universe with the early stages of galaxy formation.

\cite{Bose2020} showed how the particular accretion history of the Milky Way can also affect its satellite galaxy population. Note that the satellite luminosity function is often used as a key probe of $\Lambda$CDM models, and thus any systematic influence on this fundamental relation is particularly important. In particular,  \cite{Bose2020} find that, at fixed mass, haloes that form earlier typically contain a larger number of ultra-faint satellites than those haloes that form later. Moreover, the radial distribution of these satellite galaxies is more centrally concentrated. It is clear that the existence and timing of massive merger events in our Galaxy are crucial for many of its `global' properties (i.e. total halo mass, stellar halo, satellites, disc formation), and it is necessary to understand these events in order to put the Milky Way in context with other similar mass galaxies.

\textit{The impact of GS/E on the disc (trans)formation:} A significant merger like the GS/E does not only heat up the pre-existing Galactic disc, it can also torque, flip, and tilt it. The pervasiveness of such disc flipping and tilting is uncovered by \citet{Dillamore2022} when analysing the host response to satellite mergers in the ARTEMIS numerical simulations. They show that both the dark matter halo of the Milky Way and its disc change shape and orientation as a result of the interaction. Even though the Galaxy manages to destroy the perturber, it tends to eventually align itself with its orbital plane. The re-aligning can happen quickly (disc flipping) and slowly (disc tilting). The exact mechanics of the disc flipping is exposed in \citet{Dodge2023} who build their intuition based on a set of tailored N-body simulations. They show that, as the GS/E progenitor falls apart in the inner regions of the Milky Way, it deposits a small amount (in relative terms) of dark matter in a strongly non-axisymmetric configuration. This non-axisymmetric component exerts a torque on the disc, forcing it to change orientation. As mentioned earlier, the existence of such a non-axisymmetric dark matter component is advocated by \citet{Han2023b} and discussed by \citet{Davies2023}. Both \citet{Dillamore2022} and \citet{Dodge2023} conclude that the Galactic disc tilting unleashed a long time ago by the GS/E is likely still ongoing today. This astonishing insight has inspired \citet{Nibauer2023} to ponder the long-term effects of the tilting disc on the properties of the stellar streams in the Milky Way halo. They show that the nearby streams can be impacted by the disc tilt in a variety of noticeable and non-trivial ways. Another striking and unexpected consequence of the GS/E's flybys is highlighted in the study of \citet{Renaud2021_realign}. In one of the VINTERGATAN simulations they consider, the galaxy assembles an extended gaseous disc around the time of a GS/E-like interaction whose plane is sufficiently misaligned with respect to the original, old disc such that the two evolve almost unconnected to each other. The passage of the GS/E progenitor first ignites star formation in the extended misaligned gaseous ring and then torques it to align with the pre-existing, main disc of the Galaxy. \citet{Renaud2021_realign} show that as a result, the Milky Way ends up hosting two neatly aligned, but independently grown stellar discs. This, they argue, could also be a viable path to produce the $\alpha$-bimodality similar to that observed in the Galaxy.

\textit{The Milky Way as a transient fossil:}
Although all signs point towards a quiescent Milky Way assembly history and early disc formation, in many ways, this is just a transient state of affairs. Indeed, we know that in a few Gyrs time, the Milky Way will experience its next major merger event when it eventually engulfs the LMC. How will this change the status quo? This question was directly pondered by \cite{Cautun2019} who used the EAGLE simulations to predict the impact of a late-time LMC-mass merger onto a Milky Way-mass galaxy. Unsurprisingly, the authors predict that the stellar halo will be radically changed. Its stellar mass and metallicity will likely shoot up to reflect the massive LMC progenitor. At present our relatively metal-poor, low-mass stellar halo is indicative of an early forming halo \citep[see e.g.][]{Deason2019}, but this signature will likely be overwhelmed by the destroyed LMC material.  Moreover, a new `Splash' will form from kicked-out disc stars. Not only will there be a more metal-rich Splash, but the disc itself will likely be thickened and perhaps distorted. Thus, the presence of the LMC (which we will describe in more detail below), is perhaps a contradiction to the overarching view of a quiescent Milky Way. In fact, as put forward by \cite{Deason2016} the Milky Way is more accurately described as a `Transient Fossil', whereby a recent accretion event can disguise the preceding formation history of the halo. We end this Section with the philosophical pondering that perhaps we are lucky to not be writing this review in $2-3$ Gyr time ---  would we even know that the GS/E existed, or what the early formation of the Milky Way entailed? \footnote{Perhaps, even more philosophically --- would we even be here?}

\subsection{The lowest metallicity stars}

Stars with the lowest metallicities are likely the oldest formed in a given environment, although the age-metallicity relationships can differ substantially depending on the mass of the host galaxy. \gaia\ is heavily relied upon to (1) make sense of the previously discovered metal-poor stars, and (2) to identify new low-metallicity stars missed by previous efforts. Before \gaia, no single spectroscopic survey managed to examine the low-metallicity regime of the Galactic halo comprehensively. As a result, the currently available sample of stars with [Fe/H]$<-2$ have strong and heterogeneous selection effects from a patchwork of different studies \citep[e.g.][]{Frebel2015}. As the \gaia\ data started to come out, efforts to consolidate the census of the metal-poor end of the stellar MDF accelerated, as exemplified by e.g. the study of \citet{Li2018}. With the DR3, \gaia\ has stepped up the game by producing the largest homogeneous catalogues of low-metallicity candidates selected using its high-resolution spectroscopy \citep[see e.g.][]{Matsuno2022, Viswanathan2023b}, and low-resolution spectro-photometry \citep[XP spectra, e.g.][]{Andrae2023, Yao2024}. 

The \gaia\ XP revolution is only beginning, but the first attempts to extract stellar atmosphere parameters and individual abundances from admittedly extremely low-resolution but homogeneous and enormous XP spectra are already bearing fruit. For example, some success has been reported in extracting $\alpha$-abundances from XP \citep[][]{Li2024}. Perhaps, this analysis can even be extended to the low-metallicity regime, where the exact behaviour of the in-situ $\alpha$-abundance remains a puzzle. According to \citet{Conroy2022}, stars with [Fe/H]$<-2$ may exhibit noticeably higher $\alpha$ values than previously thought, moving above the established $\alpha$ plateau. While the sensitivity of the XP spectra to changes in $\alpha$-abundance comes as a surprise, spectral variations due to changes in carbon abundance are less subtle and thus can be exploited \citep[see e.g.][]{Witten2022}. To this end, the first all-sky, homogeneous sample of carbon-enhanced metal-poor (CEMP) stars is built by \citet{Lucey2023} using \gaia\ DR3 XP spectra. CEMPs come in several flavours: those that show signs of $r$ and/or $s$-process enhancement and those that do not \citep[][]{Aoki2007}. The $r/s$-enhanced CEMPs are most common and likely linked to pollution by companions in binary systems \citep[][]{Lucatello2005, Starkenburg2014}. Instead, CEMP-no stars appear to be mostly single \citep[but see][]{Hansen2016, Arentsen2019} and thus their carbon-enhancement is likely primordial, and linked to the very first bouts of star formation in the Universe \citep[][]{Frebel2015}. Using \gaia's astrometry to shed light on the orbital properties of confirmed CEMP stars reveals that many of them are typical halo denizens and can be associated with known halo sub-structures \citep[such as GS/E, see][]{Zepeda2023}. 

\begin{figure*}
\includegraphics[width=\linewidth]{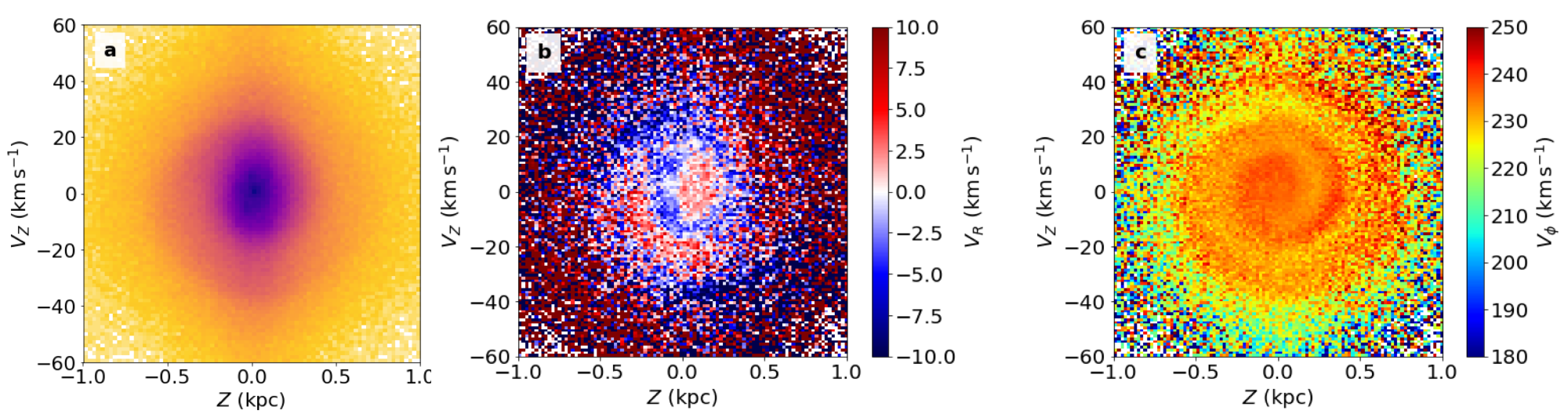}
\caption[]{\gaia\ DR2 phase-space spiral in the $z-V_z$ plane. The density is coloured according to (a) number density, (b) median $V_r$, and (c) median $V_\phi$. [Reproduced from \cite{Antoja2018}].}
\label{fig:gaia_snail}
\end{figure*}

Plotting global orbital properties of the low-metallicity stars discovered recently reveals a small but noticeable angular momentum asymmetry, with more stars found on prograde orbits at all metallicities, going down as low as [Fe/H]$\approx-6$ \citep[][]{Sestito2019, Dimatteo2020}. Note, however, that focusing on metal-poor stars with $r$-process enhancement, \citet{Roederer2018} do not see such an angular momentum asymmetry. Many of the low-metallicity stars on prograde orbits also spend most of their time close to the Galactic disc, at low $|z|$. These unexpected trends have inspired several hypotheses, including the suggestion that in the Milky Way, the stellar disc may have formed exceedingly early and at extremely low metallicities \citep[e.g.][]{Dimatteo2020}. Testing this idea against numerical simulations however reveals that in the Milky Way-like progenitors, dominant stellar discs form sufficiently late, at noticeably higher metallicities \citep[e.g.][]{Belokurov2022, Semenov2023, Chandra2023}. Note that, as discussed earlier, the MW itself appears to be somewhat of an outlier: a dominant stellar disc emerges in the Galaxy at the earliest possible epochs, i.e. around $3<z<5$, but at metallicities much above the metal-poor regime, i.e. [Fe/H]$>-1.3$. At higher redshifts, the Galaxy is in its kinematically-hot, bursty and messy Aurora state.
\cite{Sestito2021} address the origin of the low-metallicity stars on prograde, low-$|z|$ orbits, and conclude that most of them must be accreted. 
It remains unclear what induced the net spin in the low-metallicity halo, is it inherent or acquired? \citet{Dillamore2023bar} demonstrate that the Galactic bar can spin up intrinsically non-rotating stellar haloes to velocities similar to those observed. In their models, a halo spun-up by the bar shows a rotation curve declining with Galactocentric radius --- this prediction can be tested with upcoming data.

\section{Latecomers to the party and the census of accreted substructure}

\subsection{Sagittarius and the Milky Way disc}

The Sgr dwarf galaxy \citep{Ibata1994} and its associated stellar stream \citep[e.g.][]{Newberg2002, Majewski2003, Belokurov2006} are an archetype of dwarf galaxy disruption. The stellar stream emanating from a known dwarf galaxy is a clear example of accretion in action, and studies of the stream have been used to measure the Galactic potential as well as the orbit and mass of the Sgr dwarf progenitor \citep[e.g.][]{Law2010, Dierickx2017, Gibbons2017, Fardal2019, Vasiliev2021}. The mere fact that the progenitor is still intact points to a fairly recent accretion event (since redshift $z \sim 1$), and this has been confirmed by most modelling efforts. 

The omniscience of the Sgr stream has been obvious for several years, most notably displayed in the famous `field-of-streams' image from SDSS \citep{Belokurov2006}. However, the \gaia\ mission has allowed a more detailed examination of this well-known stream \citep{Antoja2020, Ibata2020, Ramos2020, Cunningham2024}.  In addition to properties of the stream itself, the \gaia\ era has provided evidence that Sgr has actually perturbed the Galactic disc during its orbit around the galaxy. Now, this is not a new idea --- several past works have shown that a Sgr-like dwarf could induce perturbations to the disc \citep[e.g.][]{Dehnen1998, Ibata1998, Quillen2009, Laporte2018}, and there has been evidence of disc perturbations from pre-\gaia\ data \citep[e.g.][]{Gomez2012, Widrow2012, Carlin2013, Schonrich2018}. However, \gaia\ provided unambiguous evidence that the galaxy is out of equilibrium, and is currently undergoing a phase-mixing process from an out-of-equilibrium state. The evidence for this phase-mixing comes from the prominent spiral feature in position and velocity discovered by \cite{Antoja2018} (see Fig. \ref{fig:gaia_snail}). This so-called \gaia\ phase-space `Snail' potentially arises from phase-mixing of past gravitational disturbances of the Milky Way disc. The spiral phase-space structures are studied in greater detail using \gaia\ DR3, where their chemical signatures can also be mapped thanks to the GSP-Spec catalogue derived from the RVS spectra \citep{gaia_cartography}.

Since its discovery, several works have attributed the \gaia\ Snail to pericentric passages of the Sgr dwarf \citep[e.g.][]{Binney2018, Khanna2019, Laporte2019, Bland-Hawthorn2021, Gandhi2022}. Is this yet another example of the co-evolution of the Milky Way disc and halo? We have previously discussed that the metal-rich `Splash' material is likely caused by the impact of the early GS/E accretion onto the proto-galaxy (see Section \ref{sec:splash}), and it also appears that the more recent Sgr accretion has influenced the Galactic disc. While it is worth noting that there are alternative hypotheses for the origin of the Snail, such as from many smaller disturbances rather than one large one \citep[e.g.][]{Hunt2022, Tremaine2023}, it remains true that in the \gaia\ era we cannot study the disc and halo in isolation, particularly when we are considering the inner halo. The co-evolution and co-existence of these components show that the accretion of substructures, particularly massive ones, can influence and shape the entire Galaxy.

\subsection{A massive LMC}

The most massive Milky Way satellites, the Large and Small Magellanic Clouds (LMC, SMC), have been known about for centuries. Visible to the naked eye in the Southern Hemisphere, their brightness is unmistakable. However, it is only in recent years that we have accepted the sheer massiveness of the Clouds, particularly the LMC. A simple abundance-matching argument already places the LMC at a significant mass: with a stellar mass of $2.7 \times 10^{9} M_\odot$ \citep{vanderMarel2002}, this roughly corresponds to a halo mass of $10^{11}M_\odot$ \citep{Moster2013}. A mass of $\sim 10^{11} M_\odot$ for the LMC also agrees with modelling efforts which aim to reproduce the significant 3D velocities of the Clouds and allow for LMC-SMC interactions \citep[e.g.][]{Besla2010, Kallivayalil2013}. Most tidal models now advocate for a massive satellite (carrying its smaller sibling) that has rapidly joined the Milky Way in the past couple of Gyr (although \citealt{Vasiliev2024} show that the observations are also consistent with the LMC being on its second passage around the Milky Way). More recent works using Magellanic `satellite-of-satellite' counts, or modelling of orbital streams in the presence of the LMC now agree that the LMC must have a mass $\gtrsim 1 \times 10^{11}M_\odot$ \citep[e.g.][]{Erkal2019, Erkal2020}

At the pericentre of the LMC ($\sim 50$ kpc), a mass of $\sim 10^{11}M_\odot$ is a significant fraction of the Milky Way halo mass out to that radius (roughly $\sim 1/3$). \cite{Gomez2015} show that a massive LMC at this proximity can significantly displace the centre-of-mass (COM) of the Milky Way. Thus, this substantial perturber in our Galaxy cannot be ignored, and the scale of the damage inflicted by the LMC is now starting to be revealed by the \gaia\ mission.

\subsubsection{LMC wake}

\begin{figure*}
\includegraphics[width=\linewidth]{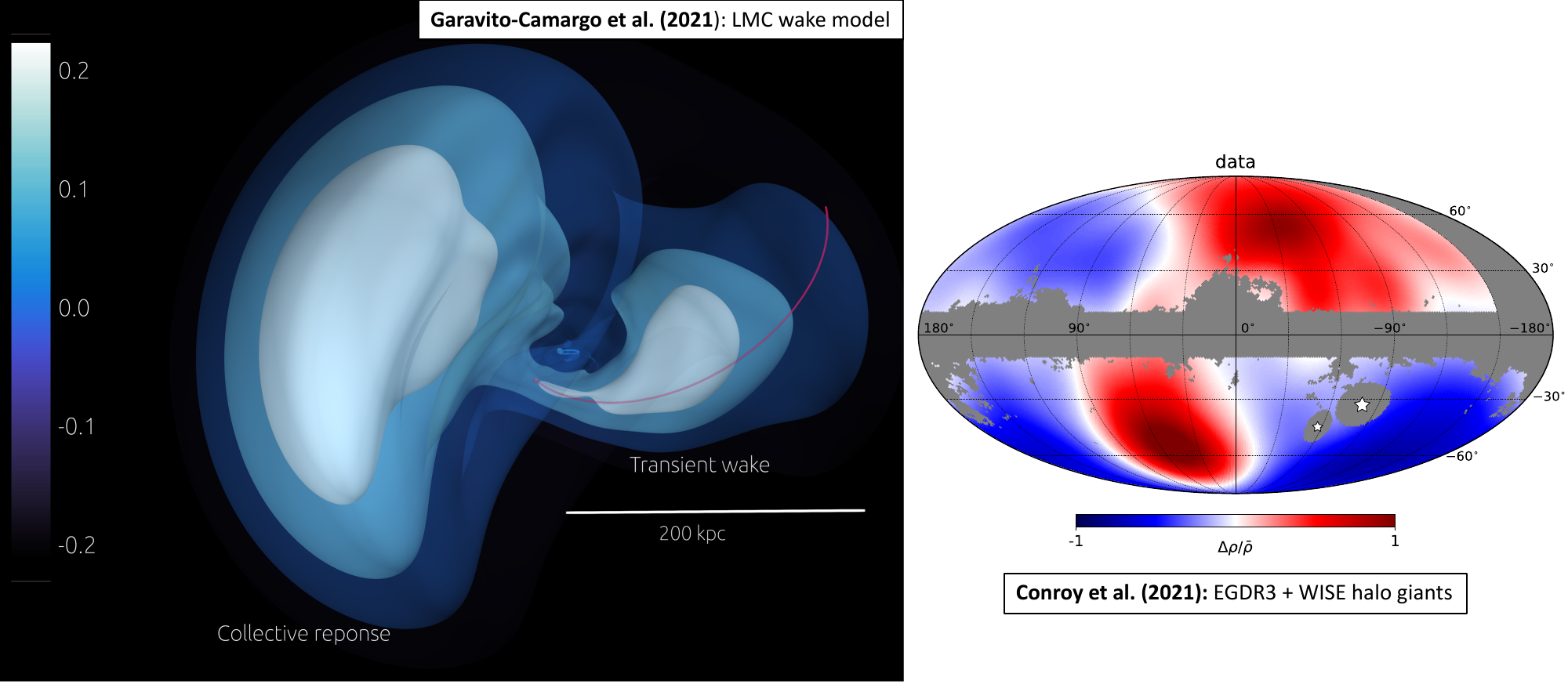}
\caption[]{\textit{Left:} A model of the LMC-induced dynamical friction (transient) wake and collective response in the Milky Way halo shown in the Galactocentric YZ plane \citep{Garavito-Camargo2021}.  The colorbar shows the density contrast (light = overdense, dark = underdense). The transient wake follows the past orbit of the LMC (indicated by the red line). The collective response appears predominantly to the North of the Galactic disc (central blue ellipse). \textit{Right:} Distribution of K giant stars in Milky Way halo at 60 kpc $<R_{\rm gal}<100$ kpc in Galactic coordinates \citep{Conroy2021}. The overdensity in the North and South-West are attributed to the collective response and local wake, respectively. [Adapted from \cite{Garavito-Camargo2021, Conroy2021}.]}
\label{fig:LMC_wake}
\end{figure*}

\cite{Garavito-Camargo2019} use $N$-body simulations to explore in detail the impact of the LMC's passage on the Milky Way halo. They find that the recent infall of this massive satellite inflicts substantial gravitational `wakes', which they decompose into a `collective' (or global) and `transient' (or local) response. The global response is reminiscent of the COM displacement discussed by \cite{Gomez2015}. Here, the presence of a massive perturber offsets the COM of the Milky Way-LMC system, but importantly, as the Milky Way is not a solid body, different parts of the Milky Way halo respond differently. For example, the outer parts of the Milky Way halo (with their longer orbital timescales) are slower to respond to this rapidly evolving scenario. The resulting `global' response is a dipole-like signature in the density and velocity space of the halo (see Fig. \ref{fig:LMC_wake}). The local response is attributed to the dynamical friction acting on the infalling LMC, which results in a collection of particles tracing the orbital path of the satellite. These signatures, in principle, should also be present in the stellar halo of the Galaxy, and this raises the intriguing possibility of detecting the wake in current and future datasets. \cite{Garavito-Camargo2019} also argues that the potential detection of the wake could have implications for the nature of the dark matter particle, as different dark matter models may predict different wake properties. Thus, there is considerable scientific motivation to detect and analyse the LMC wake in the Milky Way halo.

Initial efforts utilizing samples of halo stars selected using \gaia\ data have already revealed some promising results. For example, \cite{Erkal2021} and \cite{Petersen2021} find evidence for velocity gradients in the outer halo ($r \gtrsim 50$ kpc) that appear to match models of the LMC-Milky Way systems with a massive LMC. In particular, \cite{Erkal2021} measures a net blueshift for stars in the  Southern hemisphere and a net redshift in the North. The signal is consistent with the mostly downward acceleration of the inner halo due to the LMC (i.e. the Global response), and by comparing the results with simulations they find that the velocity gradient suggests an LMC mass of $\sim 1.5 \times 10^{11}M_\odot$. \cite{Conroy2021} use a sample of K giants selected using WISE and \gaia\ DR3 to create a density map of the (global) outer halo (see right panel of Fig. \ref{fig:LMC_wake}). Remarkably, the density of outer halo stars resembles the LMC wake structure, with signatures of both a Northern overdensity (i.e. collective response) and a local wake. Interestingly, the density contrast reported by \cite{Conroy2021} is \textit{stronger} than predicted by the models. 

It is worth pointing out that care needs to be taken in order to disentangle the `true' LMC wake signal from other potential contributions, by, for example, other substructures. \cite{Cunningham2020} show that spherical harmonic expansion can be a useful tool to disentangle perturbations on large scales, such as the LMC wake. However, they caution that stellar debris from recent, massive accretion events can complicate the analysis. An intriguing outer halo overdensity, that could potentially signify local LMC-wake material, is dubbed the `Pisces Plume' \citep[][]{Belokurov2019}. However, with limited data it is currently unclear whether this material is related to (1) the LMC-induced wake, (2) SMC stellar debris (stripped from interactions with the LMC itself), or (3) a poorly known outer wrap of the Sgr stream! The outer halo is ripe for uncovering the details of the predicted LMC-wake, and other known and unknown substructures. Future data releases of \gaia\ coupled with wide-area spectroscopic surveys such as DESI, WEAVE, and 4MOST \citep{Cooper2023, Jin2023, deJong2019} are ideally suited to investigate these outer regions in detail. Indeed, large samples of chemo-dynamical data are required to pin down the LMC-wake signal and disentangle/uncover contributions from other (likely recent) accretion events in the outer Milky Way halo.

\subsubsection{The Milky Way-LMC system}

The appreciation of the LMC's impact on the Milky Way has spurned a new era of dynamical modelling: we now need to consider the `Milky Way-LMC system' (or even the `Milky Way-LMC-Sgr system') rather than just a Milky Way host halo. The fundamental assumptions that govern many modelling techniques, such as equilibrium, static potentials, spherical or axisymmetric symmetry, etc, are no longer valid. What are the implications of this for the many decades of work aiming to measure the total mass of the Galaxy using halo tracer populations \citep[e.g.][]{Wilkinson1999, Xue2008, Deason2012, Callingham2019, Eadie2019}? A cautionary tale was spun by \cite{Erkal2021} who showed that mass estimates based on equilibrium modelling of tracers that ignore the presence of the LMC are typically biased high, and could even lead to an overestimate of up to 50 percent! On a more positive note, \cite{Deason2021} showed that by allowing for a small perturbation to the velocities of the halo stars a mass profile can be uncovered in a Milky Way-LMC potential. However, this depends on the radial range covered by the data and the region of the sky probed, as these authors focused on a region covered by SDSS where the predicted velocity gradients in the halo are small. \cite{Deason2021} also argued that systematic effects due to substructures, such as shells and clouds, can cause an even bigger effect than the LMC on mass-modelling efforts. Thus, one could argue that despite increasing sample sizes of multi-dimensional data, our aim to measure the Milky Way potential \textit{accurately}, at least using classical methods, is perhaps waning.

It is worth remarking in the post-\gaia\ era that we likely need to reassess what the `total mass of the Milky Way' even means. Particularly in light of the messiness caused by the LMC and, to a lesser extent, Sgr. Detailed modelling of the Milky Way-LMC-Sgr system \citep[][]{Vasiliev2021} has indicated that considering these major components together can prove vital. Indeed, recent stream-modelling techniques have shown that the LMC is required in order to produce stream tracks that are consistent with the data \citep[e.g.][]{Erkal2019, Vasiliev2021}. However, while tailored models are valuable there are still systematic effects revealed from cosmological simulations of stellar haloes that need to be taken into account in modelling efforts. How do we reconcile these two approaches? Perhaps the most promising avenue is the use of `Basis-Function-Expansion' (BFE) methods \citep{Lowing2011, Sanders2020, Garavito-Camargo2021, Petersen2022}. Here, a time-evolving potential from an $N$-body simulation is approximated using a small number of basis functions. Thus, at a much lower cost than the original simulation, subhalo orbits can be explored in this potential without the need to continually re-simulate the host. In principle, these methods could be used to include the cosmological context of subhalo accretion, whilst retaining the `realistic' potential of the Milky Way-LMC system. For example, Bayesian inference could be applied to a large suite of such models in order to infer global properties of the Galactic potential from the phase-space distribution of halo stars. The BFE models could even provide the framework to combine different dynamical modelling techniques, either from multiple streams and/or halo stars. With the exquisite data at hand, which will only increase in number and quality in the coming years, we clearly need to develop more sophisticated modelling approaches in order to keep pace with the data.

\subsection{Census of accreted substructures}
\label{sec:census}
\begin{figure*}
\includegraphics[width=\linewidth]{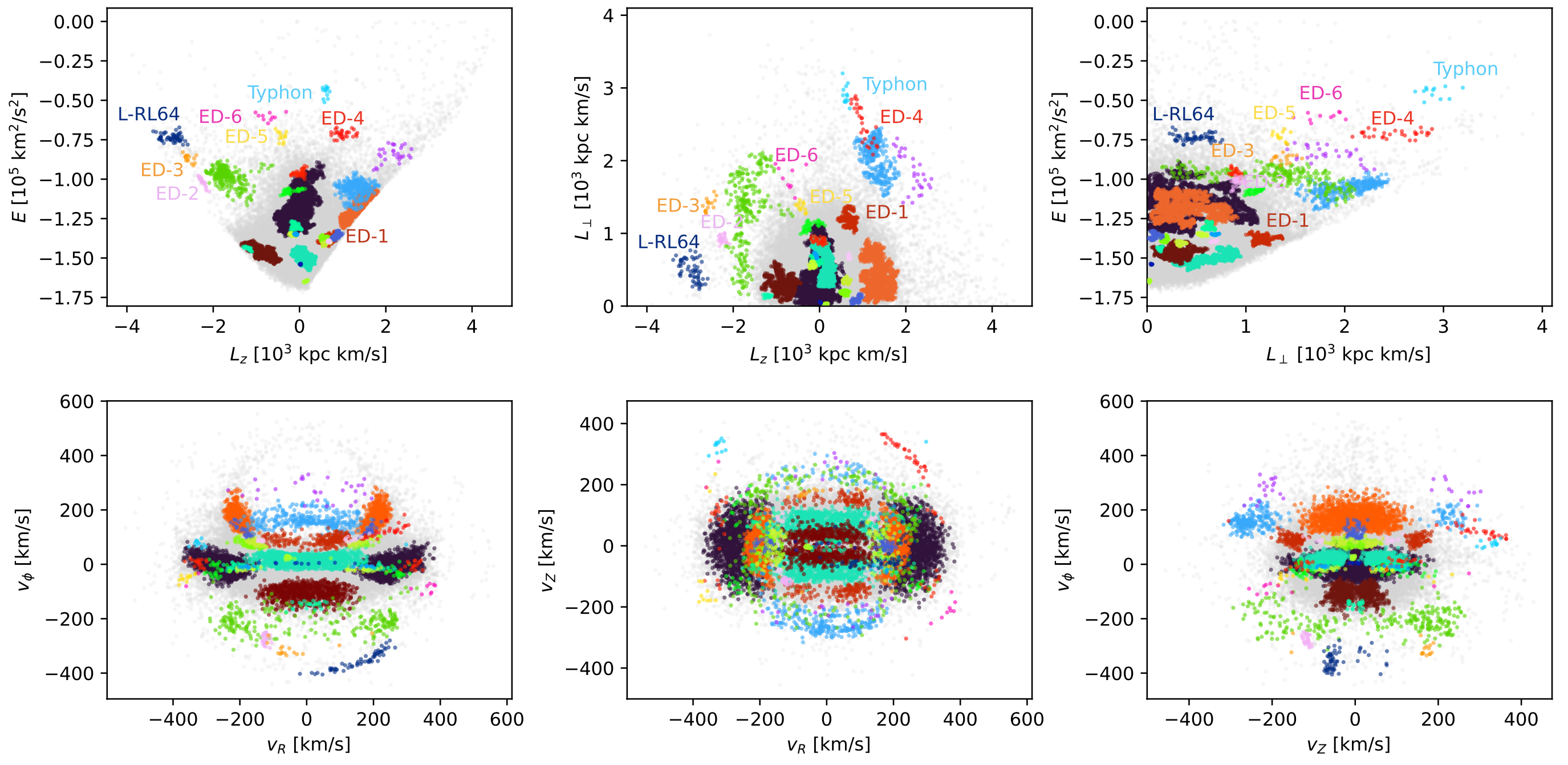}
\caption[]{Dynamical substructure identified by \citet{Dodd2023} in the local stellar halo using the \gaia\ DR3 data. {\it Top:} Projections of the energy $E$ and angular momentum $L$ space. {\it Bottom:} Projections of the velocity space in cylindrical polars. [Reproduced from \citet{Dodd2023}].}
\label{fig:Dodd2023_clumps}
\end{figure*}
\begin{figure}
\includegraphics[width=\linewidth]{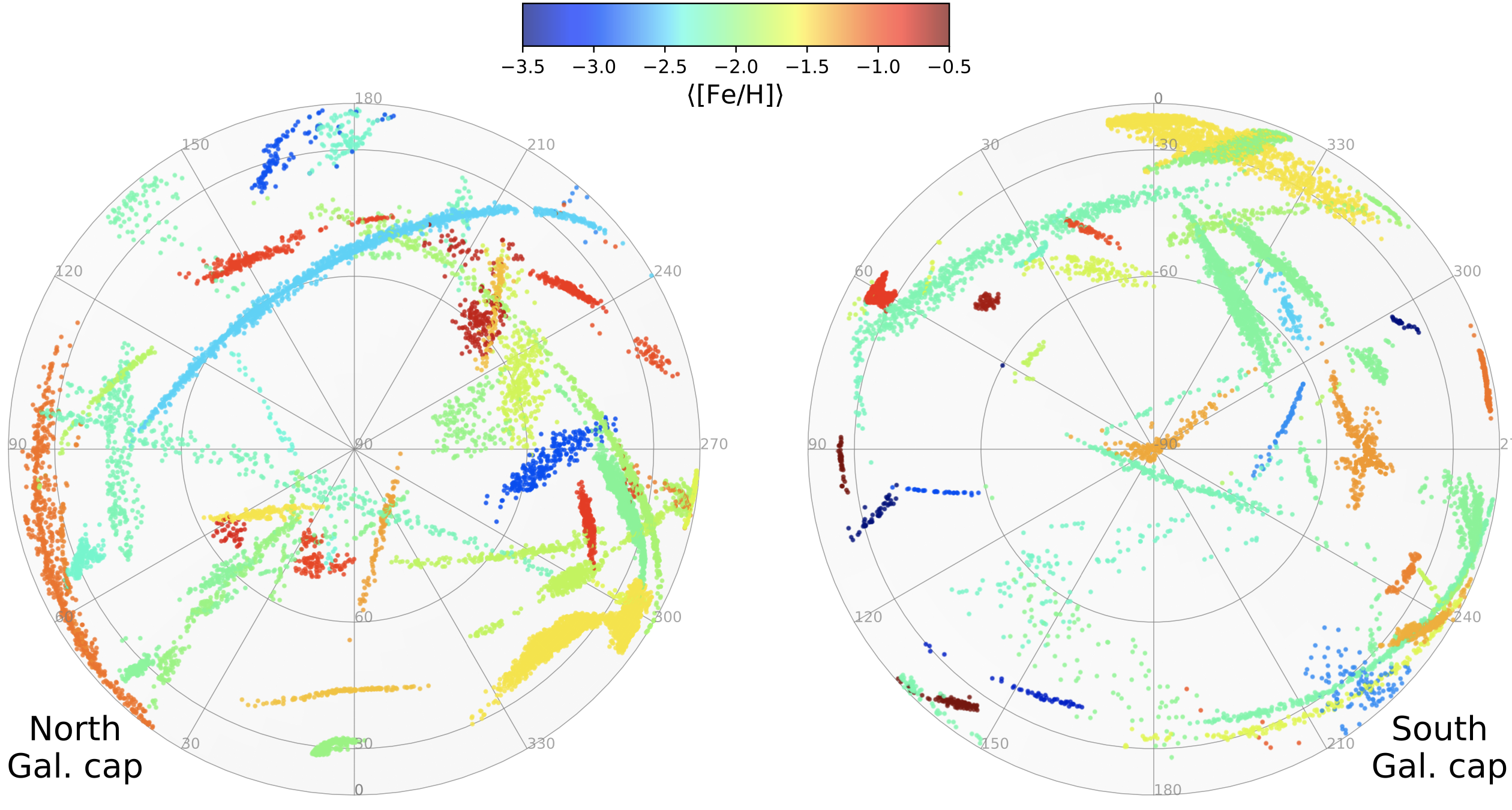}
\caption[]{Streams discovered using \texttt{STREAMFINDER} algorithm, colour-coded according to their metallicity. [Reproduced from \citet{Ibata2023}].}
\label{fig:Ibata2023}
\end{figure}

The identification by \citet{Helmi1999} of a handful of stars belonging to a stellar stream passing through the Solar neighourhood is a good example of the type of archaeological experiment that had appeared trailblazing and challenging before \gaia, but has been made effortless and almost mundane thanks to \gaia\ today. Since the pioneering study of \citet{Helmi1999} the methods to detect substructure have been sharpened \citep[see e.g.][]{HdZ2000,Harding2001,Knebe2005,Meza2005,Brown2005,BJ2005,Arifyanto2006,McMillan2008,Johnston2008,Gomez2010,Bovy2011,Sanderson2015,Lisanti2015,ANTOJA2015} whilst the available datasets have grown in size and richness \citep[e.g.][]{Yanny2000,Vivas2001,Newberg2002,Majewski2003,Clewley2006,Kepley2007,Klement2008,Morrison2009,Klement2009,Schlaufman2009,Smith2009,Xue2011,Ra2015,Janesh2016} but ultimately, the community had to wait for the arrival of the \gaia\ data to reveal the true wealth of accreted debris. This is particularly true for merger remnants that do not appear spatially coherent, although the detection of long, narrow streams has also evolved to a new level of sophistication thanks to the \gaia\ data \citep[see e.g.][]{Malhan2018_method,Malhan2018_charting,Ibata2019_abyss,Ibata2023}.

\begin{figure*}
\includegraphics[width=\linewidth]{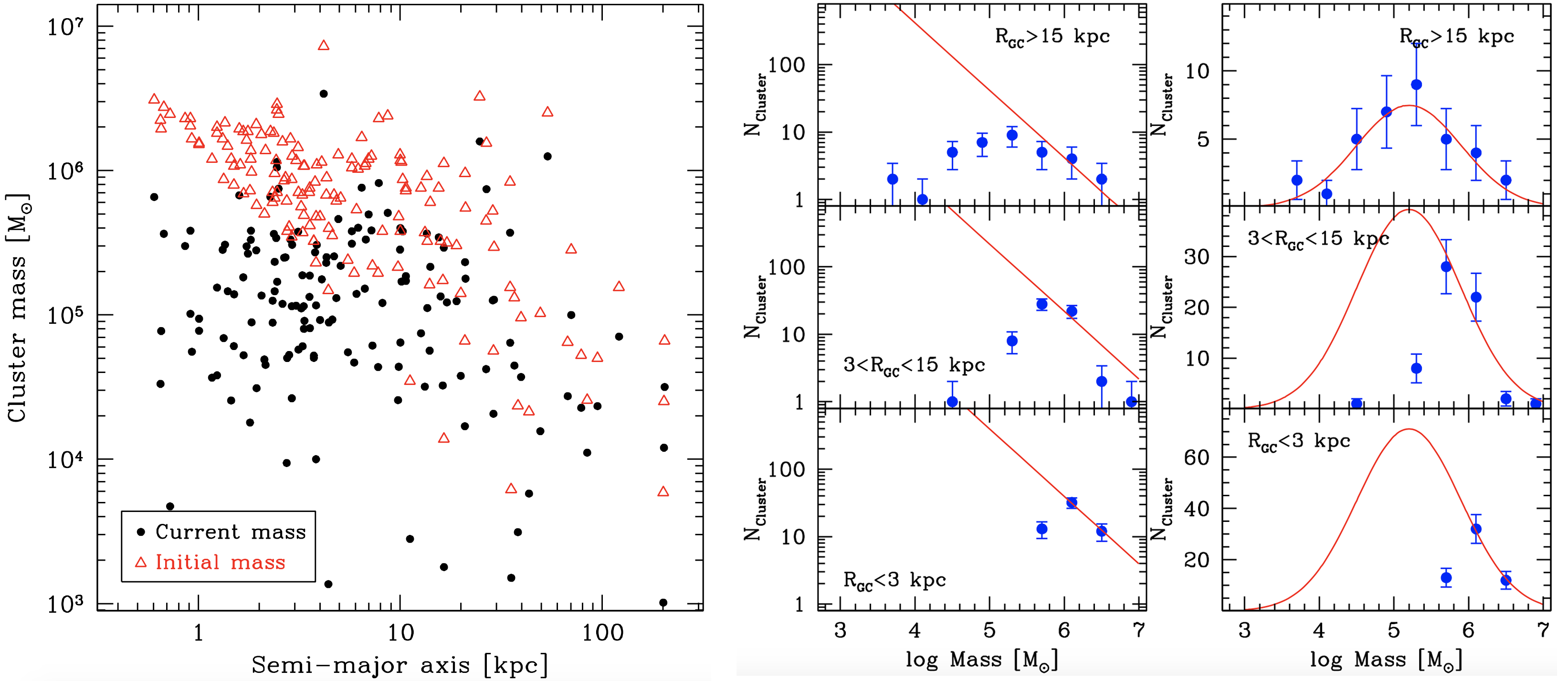}
\caption[]{{\it Left:} Current (black) and initial (red) Globular Cluster masses as a function of their orbital size. Assuming that the GC initial mass function is not a strong function of Galacto-centric radius, a large number of low-mass GC with apocentres inside $\sim$10 kpc appears to have been destroyed by the Galaxy. {\it Middle and Right columns:} Current (blue) and initial (red) mass functions of the Milky Way GCs in three different bins of Galacto-centric radius. Two different parametric forms for the initial mass functions are shown: power-law (middle) and log-normal (right). The integral of the difference between the current and the initial mass functions gives an estimate of the total number of GCs destroyed. [Adapted from \citet{Baumgardt2019}].}
\label{fig:Baumgardt_gcs}
\end{figure*}

Out of the box, the view on the halo substructure with \gaia\ DR1 appears underwhelming due to the minuscule number of halo stars in the shallow TGAS catalogue \citep[][]{Helmi2017,Myeong2017}.  Using the deeper SDSS-\gaia\ instead (see also Section~\ref{sec:gdr1}), shows the promise of the modern space astrometry: the halo in the Solar neighourhood is riddled with pieces of accretion events \citep[see][]{Myeong2018_substructure}. This study uncovers $\lesssim 10$ new clumps in velocity space, of which at least one, namely S2, had been known before \citep[][]{Helmi1999}. However, the integrals of motion space are better suited for substructure detection because phase-mixed debris can be smeared into arcs in velocity space; worse still, multiple wraps of the same stream might appear as distinct clumps. The immediate and clear improvement is demonstrated by \citet{Myeong2018} who use SDSS-\gaia\ DR1 and translate into the action space, detecting a factor of $\sim3$ more distinct sub-structures compared to their previous analysis of the same data \citep[see also][for a recent discussion of the pragmatic advantages of using action space]{Wu2022}. Once the \gaia\ DR2 was delivered, the hunt for the nearby halo merger remnants begun in earnest \citep[][]{Koppelman2018,Roederer2018,Koppelman2019,Ohare2020,Borsato2020,Necib2020,Limberg2021,Lovdal2022,Tenachi2022,Ruiz-Lara2022,Malhan2022,Viswanathan2023,Dodd2023,Ye2024}.

What is the origin of the multitude of halo substructures discovered by \gaia\ in the form of IOM clumps (see Fig. \ref{fig:Dodd2023_clumps}) or as coherent narrow streams (see Fig. \ref{fig:Ibata2023})? More specifically, how many of these sub-structures are tidal remnants of destroyed star clusters and how many come from dwarf galaxies? A direct estimate of the total number of star cluster streams can be found in \citet{Baumgardt2019}. These authors show that the initial mass function of the survived and currently observable Milky Way globular clusters is a very strong function of their Galacto-centric distance (see Fig. \ref{fig:Baumgardt_gcs}). Assuming that the cluster initial mass function should not depend strongly on the current location in the Galaxy, the total number of destroyed GCs can be estimated. Such an estimate depends on the postulated shape of the initial GC mass function. \citet{Baumgardt2019} consider two possibilities: the log-normal mass function, resembling the overall current distribution of the GC masses, and the power-law mass function, usually preferred as a model of the cluster mass distribution at birth. In the former case, roughly as many GCs are destroyed as there are currently detected ($N \approx200$), but in the latter, the total initial number is much larger ($N \approx10,000$). Irrespective of the initial mass function parameterization, the bulk of the destroyed GCs are located within the Solar radius, with a large fraction within $\approx5$ kpc from the Galactic centre, in the region matching in extent the in-situ Aurora component of the Milky Way.

Many of the stellar streams recently discovered with the {\tt STREAMFINDER} algorithm \citep[][and Fig~\ref{fig:Ibata2023}]{Ibata2021} have small intrinsic widths (by design) and are therefore suspected to have originated from disrupted or disrupting globular clusters \citep[see also][]{Martin2022_26}. Of these, the C-19 system is 
particularly striking. It is kinematically cold \citep[although see][for a follow-up study]{Yuan2022}, and displays chemical anomalies characteristic of GCs, with a large spread and correlations amongst abundances of Mg, Na and Al \citep[][]{Martin2022,Yuan2022}. What makes it stand out compared to known GCs is its extreme metallicity: at [Fe/H]$\approx-3.4$, C-19 would be the most metal-poor GC known to date in the Milky Way, by far \citep[][]{Martin2022}. C-19 could plausibly be a harbinger of a previously undetected population of metal-poor GCs, indicating that the so-called GC metallicity floor is, at least in part, due to a selection bias. Moreover, as pointed out by \citet{Martin2022}, given the typical stellar metallicity distribution function, the fractional contribution of GCs to star-formation in the C-19's host galaxy must be high. This is in agreement with the most recent constraints in the ancient Milky Way \citep[][]{BK2023_nitrogen} and in high-redshift galaxies \citep[][]{Mowla2024}.

\begin{figure}
\includegraphics[width=\linewidth]{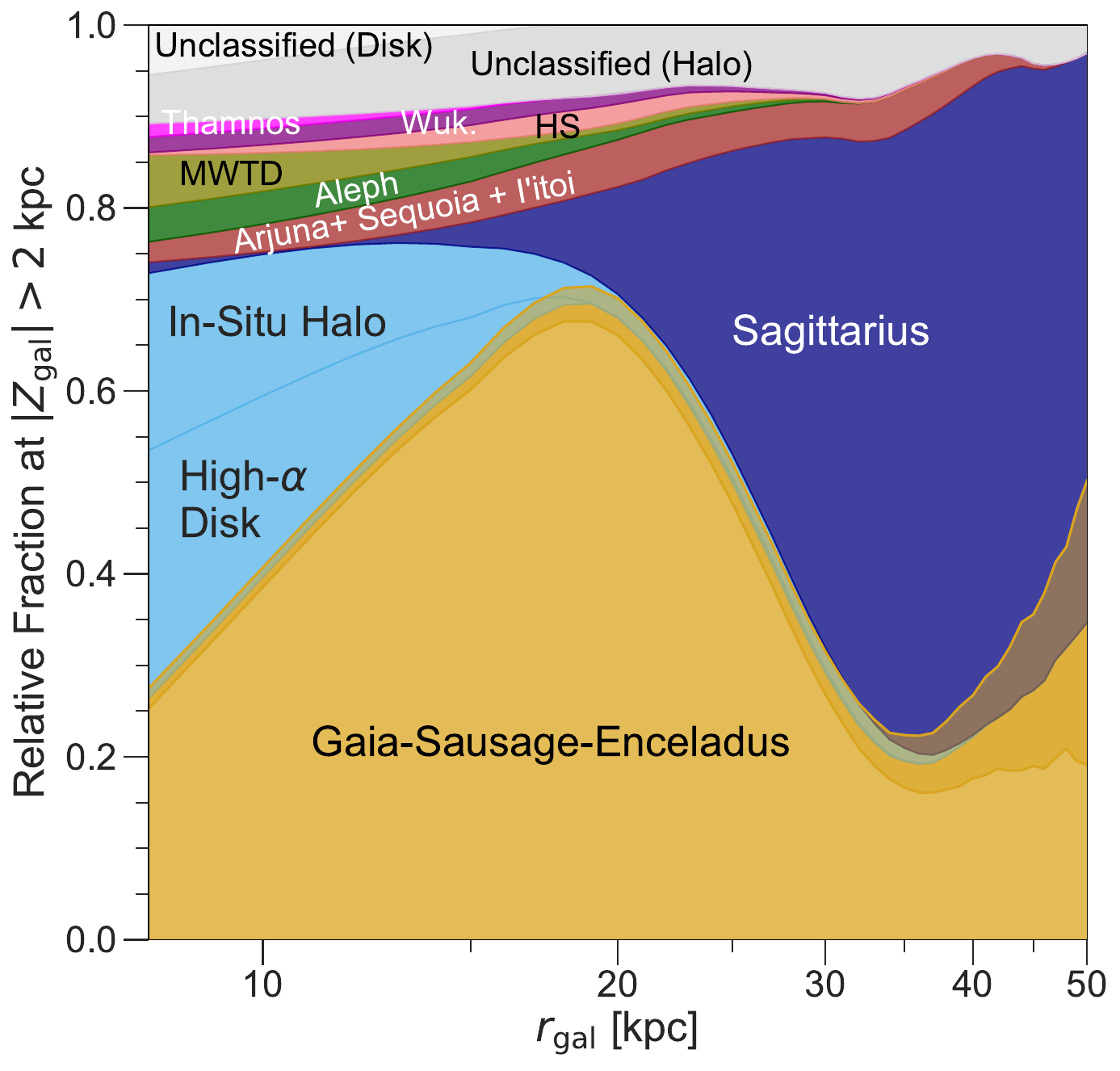}
\caption[]{Relative fraction of structures in the Milky Way as a function of Galactocentric distance. These fractions were estimated using \gaia\ and H3 survey data. Of the accreted structures, GS/E dominates in the inner halo, and Sgr dominates the outer halo. No other substructure contributes more than $\sim 5 \%$ beyond $r \gtrsim 5$ kpc.  [Reproduced from \cite{Naidu2020}]}
\label{fig:Naidu2020_fSH}
\end{figure}

In the aftermath of the \gaia\ data boon it is worth considering if we now have a good handle on \textit{all} (or at least most) of the relevant accreted structures in the halo. The short answer is almost certainly not, but we have likely uncovered the most significant progenitors by \textit{mass}. This deduction is clearly laid out by \cite{Naidu2020}, who use \gaia\ and spectroscopic data from the H3 survey to estimate the relative fraction of structures in the halo as a function of distance (see Fig. \ref{fig:Naidu2020_fSH}.) The authors argue that the GS/E dominates in the inner halo ($r \lesssim 20$ kpc), and the majority of stars in the outer halo belong to Sgr. Of the other structures considered in \cite{Naidu2020}, none contribute more than 5\% beyond $r \gtrsim 5$ kpc!

Have we discovered everything that `matters' in the stellar halo? By mass contribution we likely have, but there is still value in finding the remaining (presumably less massive) destroyed dwarfs. After all, the stellar halo itself is an argument for `quality over quantity', as only a small fraction ($\lesssim 1\%$) of the Galaxy's stellar content resides in this component. Similarly, although they may lack in stellar contribution, the lower mass stellar halo progenitors are compelling in their own right. For example, the surviving satellite luminosity function has proved to be a key test for the $\Lambda$CDM paradigm. The number counts of dwarf galaxies is intimately linked to several fundamental theories in cosmology and galaxy formation, such as the nature of dark matter, the epoch of reionization, and the early star formation in the lowest mass systems \citep[e.g.][]{Bullock2000, Lovell2014, Bose2018}. While the Milky Way satellite system is rightly viewed as a vital test for these theories, 
it is still only one `data point', and additional lines of evidence are required, either from external galaxies or from the Milky Way itself. This naturally leads to the luminosity function of \textit{destroyed} dwarf galaxies in the Milky Way. Could this be a crucial piece of observational evidence? First, we would need to overcome an important caveat: the stellar halo is seemingly overwhelmed by the GS/E and/or Sgr!

\begin{figure}
\includegraphics[width=\linewidth]{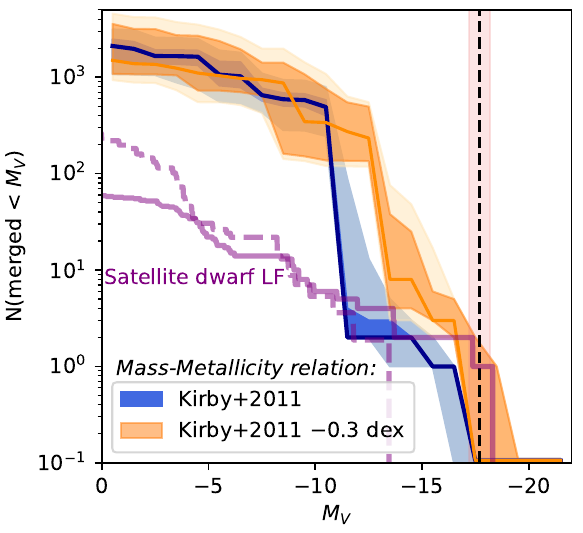}
\vspace{-15pt}
\caption[]{The estimated cumulative number of destroyed dwarfs in the Milky Way halo assuming a total stellar halo luminosity $M_{V} = -17.7 \pm 0.5$. The dark(light) shaded regions show the 16-84(1-99) percentiles, and the solid lines are the medians. The fiducial $z=0$ \citep{Kirby2011} stellar mass-metallicity relation is assumed with (orange) and without (blue) a -0.3 dex offset. This offset in the stellar mass-metallicity relation has been postulated to be more applicable to destroyed dwarfs \citep{Naidu2022}. The surviving dwarf satellite luminosity function is shown in purple (dashed line is the completeness-corrected luminosity function derived by \cite{Drlica-Wagner2020}, which does not include the LMC, SMC, or Sgr). [Reproduced from \cite{Deason2023}].
}
\label{fig:MDF-LF}
\vspace{-13pt}
\end{figure}

Theoretical predictions for the expected number of destroyed dwarf galaxies in Milky Way-mass haloes are relatively scarce, but notably \cite{Fattahi2020} show that the number of destroyed dwarfs generally exceeds the number of surviving dwarfs at all mass scales (see Figure 1 in \citealt{Fattahi2020}). Moreover, as expected, there can be significant halo-to-halo scatter due to varying mass accretion histories (see also \citealt{Deason2023}). Attempts to quantify the mass spectrum of accreted dwarfs in the Milky Way data are few and far between, but notably dynamical grouping \citep[e.g.][]{Callingham2022} and/or the use of metallicity distribution functions or chemical planes \citep[e.g.][]{Cunningham2022, Deason2023} are likely the best tools we have moving forward. \cite{Deason2023} use GMM modelling of the stellar halo MDF to estimate the luminosity function of destroyed dwarfs in the Milky Way (see Fig. \ref{fig:MDF-LF}). They make use of the stellar mass-metallicity relation and assume Gaussian MDF distributions for individual progenitors, thus assuming the overall stellar halo MDF is a mixture of MDFs from
smaller galaxies. Applying this method to a hodge podge of spectroscopic data in the Milky Way, complemented by \gaia\ DR3 astrometry, indicates that that the Milky Way stellar halo has $N \sim 1-3$ massive progenitors (with $L > 10^8L_\odot$) within 10 kpc, and likely several hundred progenitors in total.

\begin{figure*}
	\includegraphics[width=\linewidth]{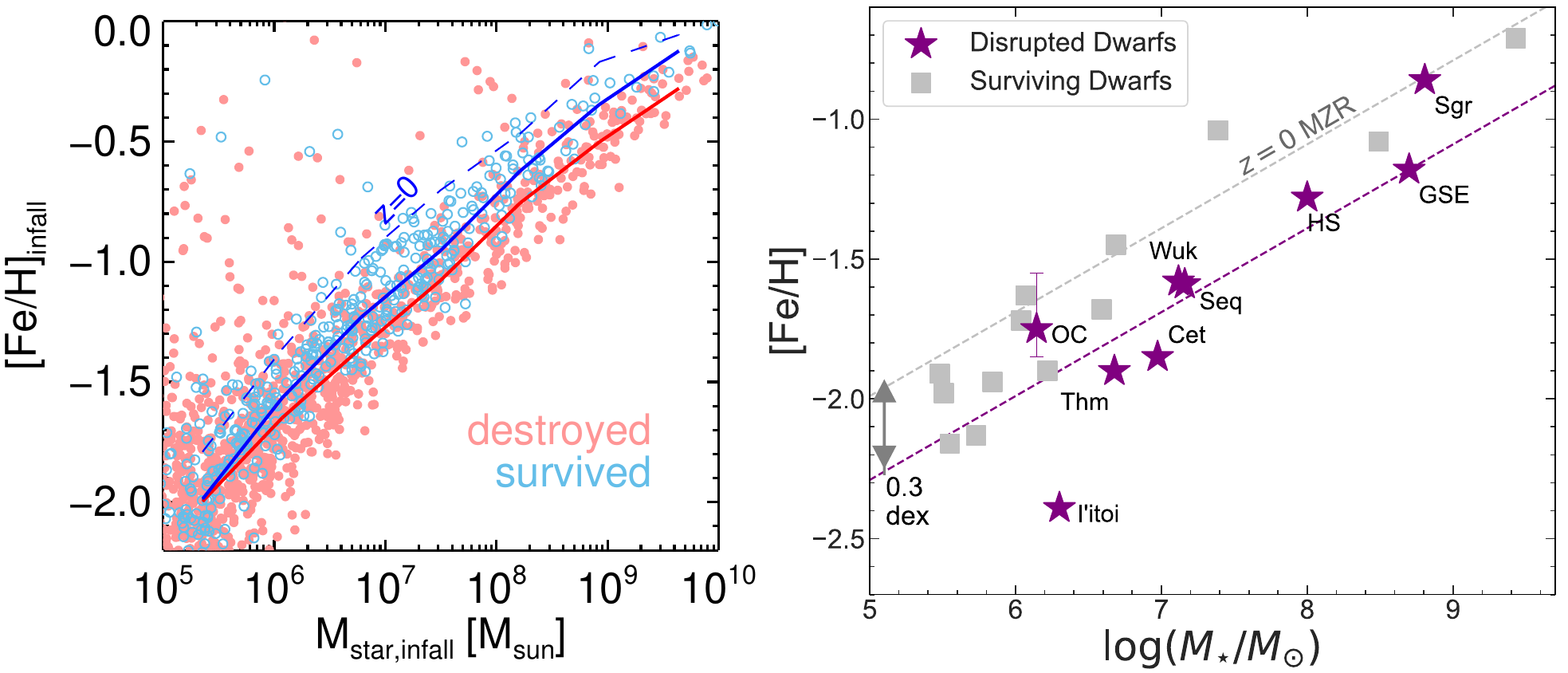}
	\caption{{\it Left panel:} [Fe/H]
          vs. stellar mass at infall for destroyed (red points) and
          surviving (blue points) dwarfs in the Auriga simulations \citep{Fattahi2020}. The average relations at fixed stellar mass are shown with the solid lines. For comparison, the dashed line shows the average relation for $z=0$ satellites. {\it Right panel:}  [Fe/H] vs. stellar mass for surviving (grey) and disrupted (purple) dwarfs in the Milky Way \citep{Naidu2022}. The grey line is the $z=0$ stellar mass metallicity from \cite{Kirby2013}. The purple line is offset from this relation by 0.3 dex, which is the median offset of the disrupted dwarfs. [Adapted from \cite{Fattahi2020, Naidu2022}].}
    \label{fig:z_mstar}
\end{figure*}

An important consideration when attempting to model the MDF of halo stars is the evolution of the MZR with time (see also earlier discussion about how this relates to the GS/E properties in Section \ref{sec:gse}). The redshift $z=0$ MZR for surviving satellites in the Milky Way is well-documented \citep{Kirby2013}, however, can we simply assume that this relation holds at all redshifts? Fig. \ref{fig:MDF-LF} shows how modelling of the stellar halo MDF is significantly impacted by the choice of MZR, and this is thus an important caveat in any such work. At the dwarf mass-scales, both observations and simulations seem to suggest that there is evolution of the MZR over time (see Fig. \ref{fig:z_mstar}). \cite{Fattahi2020} and \cite{Naidu2022} show that destroyed dwarfs are metal-poorer (by $\sim 0.3$ dex) than their surviving dwarf counterparts at fixed stellar mass. This difference is likely owing to the different star formation timescales of the two populations. Destroyed dwarfs are typically accreted earlier than surviving dwarfs, and thus to get to the same stellar mass as their $z=0$ counterparts, destroyed dwarfs must have formed their stars on a shorter timescale, and thus have less time to build-up their metals. In support of this picture, \cite{Naidu2022} also show that the destroyed dwarfs are more $\alpha$-enhanced than surviving dwarfs at fixed stellar mass. However, it is worth noting that the nature of some of the `destroyed dwarfs' used in the \cite{Naidu2022} work are still under debate (e.g. some of these apparently distinct substructures may be part of the GS/E). Thus, while there is reason to believe there is indeed evolution of the MZR at the dwarf mass scale, there is, as yet, no robust observational measure of this evolving MZR.

In order to probe down to the lowest mass scales and bypass the overwhelming signal of massive progenitors such as the GS/E we need (1) very large samples of halo stars with dynamical and/or chemical measurements (e.g. \citealt{Deason2023} argue that sample sizes of order $ \sim 10^5-10^6$ are needed to probe down to the ultra-faint dwarf mass scales using the stellar halo MDF), (2) to dig deeper into the inner halo of the Galaxy which harbours the earliest accreted and the least massive dwarfs \citep[see e.g.][]{Starkenburg2017_sims,Elbadry2018}, (3) to probe the outer halo which is likely populated by low-mass, not yet fully phase-mixed systems \citep[e.g.][]{Fattahi2020}, and (4) a focus on the lowest-metallicity halo stars (owing to the MZR). In principle, we need both large samples and low-metallicity stars, as the most massive dwarf galaxies can still contribute a significant amount of metal-poor stars \citep{Deason2016}. Fortunately, the coming years are ripe for increasing samples of distant halo stars owing to wide-field spectroscopic surveys such as DESI, WEAVE and 4MOST combined with \gaia. Furthermore, efforts are being made to exclusively build samples of very metal-poor stars \citep{Starkenburg2017}, which could prove vital for uncovering the lowest mass destroyed dwarfs. \gaia's own spectro-photometric data is already transforming our view of the low-metallicity end of the Galaxy's MDF. \gaia\ DR3's XP measurements have been used successfully by multiple groups to identify unprecedentedly large numbers of stars with [Fe/H]$<-2$ \citep[e.g.][]{Andrae2023,Martin2023,Zhang2023, Yao2024}. Time will tell whether we can harness the upcoming datasets to provide a `destroyed' version of the fundamental Galactic satellite luminosity function, but clearly \gaia\, and its potential future incarnations (e.g. \gaia NIR, \citealt{Hobbs2021}), will certainly lie at the heart of such efforts.

\section{Conclusions and future outlook}
\label{sec:concl}

The \gaia\ era has revolutionized the field of Galactic Archaeology. It has changed our view on how the Milky Way assembled, it has changed how we model the Galaxy, and it has opened our eyes to new research directions and explorations. However, the era is not over yet! In the coming years, the final \gaia\ data releases will become available and our \gaia\ all-sky map of the Galaxy will be complete. This, coupled with the complementary spectroscopic surveys such as DESI, WEAVE, 4MOST, SDSS-V, etc will provide the next phase in our Galactic exploration. Looking further ahead, pushing these missions to fainter magnitudes, and thus greater distances will be paramount (e.g. \gaia NIR).

In the wise words of Donald Lynden-Bell, one should always try to \textit{``follow the data''}. That, given the quality and abundance of data at our fingertips, should be advice well heeded. However, the interpretation of this data requires detailed modelling and simulations that will need to keep pace with the data in order to achieve the ultimate goal: a multi-dimensional, star-by-star `replay' of the Galaxy's formation.
\\
\\
\noindent
Below we list some of the most important and puzzling unanswered questions in Galactic Archaeology that future datasets and modelling efforts will be striving to address:

\begin{itemize}

\item What is the luminosity function of destroyed dwarfs in the Milky Way down to the lowest mass scales? Does this agree with the $\Lambda$CDM predictions?

\item How instrumental was the GS/E in the Galaxy's evolution? When did it merge, how much material did it bring, and how does it link to the disc, bulge, and bar formation?

  \item Why did the Milky Way disc form so early? Can this be reproduced in cosmological simulations?

  \item How can we distinguish the Galactic building blocks in the early pre-disc era of the Milky Way?

    \item What are the high-redshift contemporaries of the Galactic building blocks? How did they evolve?

      \item What is the origin of the stellar halo spin? Does this relate to the spinning up of the Milky Way disc?

\item Does the LMC-induced wake in the Galactic halo exist? What can this tell us about the nature of dark matter?

  \item What happened to the `satellites of satellites'? i.e. the satellites associated with the pre-infall massive dwarfs, like the LMC and GS/E. Can we distinguish this population from the `singly' accreted satellites? 
  \end{itemize}

\section*{Acknowledgments} 
AD is supported by a Royal Society
University Research Fellowship. AD acknowledges support from the Leverhulme Trust and the Science and Technology Facilities Council (STFC) [grant numbers
ST/X001075/1, ST/T000244/1]. VB acknowledges support from the Leverhulme Research Project Grant RPG-2021-205: ``The Faint Universe Made Visible with Machine Learning''.

We thank an anonymous referee for providing a constructive report.


\bibliographystyle{elsarticle-harv} 
\bibliography{review}






\end{document}